%% file: main.tex
\RequirePackage{fix-cm}
\documentclass[twocolumn]{svjour3}          % twocolumn
\usepackage{multicol}
\usepackage{dcolumn}
\usepackage{xspace}
\usepackage{amsmath}
\usepackage{graphicx}
\usepackage{amssymb}
\usepackage{hyperref}
\usepackage[T1]{fontenc} % if needed

\newcommand{\geantFour}{{\sc Geant4}\xspace}
\newcommand{\pythiaSix}{{\sc PYTHIA6}\xspace}
\newcommand{\fastjet}{{\sc FastJet}\xspace}

\newcommand{\xv}{\ensuremath{\mathbf{x}}\xspace}
\newcommand{\yv}{\ensuremath{\mathbf{y}}\xspace}
\newcommand{\ypv}{\ensuremath{\mathbf{y'}}\xspace}
\newcommand{\cv}{\ensuremath{\mathbf{c}}\xspace}
\newcommand{\zv}{\ensuremath{\mathbf{z}}\xspace}
\newcommand{\ind}{\ensuremath{\mathbf{1}}\xspace}

\def\figwidth{0.4\textwidth}

\newcommand{\E}{\mathbb{E}}

\newcommand{\citedata}{%
\cite{%qcd0,%
qcd30,%
qcd50,%
qcd80,%
qcd120,%
qcd170,%
qcd300,%
qcd470,%
qcd600,%
qcd800,%
qcd1000,%
qcd1400,%
qcd1800}%
\xspace%
}

\title{\boldmath Fast and accurate simulation of particle detectors using generative
  adversarial networks}

\author{Pasquale Musella \and Francesco Pandolfi}

\institute{Pasquale Musella \at
  ETH Institute for Particle Physics and Astrophysics,\\
  Otto-Stern-Weg 5, Zurich, Switzerland\\
  \email{pasquale.musella@cern.ch}
  \and
  Francesco Pandolfi \at
  INFN - Sezione di Roma\\
  Piazzale Aldo Moro 2, Rome, Italy.\\
  \email{francesco.pandolfi@cern.ch}
}

\begin{document} 
%\onecolumn
\maketitle
%\twocolumn
%\flushbottom

\begin{abstract}
  Deep generative models parametrised by neural networks have recently started to provide
  accurate results in modeling natural images. In particular, generative adversarial
  networks provide an unsupervised solution to this problem.
  In this work we apply this kind of technique to the simulation of particle-detector
  response to hadronic jets. We show that deep neural networks can achieve high-fidelity
  in this task, while attaining a speed increase of several orders of magnitude with respect
  to traditional algorithms.

  \keywords{Generative Adversarial Networks \and Deep learning \and High Energy Physics
    \and Simulation \and Fast simulation \and Jet images \and CERN open data.}
\end{abstract}

\section{Introduction}
\label{sec:introduction}

The extraction of  results from high energy physics data crucially relies on accurate
models of particle detectors, and on complex algorithms that infer the
properties of incoming particles from signals recorded in electronic sensors.
Numerical models, based on Monte Carlo methods, are used to simulate the interaction
between elementary particles and matter.\\ In particular, the \geantFour
toolkit~\cite{geant4_2003} features state-of-the art models and is employed to
simulate particle detectors at the CERN LHC. 
Reconstruction algorithms routinely used at collider experiments (see
e.g.~\cite{ATLAS_exp} and~\cite{CMS_exp}) are based on
estimators of particle trajectories and energy deposits. This information is subsequently
aggregated in order to reconstruct energy, type and direction of final state particles
produced by the collision of the primary beams. 

The CERN LHC complex will undergo a series of upgrades~\cite{HL_LHC} over the next ten years that
will allow collecting a dataset roughly 30 times larger than the one currently available.
The number of simultaneous interactions per bunch crossing in such a future dataset will increase by a
factor of about 4, compared to the present levels. It is estimated~\cite{HSF_whitepaper} that,
because of the larger volume and complexity of the data
the shortfall between needs and bare technology gains is about 4-fold
in computing power, if one assumes constant funding. 
This gap should therefore be bridged with faster, more efficient algorithms
for particle detector simulation and data reconstruction.

% a 4-fold
%improvement in the algorithm performance, at constant funding, will be needed for the LHC
%computing infrastructure to keep up with the larger volume and complexity of the
%data. This provides a very strong motivation for the development of fast algorithms for
%particle detector simulation and data reconstruction.

%In this work, we develop a generative model parametrised by a deep neural network that
%is capable of predicting the total experimental response to a hadronic jet, 
%which consists of both the simulation of the passage of particles through the 
%detectors, and the effect of the reconstruction and clustering algorithms.
In this work, we develop a generative model parametrised by a deep neural network, that
is capable of predicting the combined effect of particle detector simulation models and reconstuction algorithms to hadronic jets.
The results are based on samples of simulated hadronic jets
produced in proton-proton collisions at $\sqrt{s} = 7$~TeV that was published by the CMS
collaboration on the CERN open data portal. The dataset~\citedata is part
of the ``level 3'' category in the the High Energy Physics (HEP) data preservation
classification~\cite{DPHEP} and it contains the result of the \geantFour simulation
of the CMS detector and of the subsequent data reconstruction algorithms used by the CMS
collaboration.

Generative adversarial networks (GANs)~\cite{goodfellow_generative_2014} are pairs
of neural networks, a generative 
model and a discriminative one, that are trained concurrently as players of a minimax
game. The task of the generative network is to produce, starting from a latent
space with a fixed distribution, samples that the discriminative model tries to separate from
samples drawn from a target dataset. It can be shown~\cite{goodfellow_generative_2014}
that with this kind of setup the generator is able to learn the distribution of the target
dataset, provided that the generative and discriminative models have enough capacity
(it should be noted that this condition has actually been proven to hold for mixtures of
neural networks~\cite{gan_mix}, not for single ones; 
the interested reader may find additional information about GANs convergence
in~\cite{gan_mix,f_gan,w_gan1,w_gan2} and references therein.

Since they were first proposed, GANs have been applied to an increasingly large number of
problems in machine learning, mostly dealing with natural-image data, but not
only. Applications of adversarial networks were also proposed in the context of HEP,
mainly with two purposes: training of robust discriminators that are
insensitive to systematic effects, or uncorrelated from observables used for signal
extraction~\cite{Louppe:2016ylz,Shimmin:2017mfk,nips17_systs}, and for event generation
and detector simulation~\cite{deOliveira:2017pjk,Paganini:2017hrr,nips17_calo}. Similar
applications were proposed in the context of  cosmic ray experiments~\cite{Erdmann:2018kuh}.

We use GANs to train a generative model that generates the reconstruction-level detector
response to a hadronic jet, conditionally on its particle level content. We represent hadronic
jets as ``gray-scale'' images of fixed size 
centred on the jet axis, where the pixel intensity reflects the fraction of jet energy
deposited in the corresponding geometrical cell.
The architecture of the networks and the problem formulation, that can be classified as a
domain mapping one, are based on the image-to-image translation work described
in~\cite{pix2pix}. 
We introduce a few differences to tailor the approach to the generation of jet images: we
explicitly model the set of non-empty pixels in the generated images, which are much
sparser than in natural images; we enforce a good modelling of the total pixel intensity,
though the combined use of feature matching~\cite{improved_gans} and of a dedicated
adversarial classifier; and we condition the generator on a number of auxiliary features.
The use of feature matching on physics-inspired features was already
introduced in the HEP context in~\cite{Paganini:2017hrr}, while conditional GANs are
common in image generation problems (see, e.g.~\cite{cgan} and references
in~\cite{pix2pix}).
Previous work on the handling of sparsity in hadronic jet images was proposed
in~\cite{Paganini:2017hrr} and~\cite{deOliveira:2017rwa}, and it was based on the
engineering of high level auxiliary features.

Related work has recently been presented in~\cite{Paganini:2017hrr}
and~\cite{Erdmann:2018kuh}. % \\
The authors of~\cite{Paganini:2017hrr} proposed to use a deep convolutional neural network
to simulate calorimeter showers, thus aiming at modelling the particle interaction with the detector
medium. The solution that we explore here allows the largest reduction in computation
time, by predicting directly the objects used at analysis level, and thus reproducing the
output of both detector simulation and reconstruction algorithms. 
This philosophy is
similar to that of the parametrised detectors simulations~\cite{delphes} that are often
used in HEP for phenomenological studies, and that are very limited in accuracy. We show
that using a deep neural network model allows attaining accuracies that are comparable to
that of the full simulation and reconstruction chain. % \\
The approach of~\cite{Erdmann:2018kuh}, that studied the application of GANs to the
generation of air-showers, is more similar to ours, as it aims at predicting the
patterns reconstructed by the detectors, conditionally on the energy and type of the
primary particles.

\section{Inputs and problem formulation}
\label{sec:inputs_and_prob}
 
For this study we use simulated samples of hadronic jets produced by the CMS collaboration
and published on the CERN open data portal. In particular we take hadronic jets produced
in proton-proton collisions at $\sqrt{s} = 7$~TeV. These events feature state-of-the-art
characteristics in terms of simulation and reconstruction algorithms in HEP.
The events were generated with the \pythiaSix event generator~\cite{pythia6}, the CMS
detector response was simulated using \geantFour~\cite{geant4_2003}. Concurrent
proton-proton interactions (``pile-up'') were simulated, roughly reproducing the LHC
running conditions of 2011. The samples contain the results of the full CMS reconstruction
chain~\cite{CMS_exp}. 

In the input dataset, hadronic jets were clustered with the anti-kt
algorithm~\cite{antikt}, using the \fastjet library~\cite{fastjet} and a distance
parameter of 0.5. We used two sets of hadronic jets: those clustered from the list of 
stable particles produced by \pythiaSix, and those clustered from
the list of reconstructed particle candidates. In the following we term ``particle-level
jets'' the former, and ``reconstructed jets'' the latter.

Following standard practices in collider experiments, we employ a cylindrical system of
coordinates. The origin of the coordinate system is set to the centre of the CMS detector,
the z axis is chosen to be parallel to the beam line, and the x axis is chosen to point
towards the centre of the LHC ring. We indicate as $\phi$ and $\theta$ the azimuthal and
polar angles, respectively, and we define the pseudorapidity $\eta$ as $\log(\cot(\theta/2))$.

In each event, we select particle-level jets with a transverse component of the momentum above
20 GeV and with an absolute value of the pseudorapidity below 2.5. A search is then
performed in the reconstructed jet collection to find reconstructed jets satisfying\\
$\sqrt{ \Delta \eta^2 + \Delta \phi^2} < 0.3$, where $\Delta \eta$ and $\Delta \phi$
are, respectively, the difference between pseudo-rapidity and azimuth of the particle-level and
reconstructed jets. Pairs of reconstructed and particle-level jets satisfying these conditions
are considered in this study. As we are interested in the particle content of jets, 
no jet energy calibrations are applied to the reconstructed jets.

Jet images are constructed by opening a window of size $\Delta \eta\times\Delta \phi
= 0.3\times 0.3$ around the particle-level jet axis. The window is split into $32\times
32$ identical square pixels and the intensity associated to each pixel is proportional to
the total transverse momentum of the jet components contained in it, divided by the
transverse momentum of the particle-level jet. Roughly 80-90\% of the jet energy is
contained in the jet window that we considered, and the jet components not contained in
the window are used to fill the closest pixel of the image border.
The choice of studying the central part of the jet was aimed at limiting the data
size, while keeping sufficiently high spacial resolution. Increasing the window to the
$0.5 \times 0.5$ region would have in fact almost doubled the image size and
thus considerably increased the training time. This technical limitation will have to be
addressed by future work, but is not expected to significantly change the conclusion of
this work.

\begin{figure}[thb]
\centering
\includegraphics[width=\figwidth]{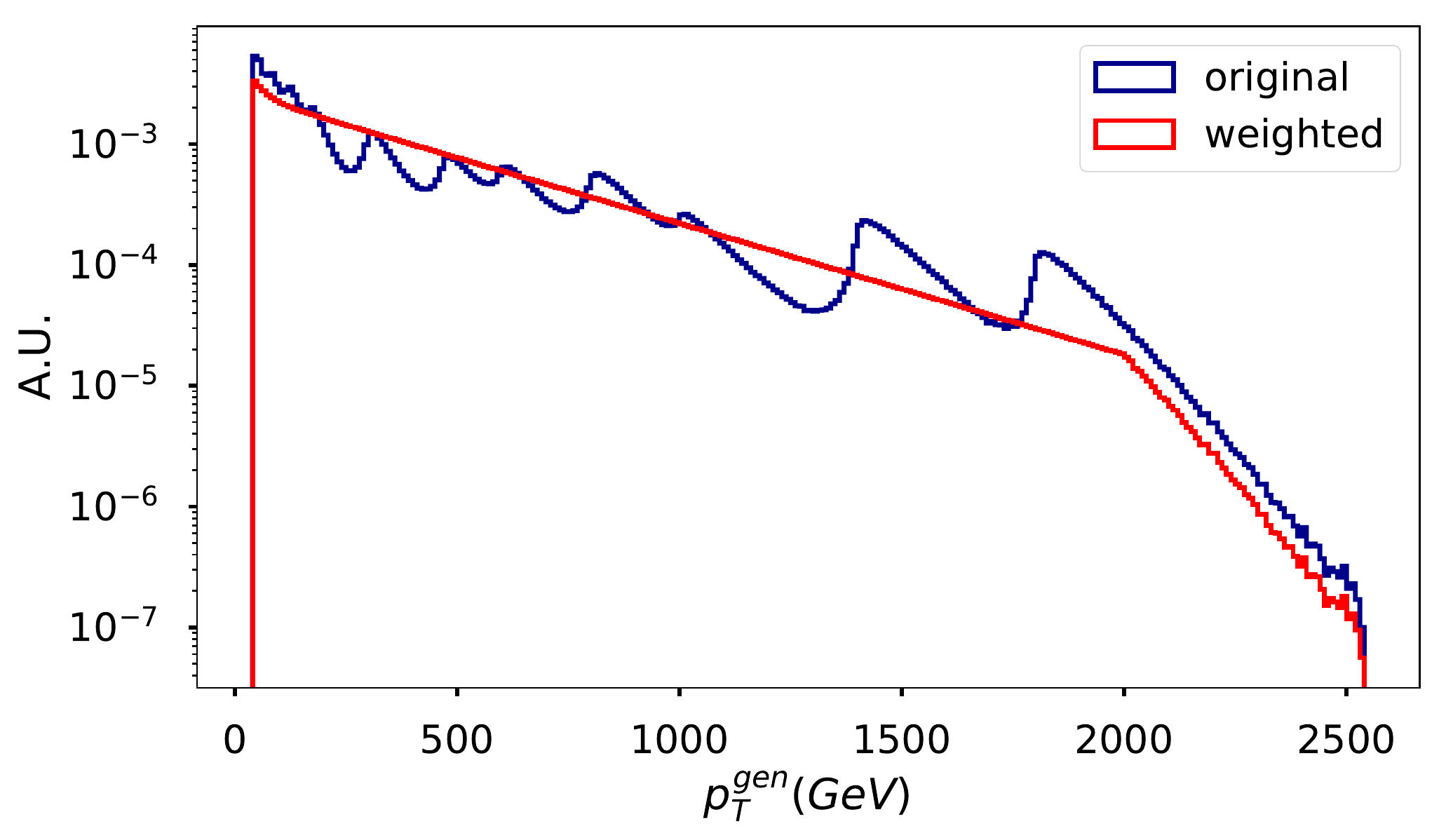}
\caption{\label{fig:gen_pt}
Distribution of the jet transverse momentum at particle level for the hadronic jets
in the input dataset.
The blue histogram shows the distribution as coming from the dataset, while the red
one shows the distribution obtained after applying the weights used for training.
}
\end{figure}

Each pair of jet images is further associated to four auxiliary features: the transverse
momentum ($p^{gen}_T$), pseudo-rapidity ($\eta^{gen}$) and azimuthal angle  ($\phi^{gen}$) of the
particle-level jet, and the number of pile-up interactions ($n_{PU}$) that were simulated in
the event under consideration. No further preprocessing or standardisation was
performed on the input data, which is available on the Zenodo information server~\cite{data}.

The distribution of $p^{gen}_T$ for the selected hadronic jets is shown by the blue
histogram in figure~\ref{fig:gen_pt}. The shape of the distributions is multi-modal
because the original dataset was split in several bins in the scale of the parton-parton
interaction. A set of weighting factors, as a function of $p^{gen}_T$ was applied to obtain
a falling distribution. The distribution obtained after applying such weighting
factors is shown by the red histogram.

\subsection{Notation and learning setup} 

We adopt the following notation: we denote by \xv and \yv the jet image at
particle and reconstruction level, respectively, we indicate with \cv the
set of auxiliary features, and we use \zv to denote a latent space of
uniformly distributed noise.
With this notation, our problem can be formulated as follows.
\begin{itemize}
\item Given:\begin{itemize} 
        \item [ ] $(\cv,\xv) \sim p_{cx}( \cv, \xv )$
        \item [ ] $\yv \sim p_y(\yv | \xv, \cv)$ 
        \item [ ] $\zv \sim p_z(\zv) = U(\zv)$;
        \end{itemize}
        where $p_{cx}$ and $p_y$ are the input-data distributions, and $U$ indicates the
        uniform distribution.
\item We want to construct a function $G$, such that\begin{itemize}
        \item[ ] $\ypv = G(\zv, \xv, \cv) \sim p_y( \ypv | \xv, \cv )$.
        \end{itemize}
\end{itemize}

We note that the \cv and \xv variables sets play, from a mathematical point of view, an
identical role in the problem, as we want to condition the output of $G$ is conditional
on the union of the two. The reason to separate them is mostly conceptual, as \xv
represents the image data associated with a jet, while \cv parametrises information about
 jet kinematics and the environment. Furthermore, the two sets of variables are treated
differently by the neural network architecture, as detailed in the following.

The function $G$ is a generative model that approximates the combined response of
particle detector simulation and reconstruction algorithms to hadronic jets.
Following the GAN paradigm, we look for a solution to this problem by introducing a
discriminative model $D$ and setting-up a minimax game between the two models, with value
function $V(G,D)$ defined as:

\begin{equation}
\begin{array}{ll}
\label{eq:VGD}
V(G,D) ~  = & \min_G  \max_D \{ \E_{(\cv,\xv,\yv)  % \sim p_c(\cv)  \cdot p_x(\xv | \cv) p_y(\yv | \xv,
      % \cv) 
     } \left[ \log( D(\yv,\xv,\cv) ) \right ] \\[1ex]
   & ~~ + ~ \E_{(\cv,\xv,\zv)  %\sim  p_c(\cv) \cdot p_x(\xv | \cv), \zv \sim p_z(\zv) 
     } \left[
        \log( 1 - D(G(\zv,\xv,\cv),\xv,\cv) ) \right] \large \}
%\right 
% & \} 
\end{array}
\end{equation} 

While the problem could in principle be solved using the GAN setup alone, we
inject additional information in order to stabilise and speed-up the
convergence. In particular, we take into account two facts:
\begin{enumerate}
\item $G$ and $\yv$ should match on average;
\item $\yv$ is very sparse (on average, roughly 3\% of the pixels have non-zero values).
\end{enumerate}

The authors of~\cite{pix2pix} show that the first requirement can be efficiently satisfied
by adding an $L_2$- or $L_1$-norm term to the loss function. We adopt this approach,
using in particular an $L_2$-norm term.

To explicitly take into account the second requirement, we modify the structure of the
generator, by increasing the depth of its output: one channel is used to model the pixel
intensity, while a second channel, to which we refer as a ``soft-mask'', models the probability
of a pixel to be non-zero. We denote the two channels as $G_0$ and $G_1$, and we modify
the generative model loss function by adding a term of this form:
\begin{equation}
    \label{eq:Pix}
    \begin{array}{ll}
      \lambda & {\rm {\cal L}_{P}}(G) ~ =  \lambda ~ \E_{ \yv=0 } \left[ - \log(G_0(\zv,\xv,\cv)) \right ] \\
                                          & + ~ \lambda ~ \E_{ \yv>0 } \left[ - \log(1 - G_0(\zv,\xv,\cv)) + 
                                            \left( G_1(\zv,\xv,\cv)-\yv \right )^2 / 2 \right ]  \\
    \end{array}
\end{equation}

where $\lambda$ is the associated hyperparameter.

To generate images, we sample the soft-mask probabilities to create a ``hard-mask''
binary stochastic layer $G_1'(\zv) = \ind_{\zv < G_1}$, where $\ind$ is the so-called indicator
function. The GAN value function in eq.~\ref{eq:VGD} becomes $V(G_2, D)$, where $G_2$ is
defined as $G_0 \cdot G_1'$, and the differentiability is preserved by replacing $G_1'$
with $G_1$ during back-propagation~\cite{hinton_bsn}.

Finally, we enforce a good modelling of the total image intensity, which is proportional
to the reconstructed jet energy, with two additions:
\begin{itemize}
\item we add an extra term to the generative model loss function, proportional to the mean
    squared error of the total image intensities:
    \begin{equation}
        \label{eq:Tot}
        \tau  {\rm {\cal L}_{T}}(G_2) ~ =  \tau ~ \E \left[ 
            \left( I( G_2(\zv,\xv,\cv) ) - I( \yv ) \right )^2 \right ]
    \end{equation}
    where the operator $I$ computes the total intensity of the jet images and $\mu$ is the
    associated hyperparameter.
\item We introduce a second discriminative model $D_T$ that receives as input the total
    reconstructed jet image intensities, and the
    auxiliary features \cv, and we set-up an additional minimax game, whose importance is
    controlled by the $\mu$ hyperparameter, with value function
    $V_T(G_2,D_T)$, defined as below:
    \begin{equation}
        \begin{array}{ll}
          \label{eq:VGD}
          \mu & V_T(G_2,D_T)  =  \\[1ex]
                         & \mu \min_{G_2}  \max_{D_T} \{ \E \left[ \log( D_T(I(\yv),I(\xv),\cv) ) \right ] \\[1ex]
                         & ~~ + ~ \E \left[
                        \log( 1 - D_T(I(G_2(\zv,\xv,\cv)),I(\xv),\cv) ) \right] \large \}
% & \} 
\end{array}
\end{equation} 
\end{itemize}

%% Two additional hyperparameters, $\tau$ and $\mu$, are introduced to tune the strength of
%% the two contributions, which read, respectively as.

%% MD summarize the loss function?

\subsection{Model architecture}

%% MD a scheme like in your slides would help

\begin{figure}[thb]
\centering
\includegraphics[width=\figwidth]{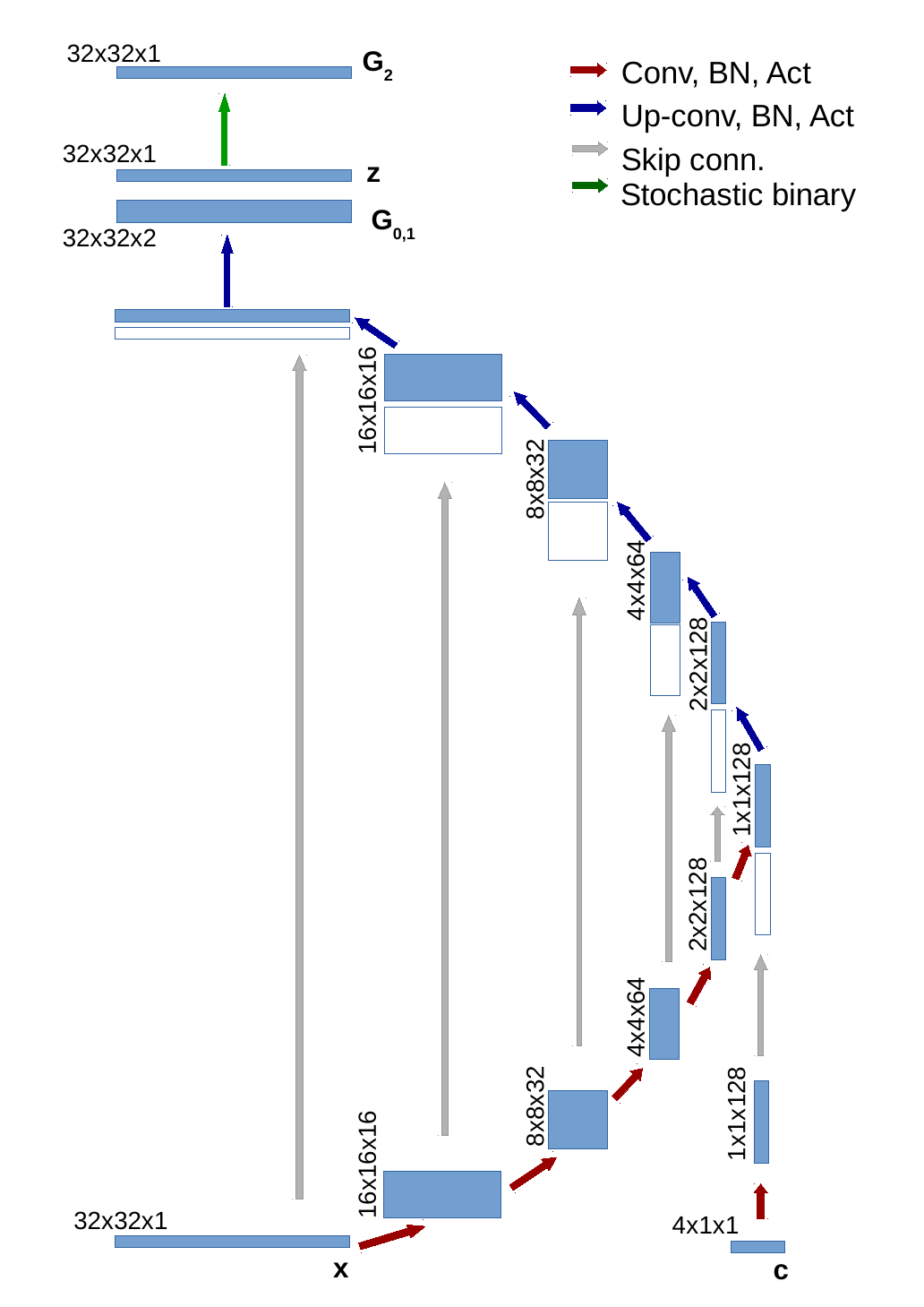}
\caption{\label{fig:unet}
Graphical summary of the generator network. Boxes shows the data representation, while
arrows represent operations. Input and output nodes annotated with bold-text labels.
On the right side of the diagram, white and blue boxes are concatenated before applying
up-convolution operations. Convolutions applied to the \xv node and its daughters have a
filter size of 3x3, while a 1x1 convolution is applied to the \cv node. All
up-convolutions use 3x3 filters. 
%% Distribution of the jet transverse momentum at particle level for the hadronic jets
%% in the input dataset.
%% The blue histogram shows the distribution as coming from the dataset, while the red
%% one shows the distribution obtained after applying the weights used for training.
}
\end{figure}

The generative model and the discriminative model $D$ are implemented as convolutional neural
networks~\cite{lecun-98}. For the generator we adopt the so called ``U-net''
architecture~\cite{Unet}, that consists of an encoding section followed by a decoding one,
with additional skip connections linking encoding and decoding layers with the same
spatial dimension. 
The input images are first fed into a batch normalisation layer~\cite{BN}, which
allows running the network on non-standardised inputs.
Afterwards, we use 5 encoding layers and 5 decoding ones. Each encoding layer
consists of a convolutional unit, followed by a 
batch-normalisation one and by a leaky ReLU activation~\cite{LReLU},
with a slope of 0.2 in the negative domain. The encoding filters size is chosen to be of
3$\times$3, with a stride of 2 %and a ``same'' padding option 
in order to reduce the
representation width. The number of filters is set to 16 for the first layer and
it is doubled at each step.
The decoding layers comprise a concatenation unit to implement the skip connections,
followed by up-convolutional units, batch normalisation and leaky ReLU activation
ones. Drop-out units are also employed in the first two layers of the decoding section.
At each step in the decoding section, the depth of the representation is halved, while its
width is doubled. This is achieved by decreasing the number of convolutional filters,
while appropriately choosing the stride and padding parameters.
Auxiliary conditional features are injected in the architecture as follows: they are first passed
through a batch normalisation unit and then into a 1$\times$1 convolution unit whose
output matches the depth of the last encoding layer; the 1$\times$1 convolutional unit
output is subsequently concatenated with that of the last encoding convolution.
Noise can be injected into the architecture at the same level. However, we obtained better
results by feeding noise only in the form of a stochastic sampling of the output
soft-mask.
Figure~\ref{fig:unet} shows a graphical summary of the generator architecture.

The discriminative model $D$ uses 4 layers, comprising 3$\times$3 convolutional filter
units, batch normalisation units and leaky ReLU activation ones. Stride and padding are
tuned in such a way that the largest field of view of the convolution layers is of
13$\times$13 pixels. The model acts as a ``patch-GAN''~\cite{pix2pix}, i.e. it is only
sensitive to the local structure of the jet images. 
The convolutional layers are followed by a fully-connected layer with a sigmoid
activation function. The auxiliary variables are treated similarly to what is done in the
generative model and are injected at the input of the fully connected layer.

The model $D_T$ takes as input the total intensities for \yv (or $G_2$) as well as
\cv, and is parametrised as a feed-forward fully connected neural network with 4
layers, using dense units, batch normalisation and leaky ReLU activations. The 
fully connected layers have widths of 64-64-32-16 and are followed by an output layer with
a sigmoid activation function.

\begin{figure}[htb]
\centering
\includegraphics[width=\figwidth]{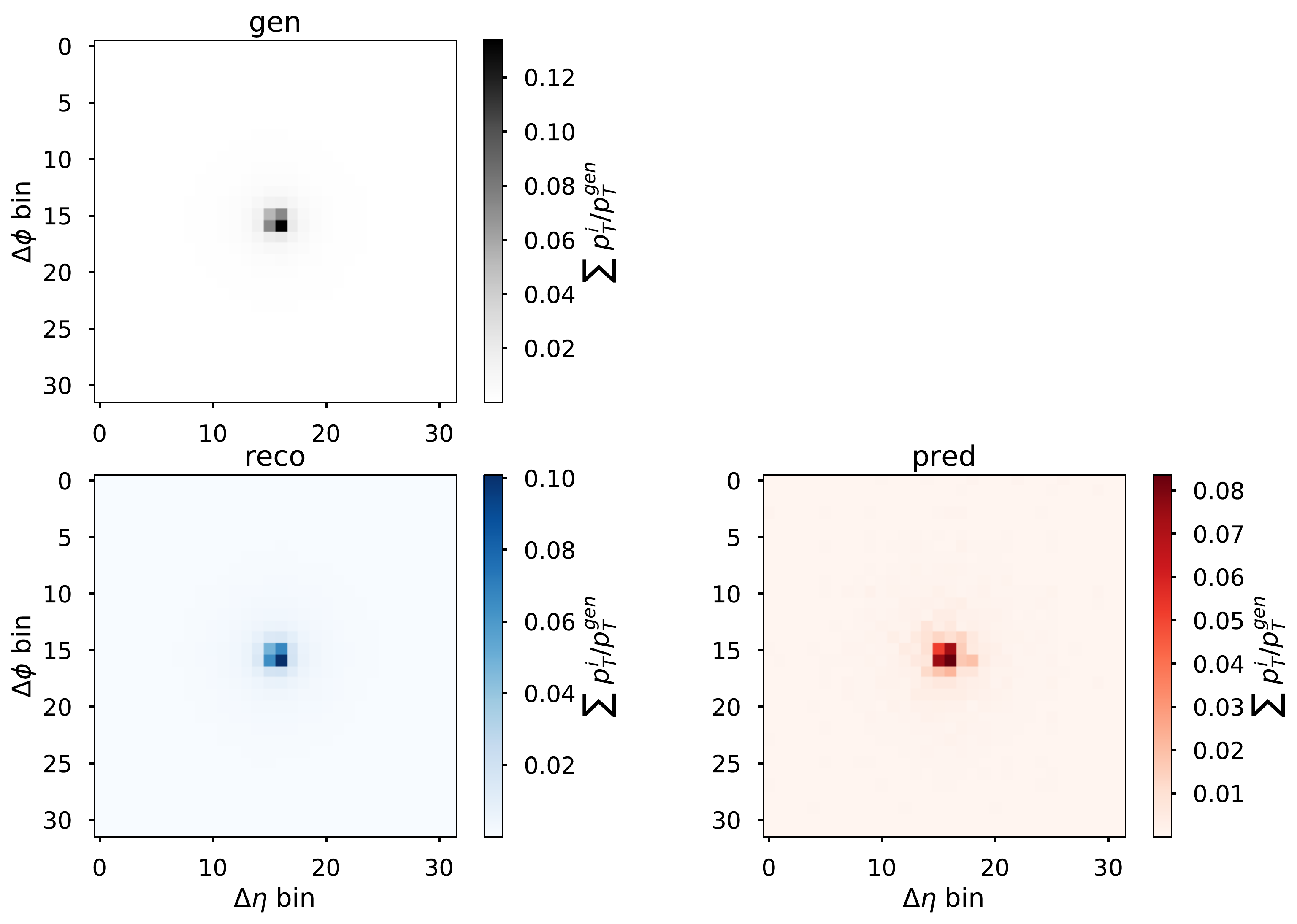}
\caption{\label{fig:avg_img}
Average jet images obtained at the particle-level (``gen''), after detector simulation and
reconstruction (``reco''), and those predicted by the generative model (``pred'').
}
\end{figure}

We did not perform a formal optimisation of the neural networks architecture, but we picked
a particular set of values after exploring the parameter space in terms of width and depth
of the networks, based on two factors: the performance of the model and the computational
times required. 

\begin{figure}[hbt]
\centering
\includegraphics[width=\figwidth]{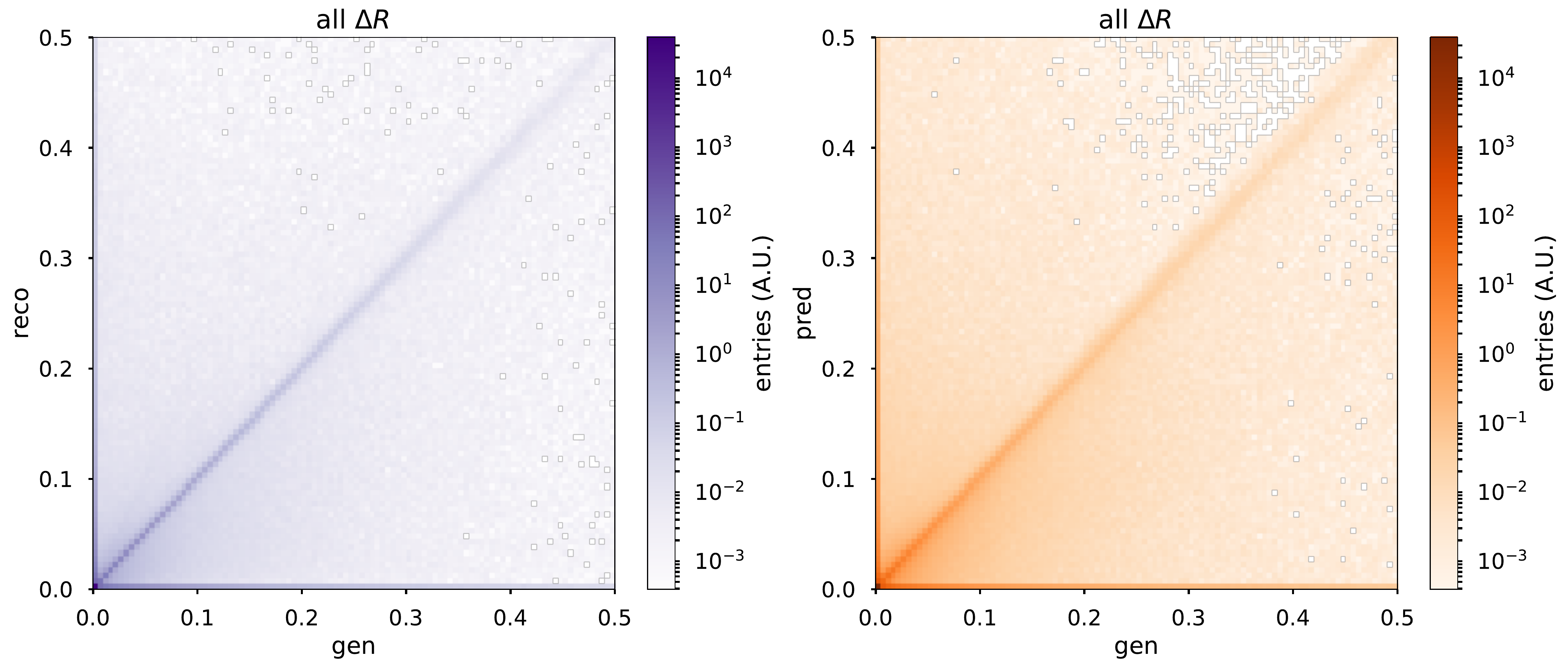} 
\caption{\label{fig:scatter}
Joint distribution of the particle-level (``gen'') and reconstruction level pixel
intensities. The left plot shows the distribution of the input data, while the right plot
is obtained using the generative model.
}
\end{figure}

A software package implementing the neural networks' instantiation and training is
openly available in~\cite{code}.

%\afterpage{\clearpage}

\section{Results}

The models were trained on two million jet images extracted from the dataset described in
section~\ref{sec:inputs_and_prob}. The TensorFlow~\cite{tf} framework, and
the Keras~\cite{keras} high level interface were used to implement and train the
models. Computing resources from the Piz Daint Cray supercomputer located at the Swiss
Centre for Supercomputing were used to obtain the results that we present here.
Two independent Adam~\cite{adam} optimisers were employed for parameter sets of the
generative and discriminative models. 
NVIDIA Pascal P100 GPUs  were used to accelerate the computations
and the models were trained for 10-20 epochs, which were sufficient to achieve
convergence. The training time for these models was around 1 hour per epoch. Inference ran  
at roughly 100Hz on Intel Xeon CPUs and at roughly 10kHz on NVIDIA Pascal P100 GPUs, which 
are to be compared to the typical time scale for event simulation and reconstruction,
i.e. $10^{-1}$--$10^{-2}$~Hz.

\begin{figure}[t]
\centering
\includegraphics[width=0.2\textwidth]{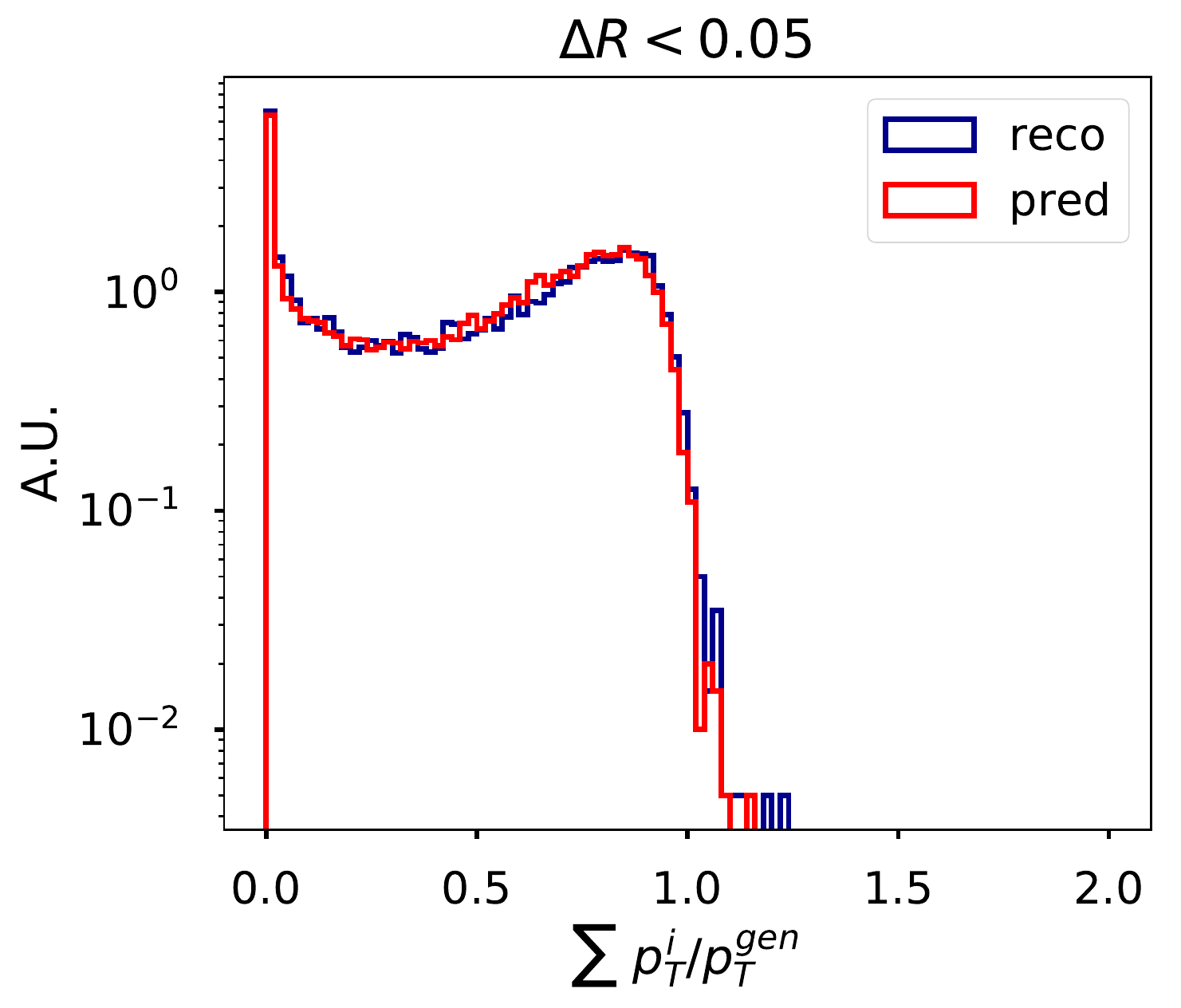} 
\includegraphics[width=0.2\textwidth]{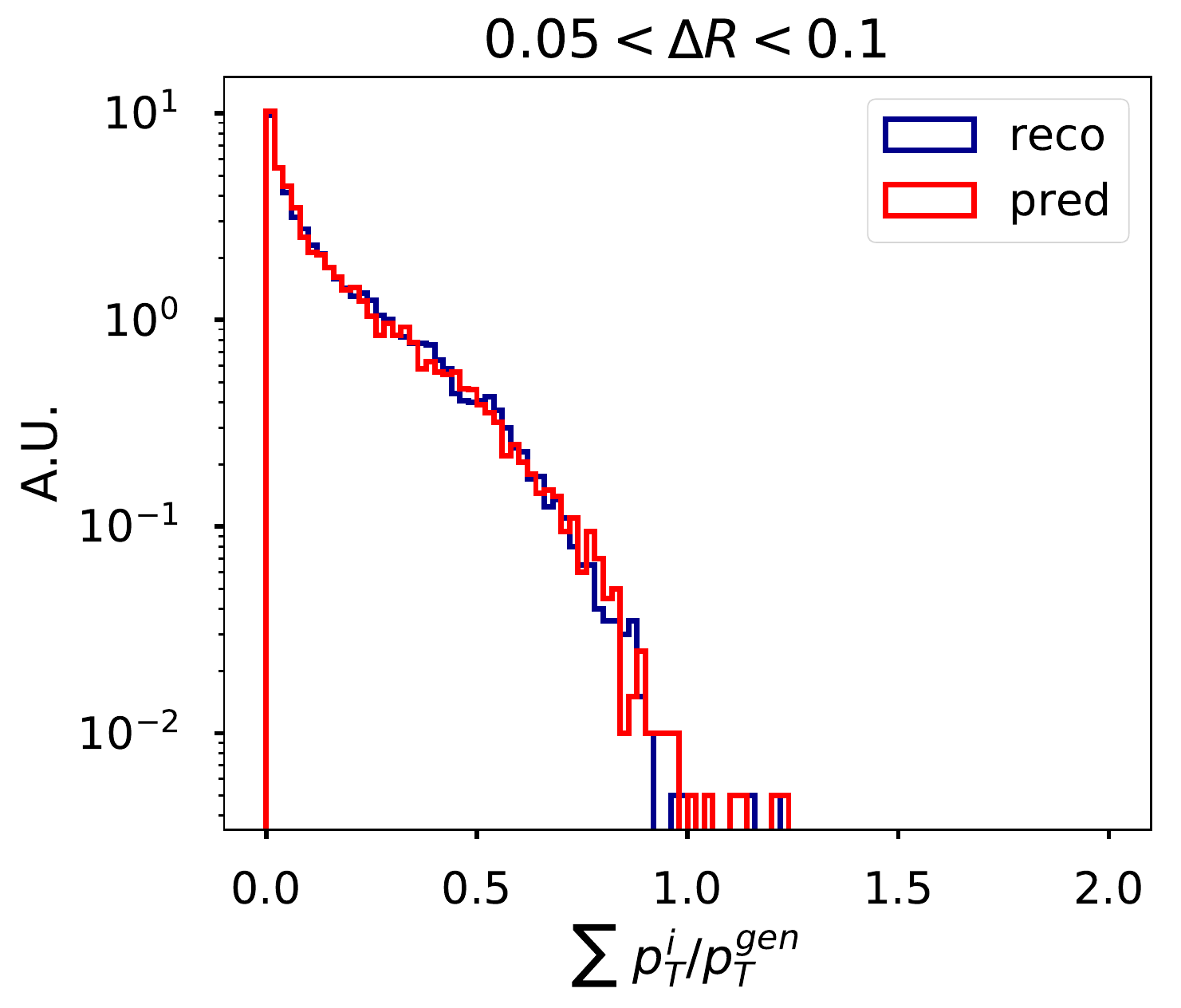} \\
\includegraphics[width=0.2\textwidth]{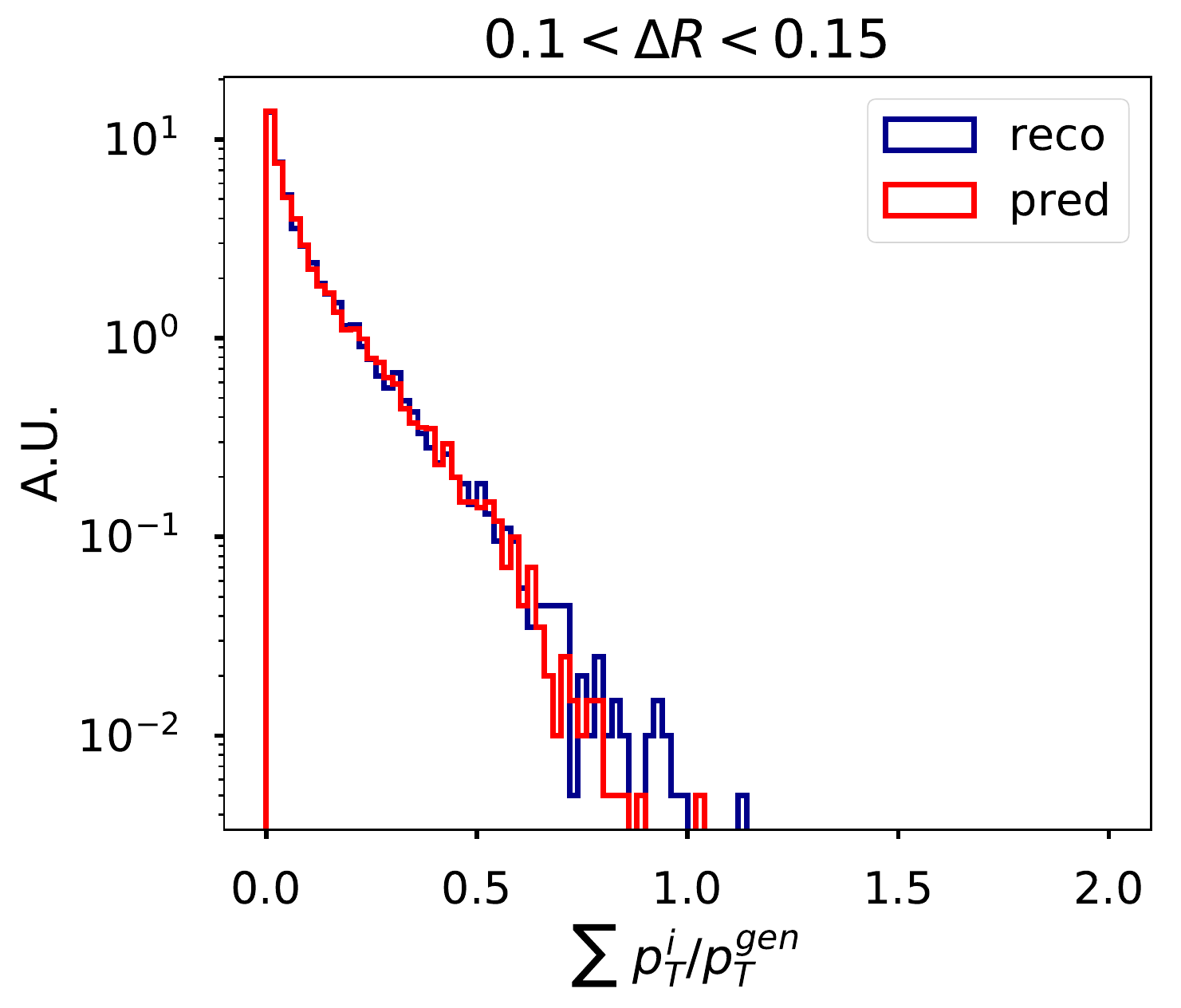} 
\includegraphics[width=0.2\textwidth]{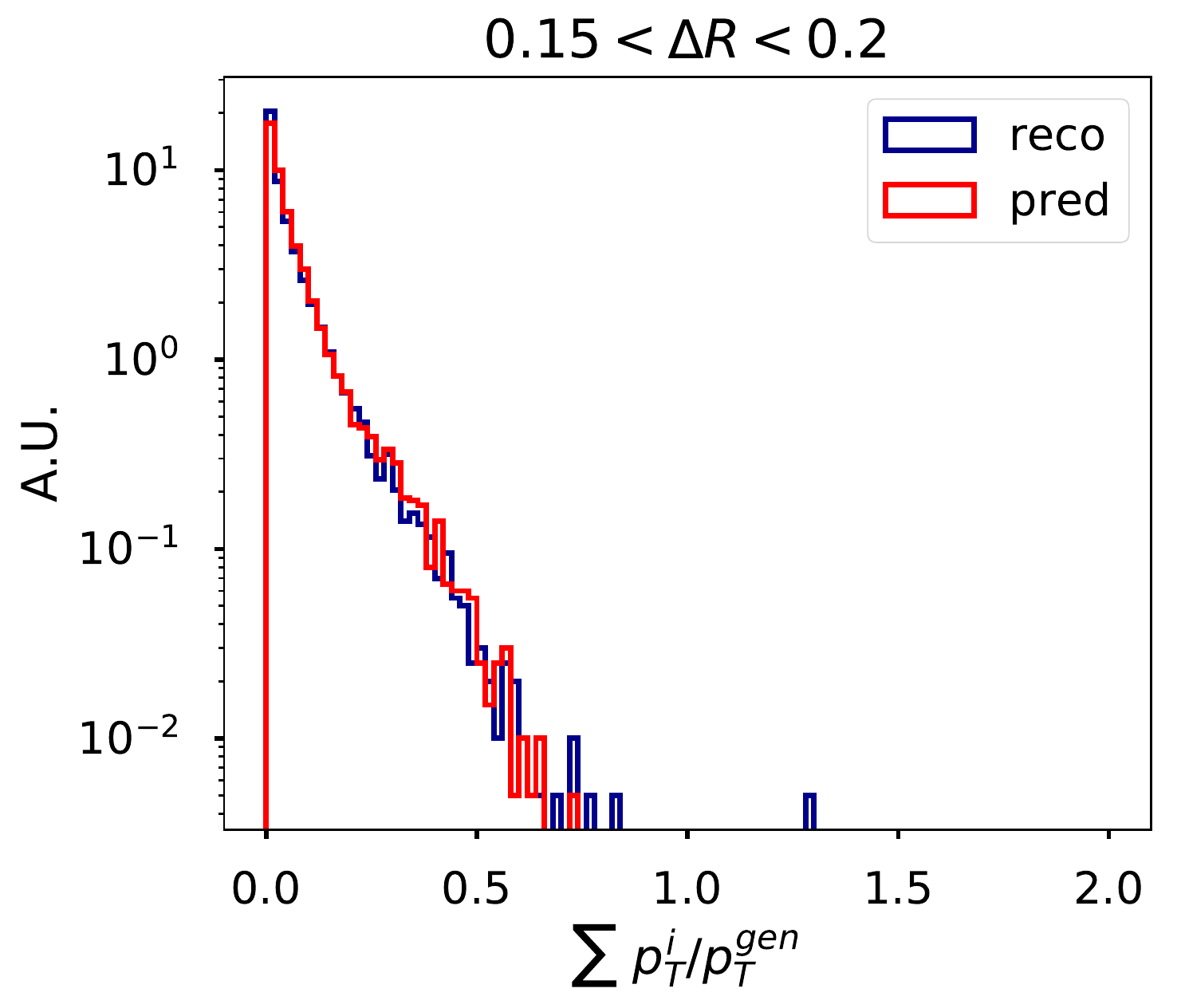} \\
\includegraphics[width=0.2\textwidth]{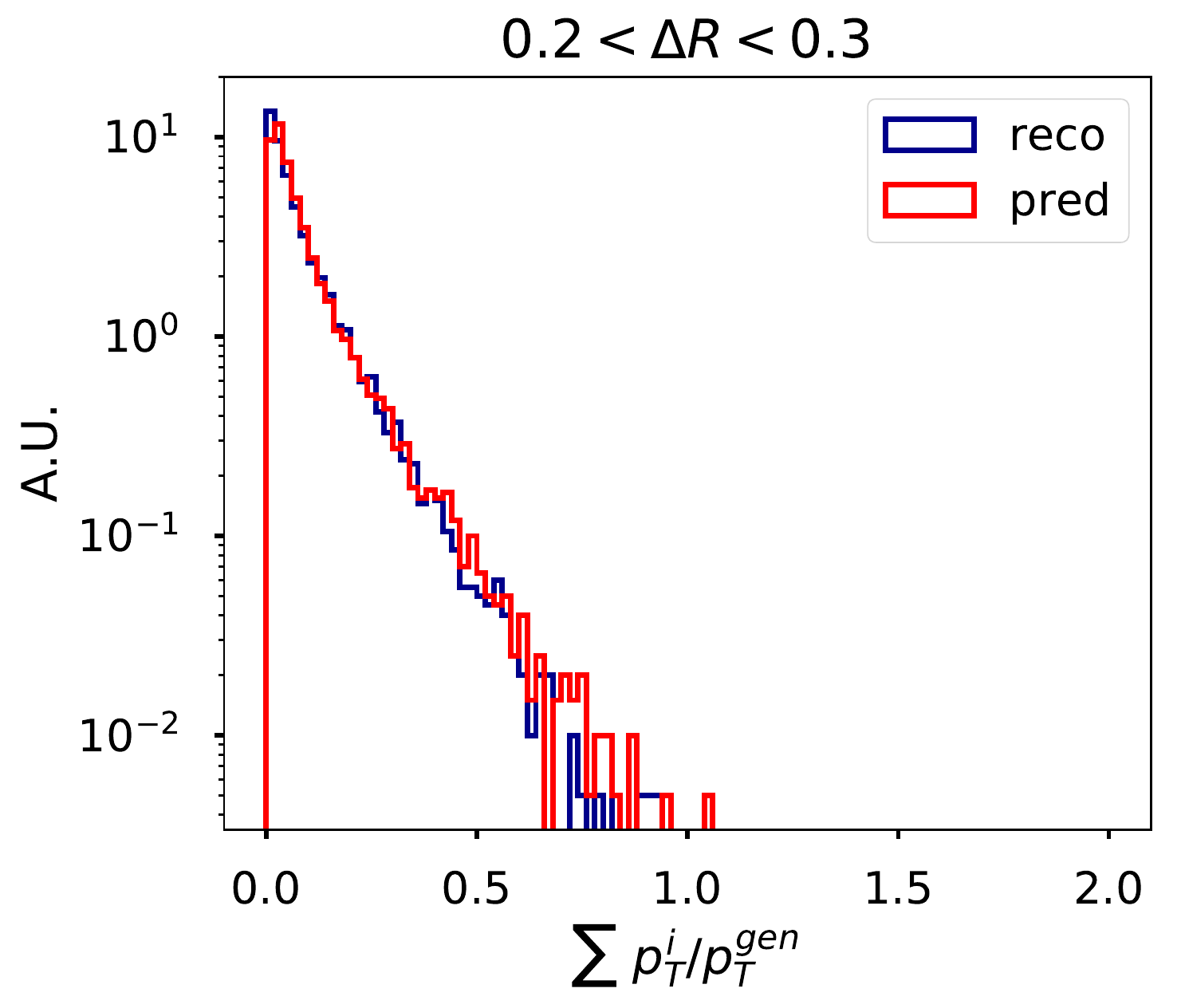} 
\includegraphics[width=0.2\textwidth]{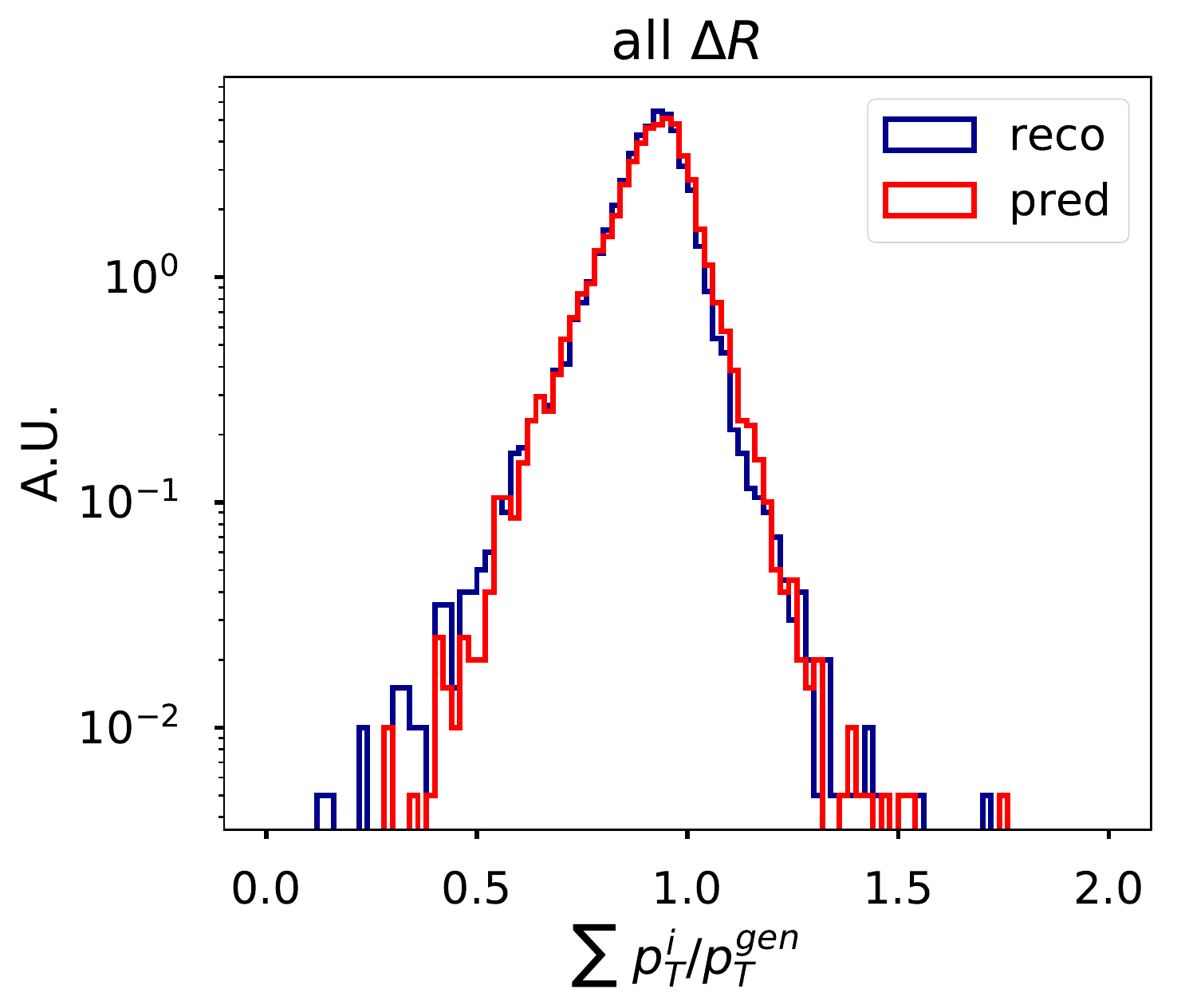}
\caption{\label{fig:sumpred}
Aggregated pixel intensities for different rings in $\Delta \eta$--$\Delta \phi$. Blue
histograms are obtained from the input data, while red ones are obtained using the
generative model. 
}
\end{figure}

\begin{figure}[t]
\centering
\includegraphics[width=0.2\textwidth]{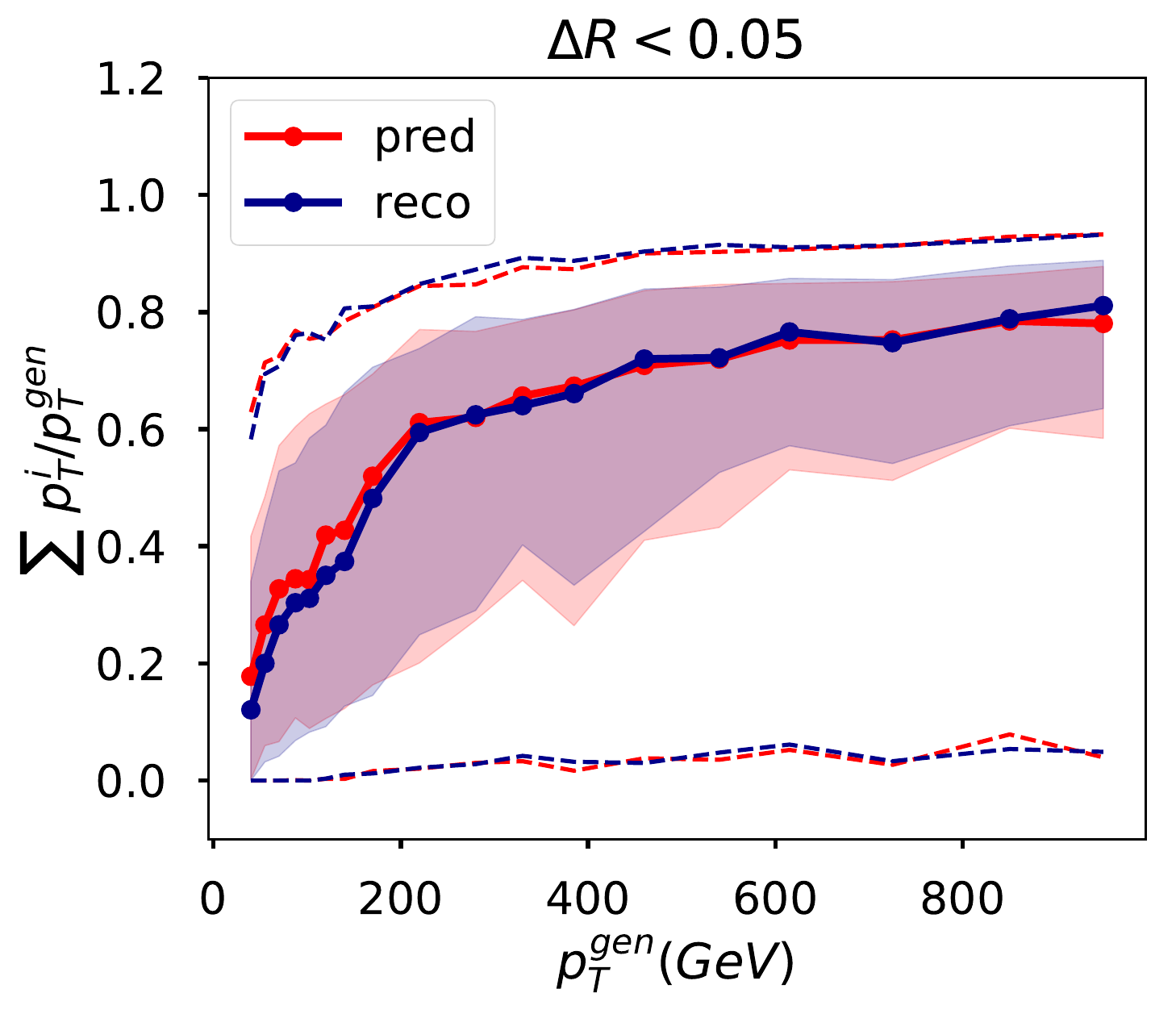} 
\includegraphics[width=0.2\textwidth]{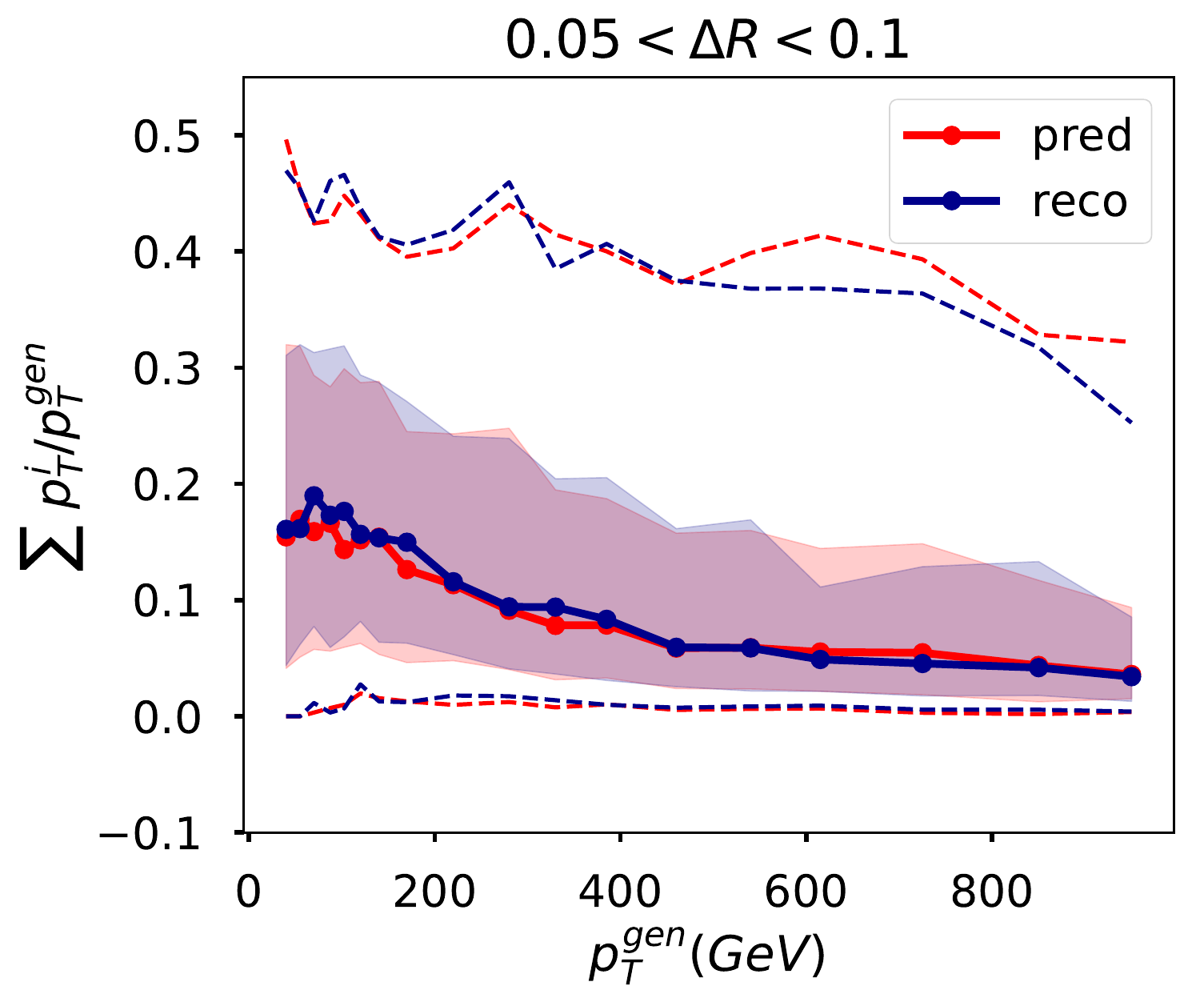} \\
\includegraphics[width=0.2\textwidth]{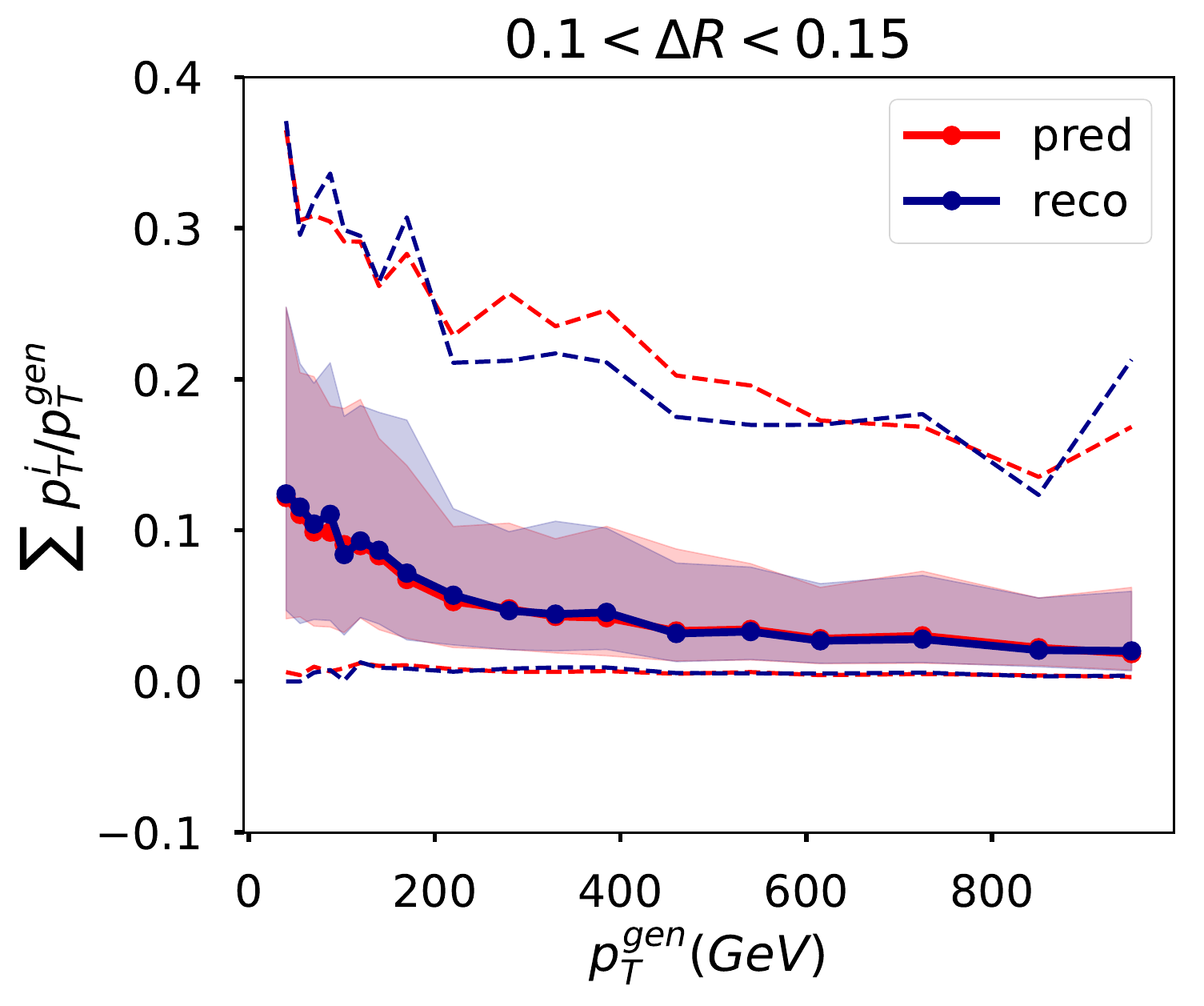} 
\includegraphics[width=0.2\textwidth]{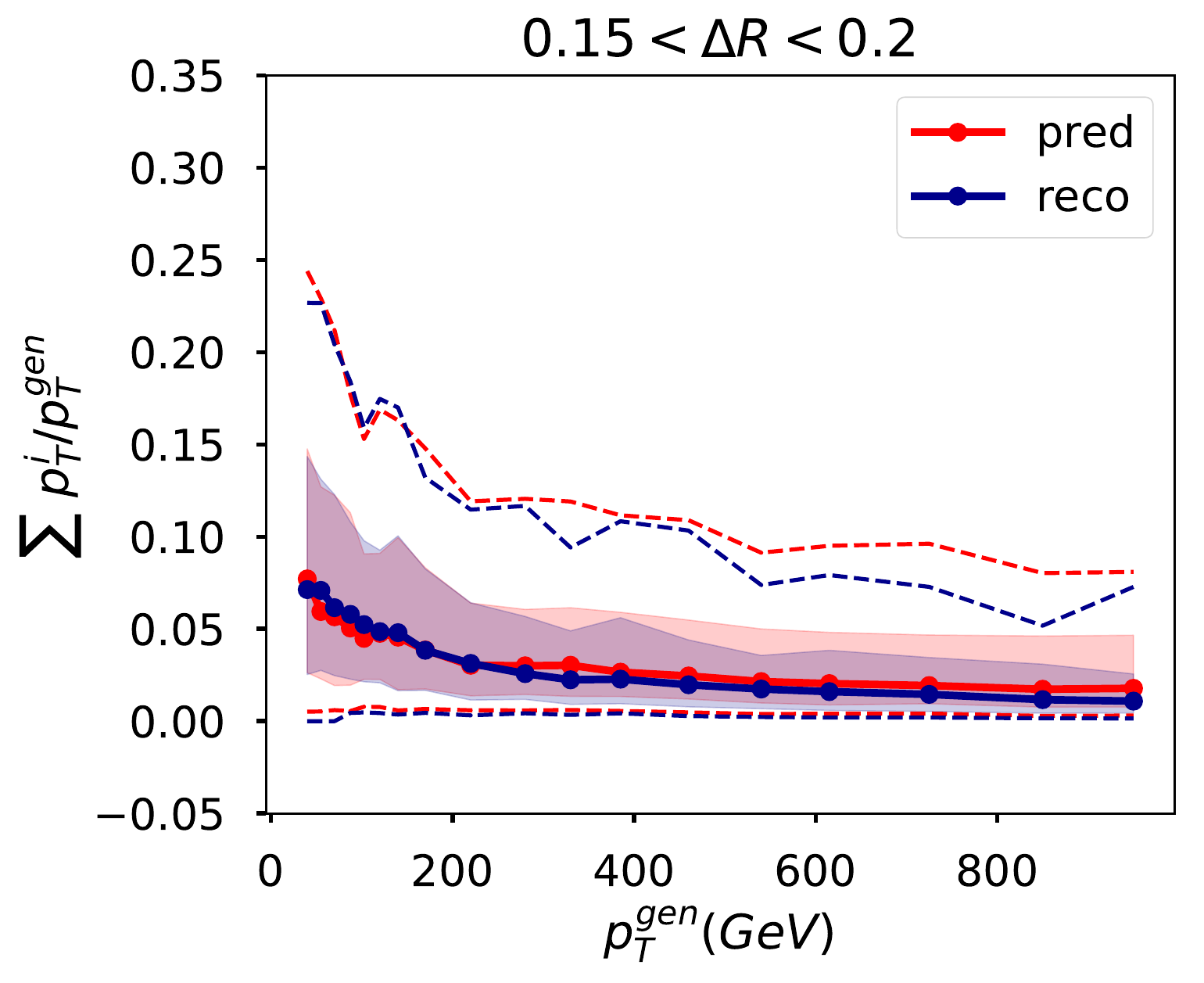} \\ 
\includegraphics[width=0.2\textwidth]{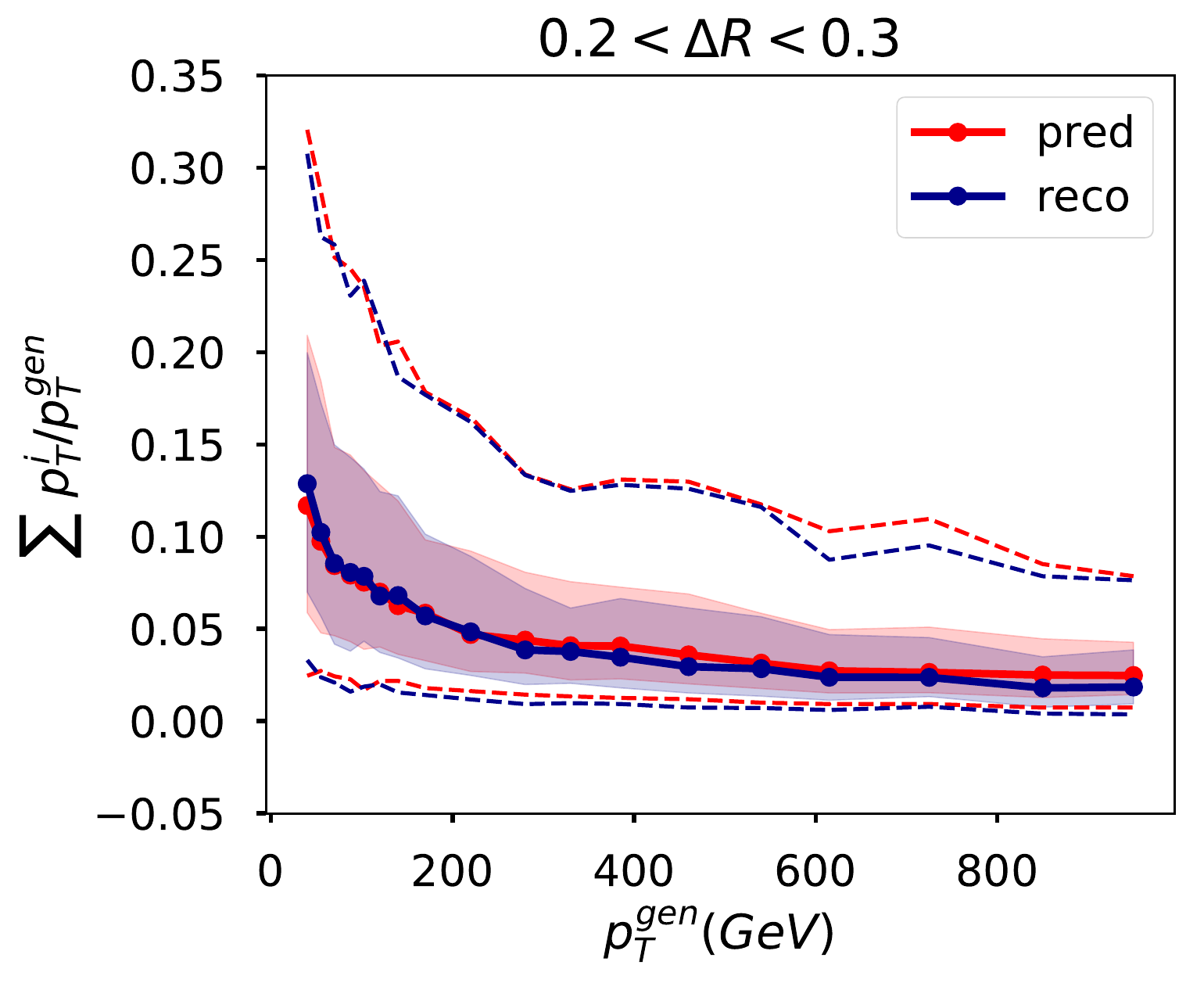} 
\includegraphics[width=0.2\textwidth]{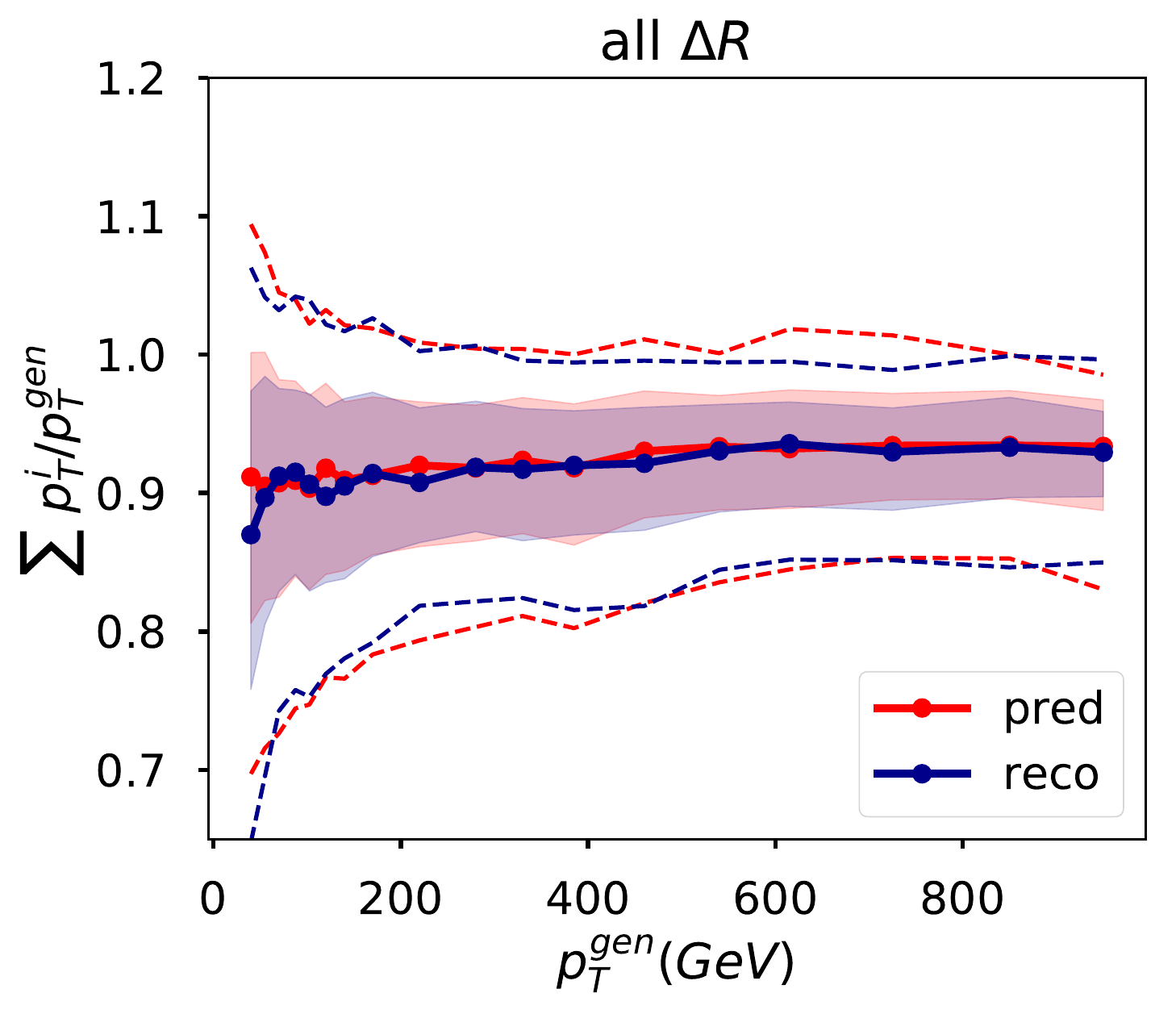}
\caption{\label{fig:ptpred}
Evolution of the aggregated pixel intensities for different rings in $\Delta
\eta$--$\Delta \phi$ as a function of the particle level jet transverse momentum. Solid
lines represent the median of the distribution, filled regions show the inter-quartile
range, while dashed lines mark the 10\% and 90\% quantiles. 
Blue lines are obtained from the input data, while red ones are obtained using the
generative model. 
}
\end{figure}

Figure~\ref{fig:avg_img} shows the average jet image obtained at particle-level by
aggregating the \yv values, as well as those obtained at reconstruction-level by
aggregating either the \xv values or the $G_2$ ones. One can observe that the average
effect of the detector simulation and reconstruction algorithms is to generally spread the
jet energy away from the core. We see that this general trend is correctly reproduced by
our set-up.

Figure~\ref{fig:scatter} shows the correlation between the particles and detector-level pixel
intensities for the input dataset and for the predicted images, obtained from a test
sample of 100000 images. Three sub-populations can be observed in the distributions:
\begin{itemize}
\item well measured jet components populate the diagonal;
\item errors in the position reconstruction of the jet components lead to energy migration
    between close-by pixels and thus contribute to the sub-populations
    located close to the horizontal and vertical axes;
\item non-reconstructed jet components manifest as empty reconstruction-level pixels that
    correspond to non-empty ones at particle level and therefore to the sub-population
    located along the horizontal axis
\item pile-up effects lead to non-empty pixels in the reconstruction level image in
    correspondence to empty particle-level pixels and so contribute to the sub-population
    close to the vertical axis. 
\end{itemize}
As can be seen from the figure, our set-up (right panel) achieves a good modelling of the
relative weight of the three sub-populations, which result from a non-trivial set of
effects. 
The sub-populations located along the diagonal, the horizontal axis and the vertical one
account for about 40\%, 25\% and 25\% of the non-empty pixels, respectively and our
algorithm is able to reproduce such numbers with a relative accuracy of roughly 30\%.

In figure~\ref{fig:sumpred} we show the distributions of the total pixel intensities
obtained integrating over rings in $\Delta R$ centred on the particle-level jets
axis. Blue histograms are obtained from the input dataset, while red ones show the
results of the generative model. Figure~\ref{fig:ptpred} shows the evolution of
the aggregated pixel intensities distribution as a function of the particle-level
jet transverse momentum. These results show that our set-up allows good modelling of
hadronic jet structure over more than two orders of magnitude in  jet transverse
momentum.

We further investigate the goodness of the learned model by evaluating its ability to
reproduce high level jet features that are typically used in physics analyses. We
concentrate, in particular on two sets of variables:
\begin{enumerate}
\item variables used in the context of quark/gluon discrimination;
\item jet substructure variables used in the context of merged jets discrimination.
\end{enumerate}

%% MD structures?
\begin{figure}[t]
\centering
\includegraphics[width=0.2\textwidth]{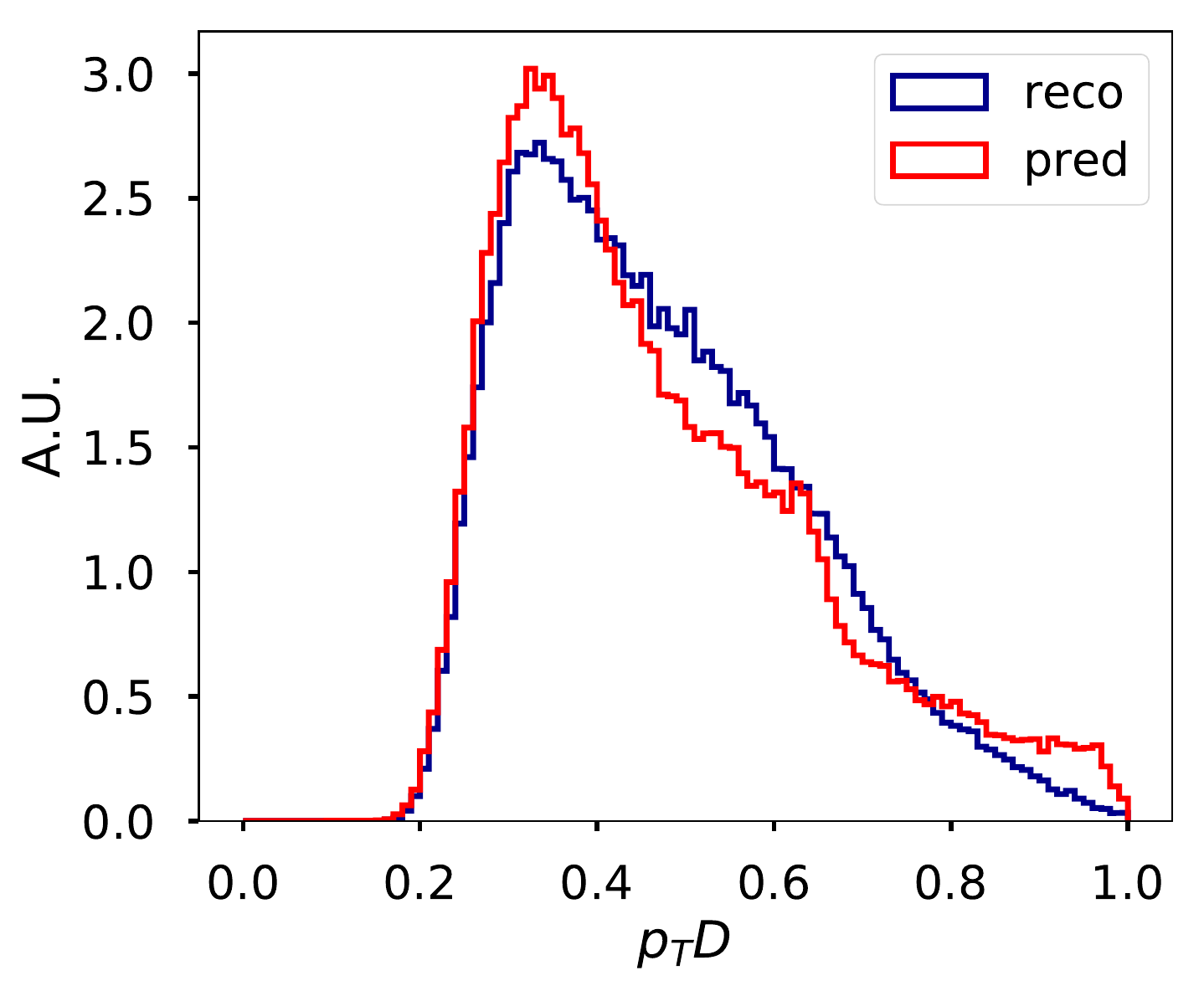} 
\includegraphics[width=0.2\textwidth]{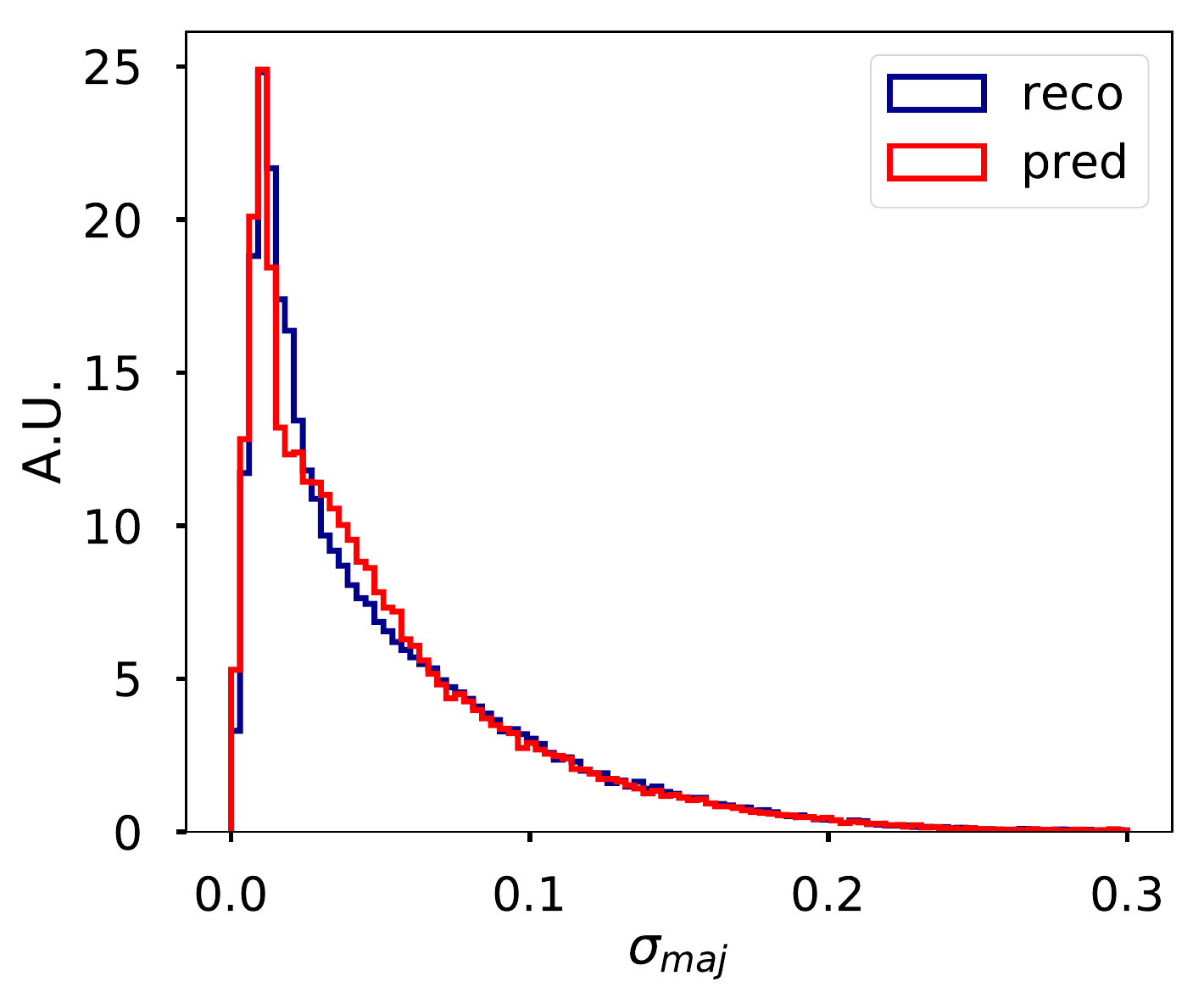} \\
\includegraphics[width=0.2\textwidth]{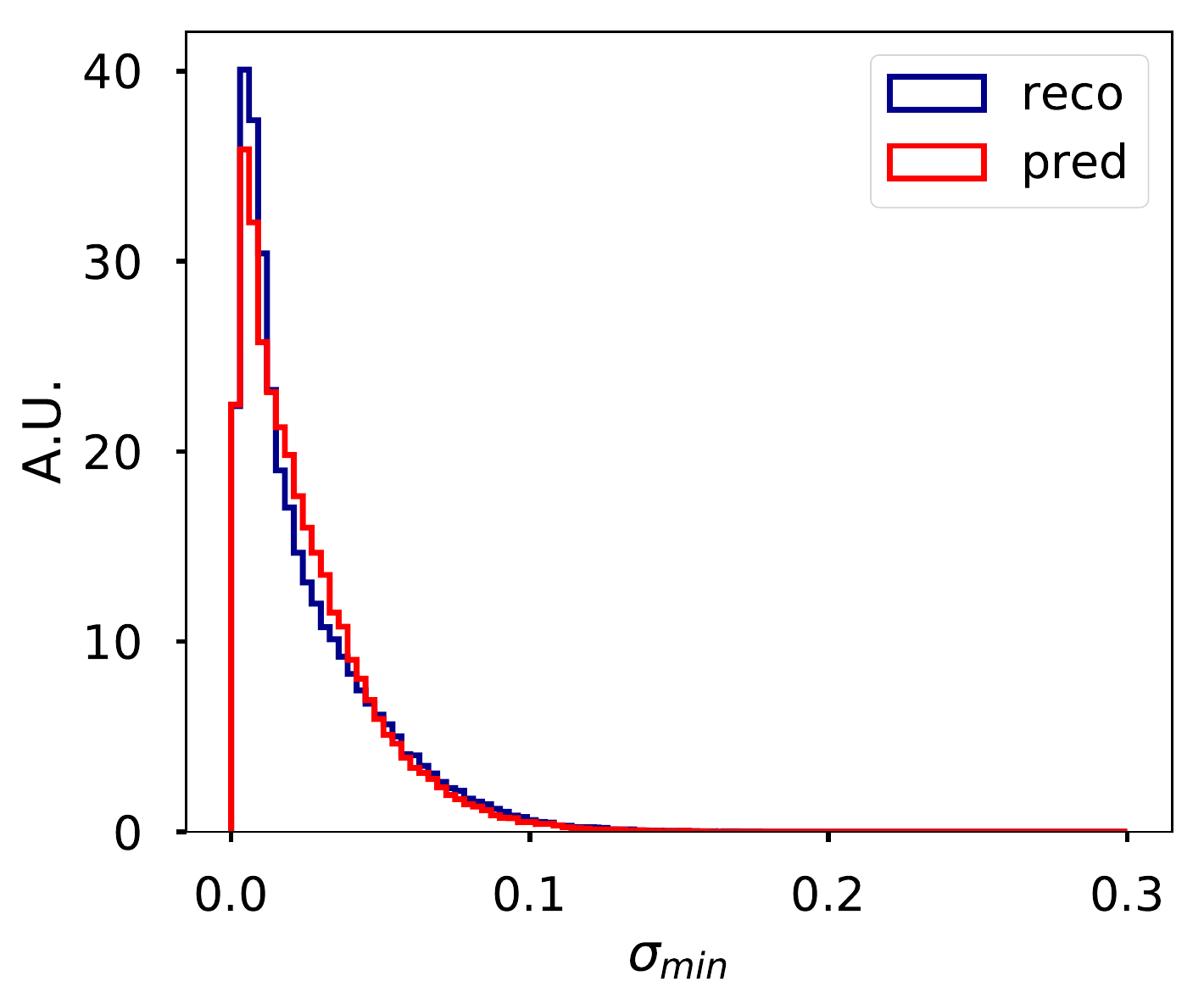} \\
\includegraphics[width=0.2\textwidth]{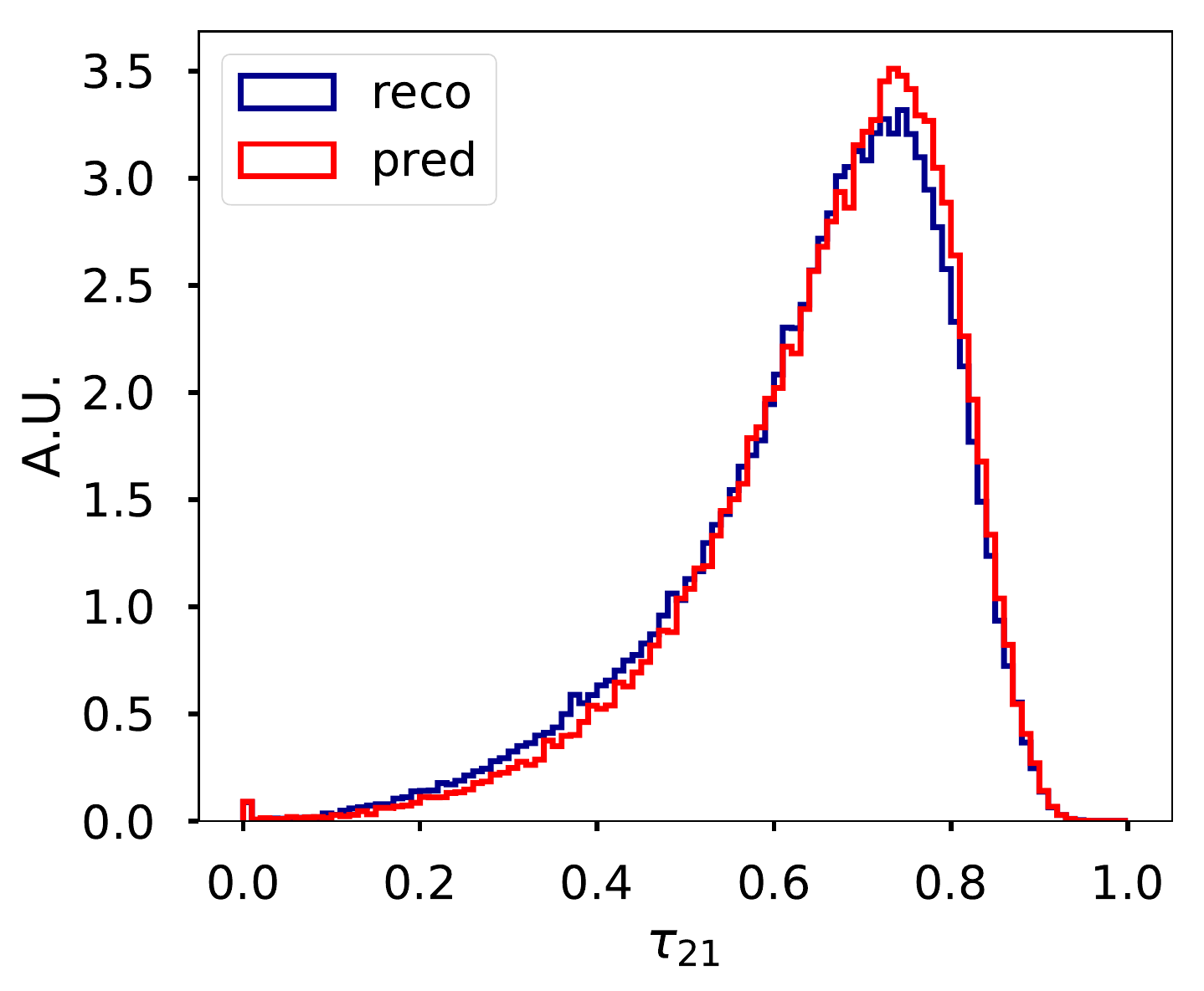} 
\includegraphics[width=0.2\textwidth]{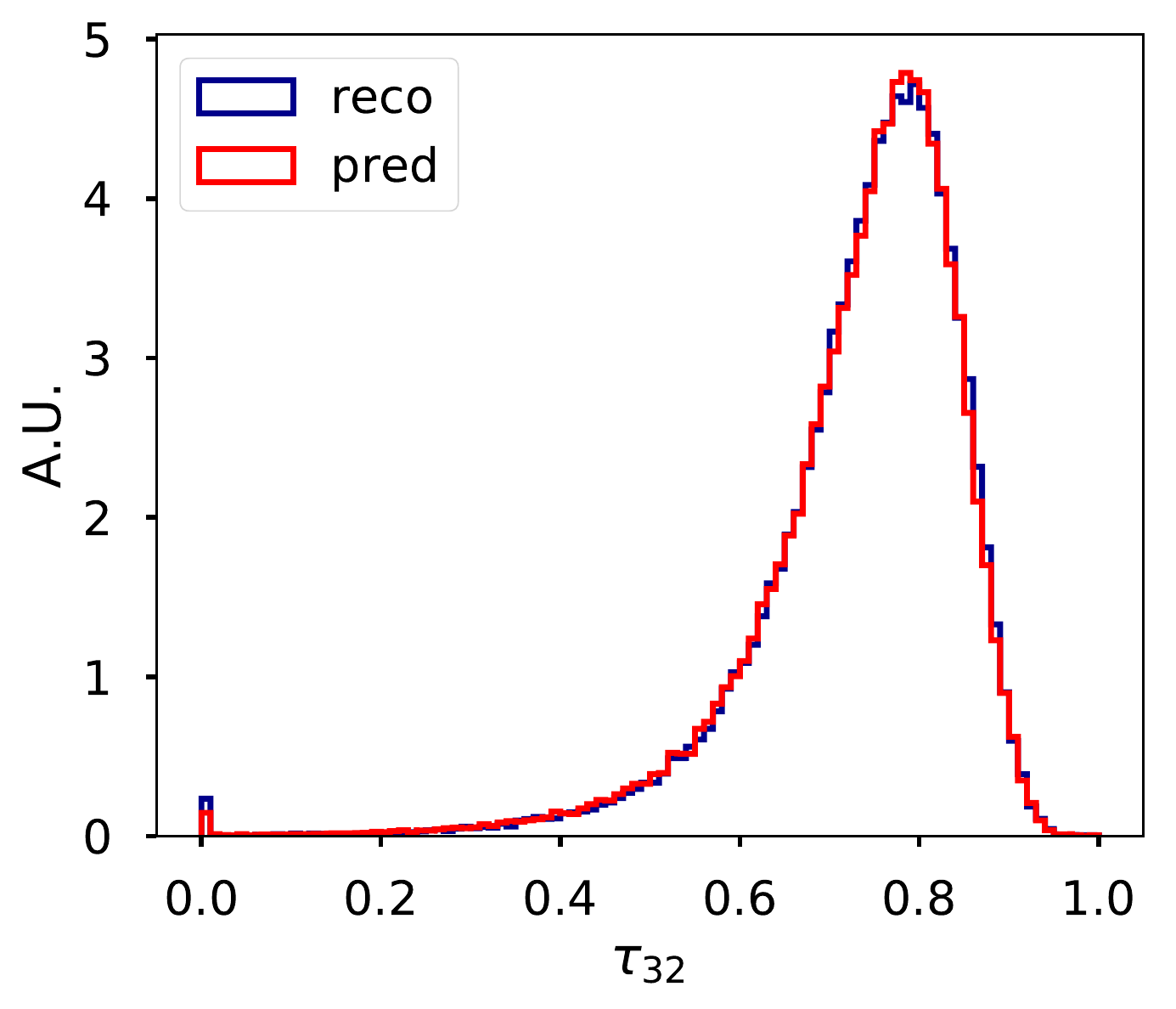} 
\caption{\label{fig:metrics}
Distribution of high level variables used for quark/gluon discrimination (first two rows) and
merged jets tagging (last row). Blue histograms are obtained from the input data, while
red ones are obtained using the generative model. 
}
\end{figure}

\begin{figure}[t]
\centering
\includegraphics[width=0.2\textwidth]{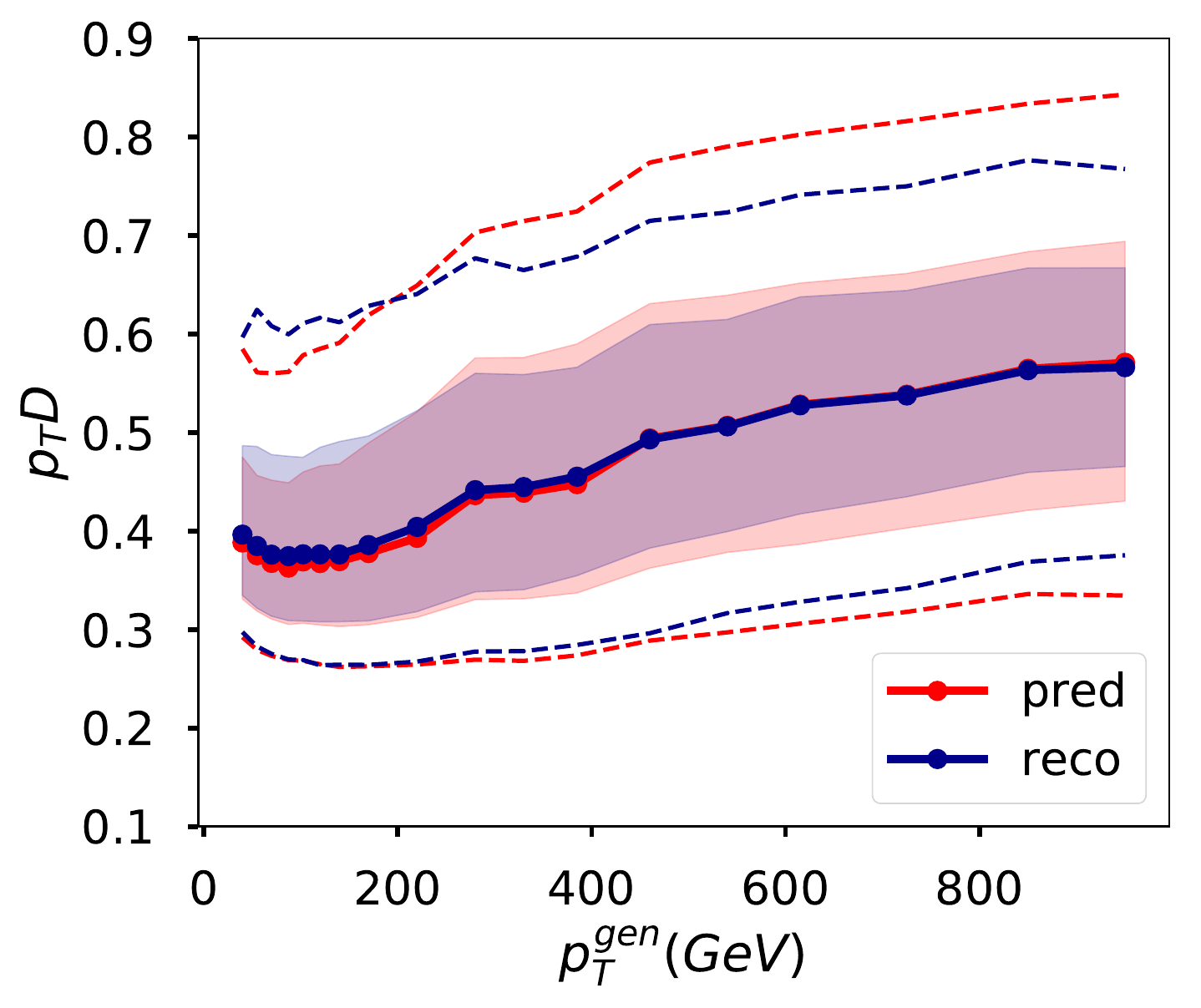} 
\includegraphics[width=0.2\textwidth]{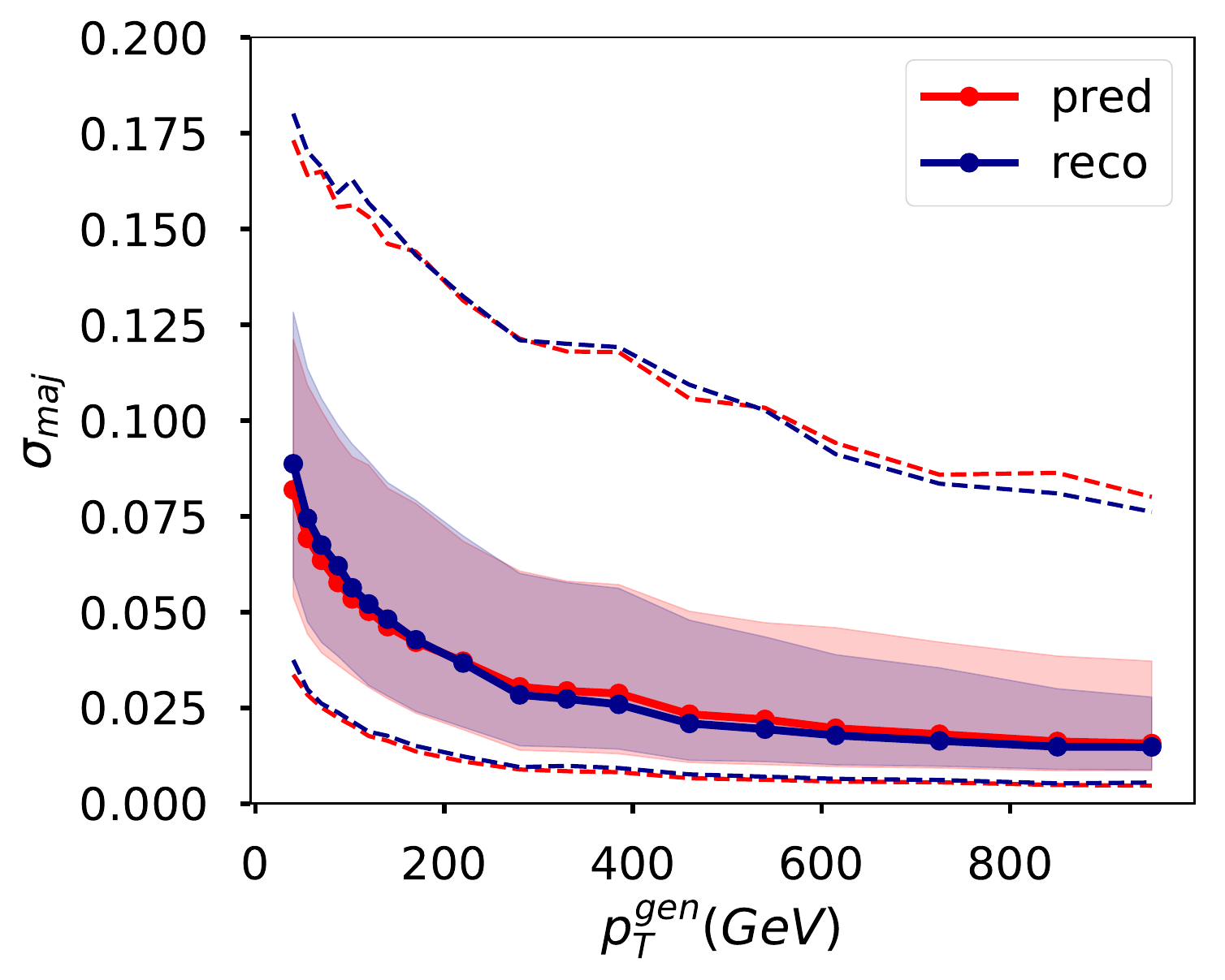} \\
\includegraphics[width=0.2\textwidth]{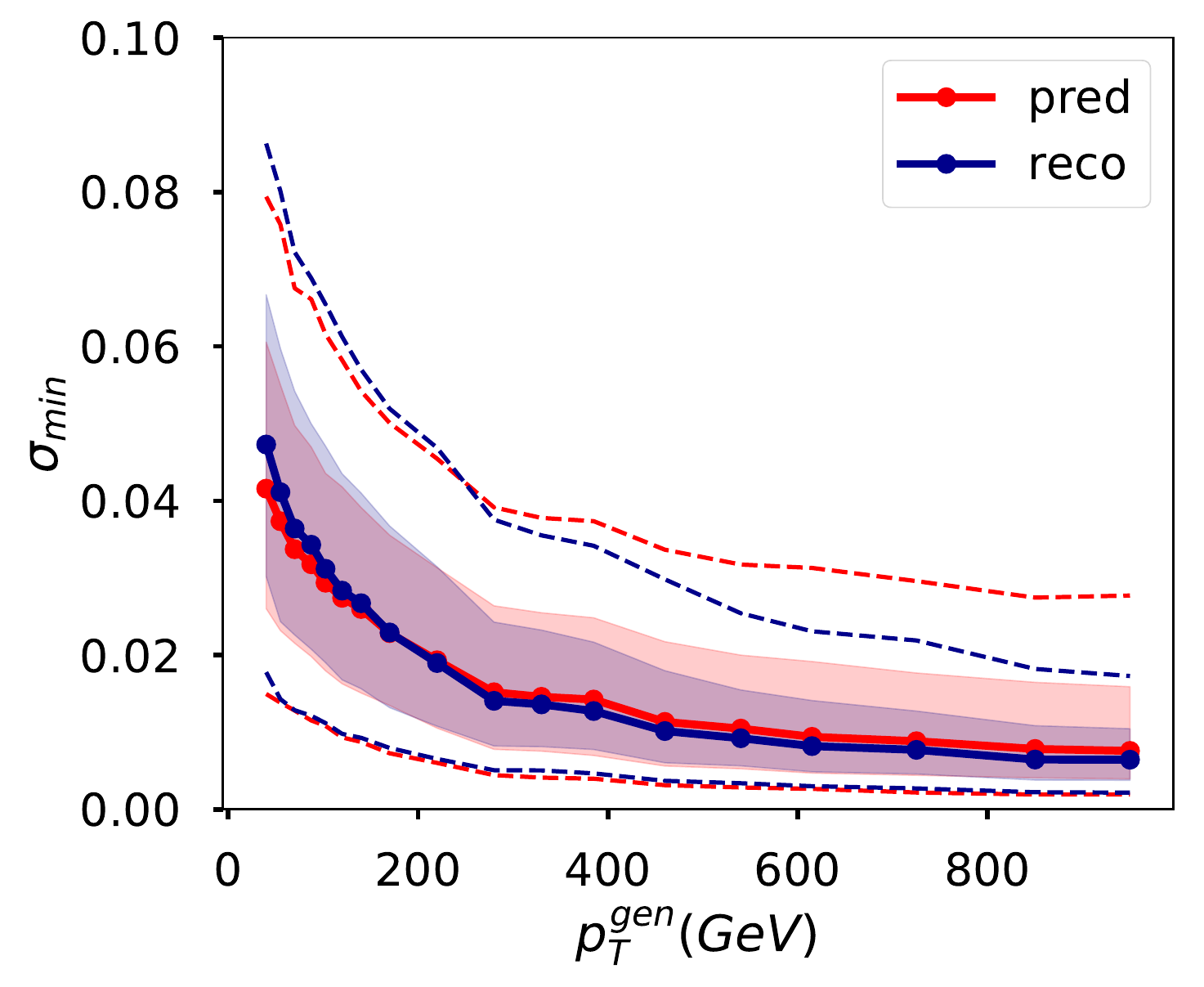} \\
\includegraphics[width=0.2\textwidth]{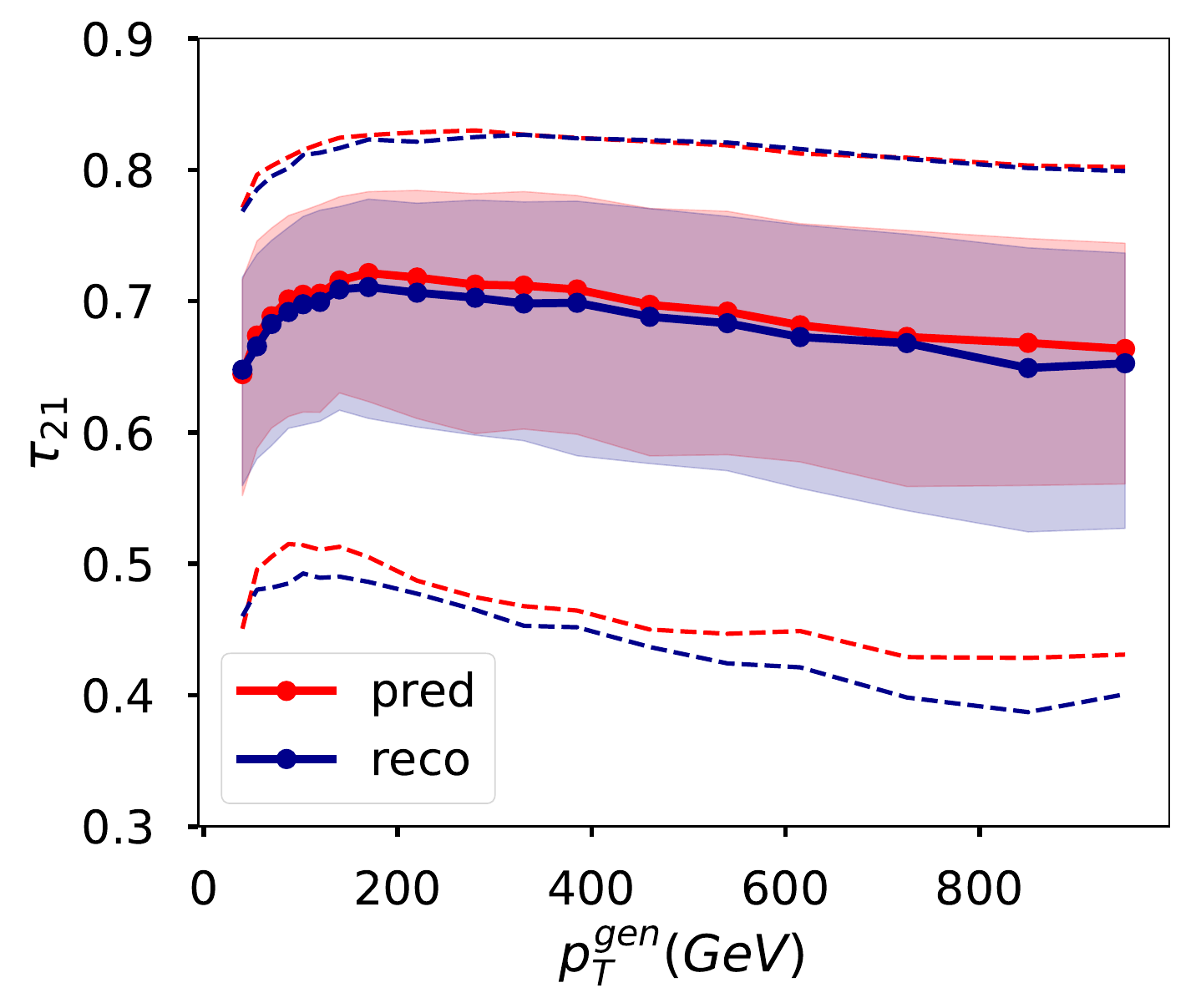} 
\includegraphics[width=0.2\textwidth]{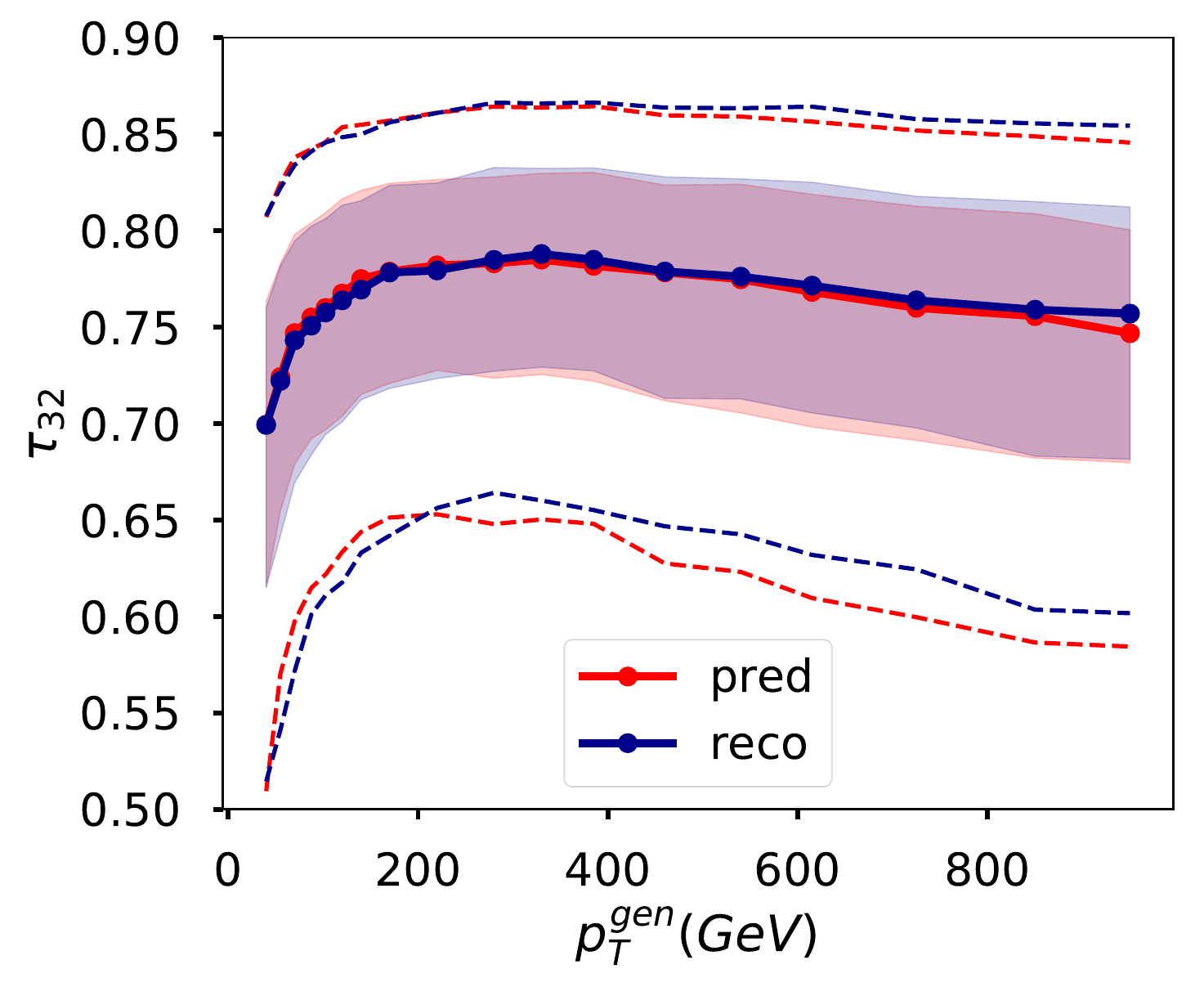} 
\caption{\label{fig:ptmetrics}
Evolution of the quark/gluon (first two rows) discrimination and merged jet tagging (last row)
variables as a function of the particle level jet transverse momentum. Solid 
lines represent the median of the distribution, filled regions show the inter-quartile
range, while dashed lines mark the 10\% and 90\% quantiles. 
Blue lines are obtained from the input data, while red ones are obtained using the
generative model. 
}
\end{figure}

From the first set, we choose the so-called major and minor axes, i.e. the square root of
the eigenvalues of the $\eta$-$\phi$ covariance matrix of the jet image, and the $p_TD$
variable, i.e. the ratio between the square root of the second and first non-central
moment of the pixel intensities~\cite{Pandolfi2013}.
From the second set, we choose the ratio between the 2- and
1-subjettines~\cite{subjett} and that between the 2- and 3-subjettiness. The subjettiness
variables were computed using the FastJet package~\cite{fastjet1,fastjet2}, approximating
each jet as a set of mass-less particles with energies and directions obtained from
the pixel intensities and positions. 

Figure~\ref{fig:metrics} shows the distribution of the quark/gluon and merged jets discrimination
variables that we considered aggregated over the test dataset, while
figure~\ref{fig:ptmetrics} shows the evolution of the distributions as a
function of the transverse momentum of the jet at particle level. The level at which these
variables are predicted by our set-up is good (in general, the marginal densities and the
value of the distributions quantiles are predicted with an accuracy of 5-10\%), even
though some mismodelling can be observed for the quark-gluon discriminating variables. In
particular, mismodelling at the 10-20\% level in the marginal density, and in the
$p^{gen}_T$ dependence of the 75\% and 90\% quantiles can be observed for the $p_TD$ and
the major and minor axes. These kind of disagreements point to the fact that the
correlation between the number of non-empty pixels and their energy sharing is not
perfectly modelled.
%% In addition, the size
%% of the major and minor axes is overestimated for jets with transverse momentum below
%% roughly 300GeV and it is underestimated above.

\subsection{Discussion}

The results that we discussed above represent a step forward in terms of accuracy
of fast simulation systems proposed in the context of collider detector physics.
We believe that three main aspects contributed to this:
\begin{itemize}
\item the use of a generative model that is designed to handle spatial correlations well, and 
    the use of a conditioning space (i.e. that of particle-level images) that encodes
    large amounts of spatial information;
\item the explicit handling of the sparsity through the soft-mask layer;
\item the use of physics-driven constraints on the total intensity of the jet images.
\end{itemize}

The method that we outline here has potential application in conjunction with fast
detector simulation models, or parametrised ones~\cite{delphes,cms_fast_sim}. In this
context, being able 
to accurately predict the output of simulation and reconstruction algorithms for objects
like hadronic jets, which are ubiquitous at the LHC, would allow to save large amounts of
the computing power, by reducing the cost of producing simulated events samples.

%% This hybrid approach that combines the adversarial setup with a parametric loss function
%% allowed stabilising the training, and speeding up the convergence. 
While a relatively stable set-up was established by tuning the model hyper-parameters,
this aspect of the work is not yet completely satisfactory, as the region of
hyper-parameters space that lead to satisfactory results was found to be relatively
narrow. A review of the loss function structure, possibly incorporating the use of
alternative formulations of the GAN game~\cite{f_gan,w_gan1,w_gan2}, and of the model
training strategy in general, will be important to allow streamlining the method, and will
the subject of future work.

Furthermore, the current approach introduces noise only through the stochastic generation
of the set of active pixels. Our attempts at injecting noise at a more fundamental level
in the generative model structure have been unsuccessful so far. Similar
problems have been reported by researchers working on natural image generation (see
e.g.~\cite{pix2pix}). A more extensive investigation of the handling of noise will also be
subject of subsequent work.

\section{Conclusions}

We have reported on a method that uses deep neural networks to learn the response of particle
detectors simulation and reconstruction algorithms. The method is based on generative
adversarial networks and it was applied to the generation of hadronic jet images at the CERN
LHC.

We trained a generative model to reproduce the combined response of state-of-the-art
simulation and reconstruction algorithms. This was possible thanks to the exploitation of
the open datasets published by the CMS collaboration under the HEP data preservation
initiative. 

Starting from proposals made for natural image processing, we devised a hybrid set-up
based on the combined use of generative adversarial networks and analytic loss functions
that is able to take into account the conditioning on auxiliary variables and
physics-driven constraints on the generation process.

Our method allows reducing the computation time required to obtain reconstruction-level
%% MD quantify?
hadronic jets from particle-level jets by several orders of magnitude, while achieving
a very good accuracy in reproducing the simulation and reconstruction algorithms
response. The model is in particular capable of reproducing the evolution of the
reconstructed jet shapes as a function of several conditional variables. Physics-driven
high level features commonly used for merged jets tagging and quark/gluon discrimination,
and their evolution, are also generally well modelled.

The results obtained with this work represent a promising step forward towards the
development of fast and accurate simulation systems that will be crucial for the future of
collider experiments in high energy physics.

\section*{Acknowledgements}

We thank the CMS collaboration for publishing state of the art simulated data under
the open access policy. We strongly support this initiative and believe that it will be
crucial to spark developments of new algorithms from which the HEP community as a whole
can profit.
We thank Dr. M. Doneg\`a, Prof. G. Dissertori, and Dr. M. Pierini, for their careful review
of this manuscript.
This work was supported by the Swiss Centre for Supercomputing under the project D78.
The final publication is available at \url{https://doi.org/10.1007/s41781-018-0015-y}.

\section*{Conflicts of interest}

On behalf of all authors, the corresponding author states that there is no conflict of
interest. 

\bibliography{biblio}
\bibliographystyle{spphys}

\newpage
\clearpage
\appendix
\input{appendix}

\end{document}

%% file: appendix.tex
\section{Comparison with traditional fast simulation methods}

In this appendix we compare the results obtained using our generative neural networks with
methods for fast simulation traditionally used for phenomenological studies in high energy
physics. 
These packages parametrise particle detectors responses, on a particle by particle basis,
in terms of gaussian energy and momentum resolutions and binomial reconstruction
probabilities.
In particular, we compared the results obtained with our GAN with the output of the
DELPHES~\cite{delphes} package. In order to do that, we used version 3.4.1 of the program
and the standard CMS datacard included in the package.

\begin{figure}[ht]
\centering
\includegraphics[width=0.2\textwidth]{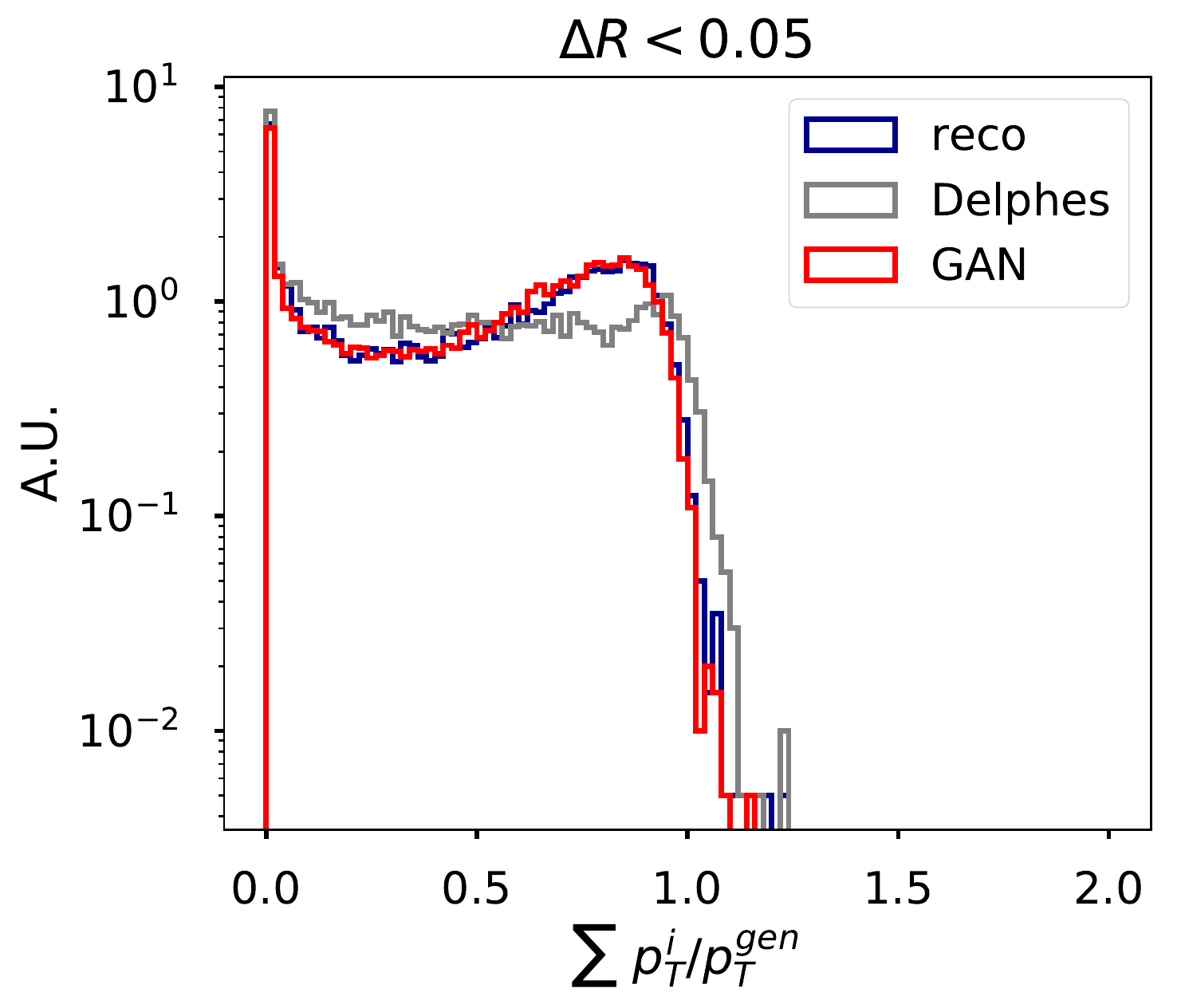} 
\includegraphics[width=0.2\textwidth]{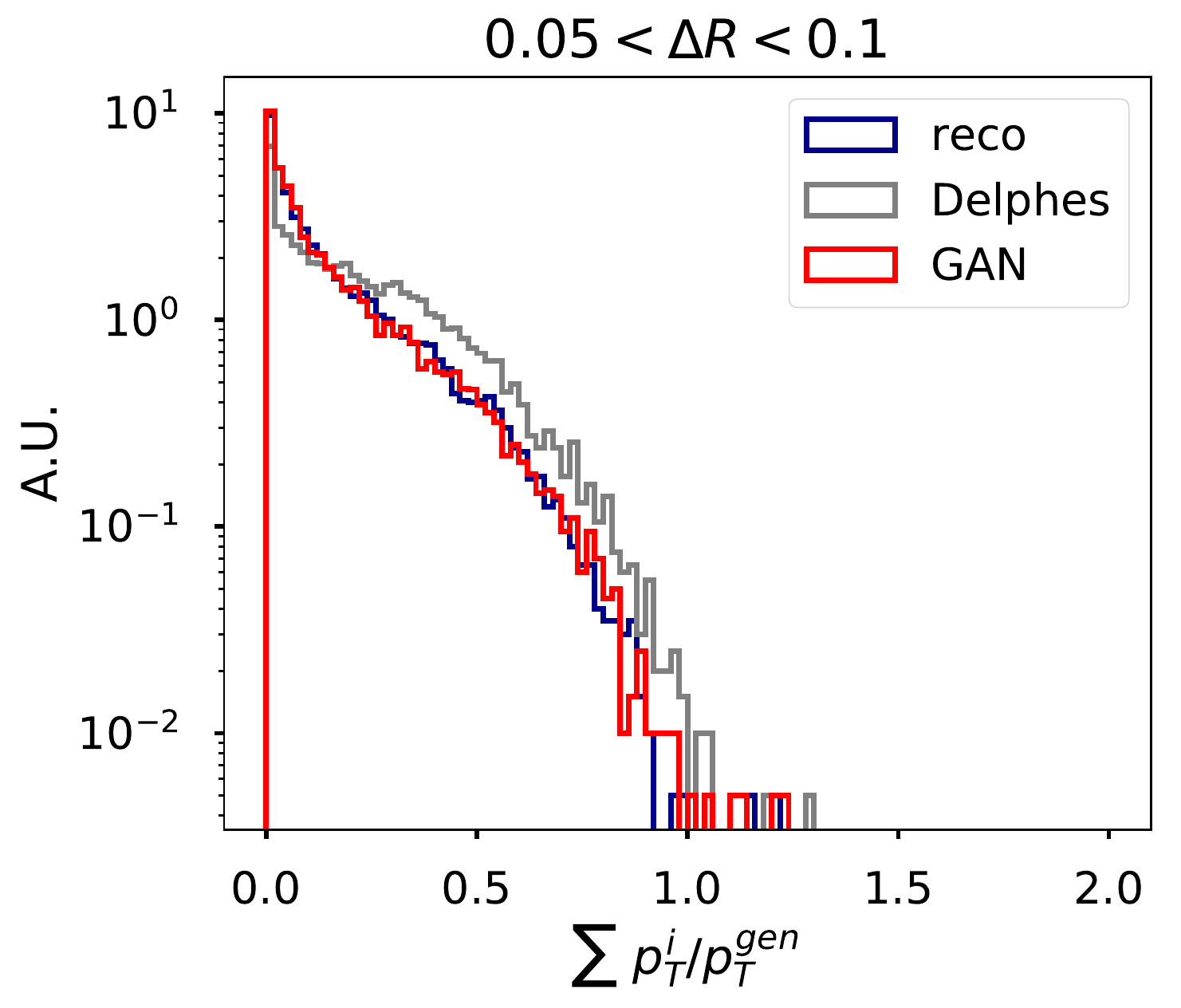} \\
\includegraphics[width=0.2\textwidth]{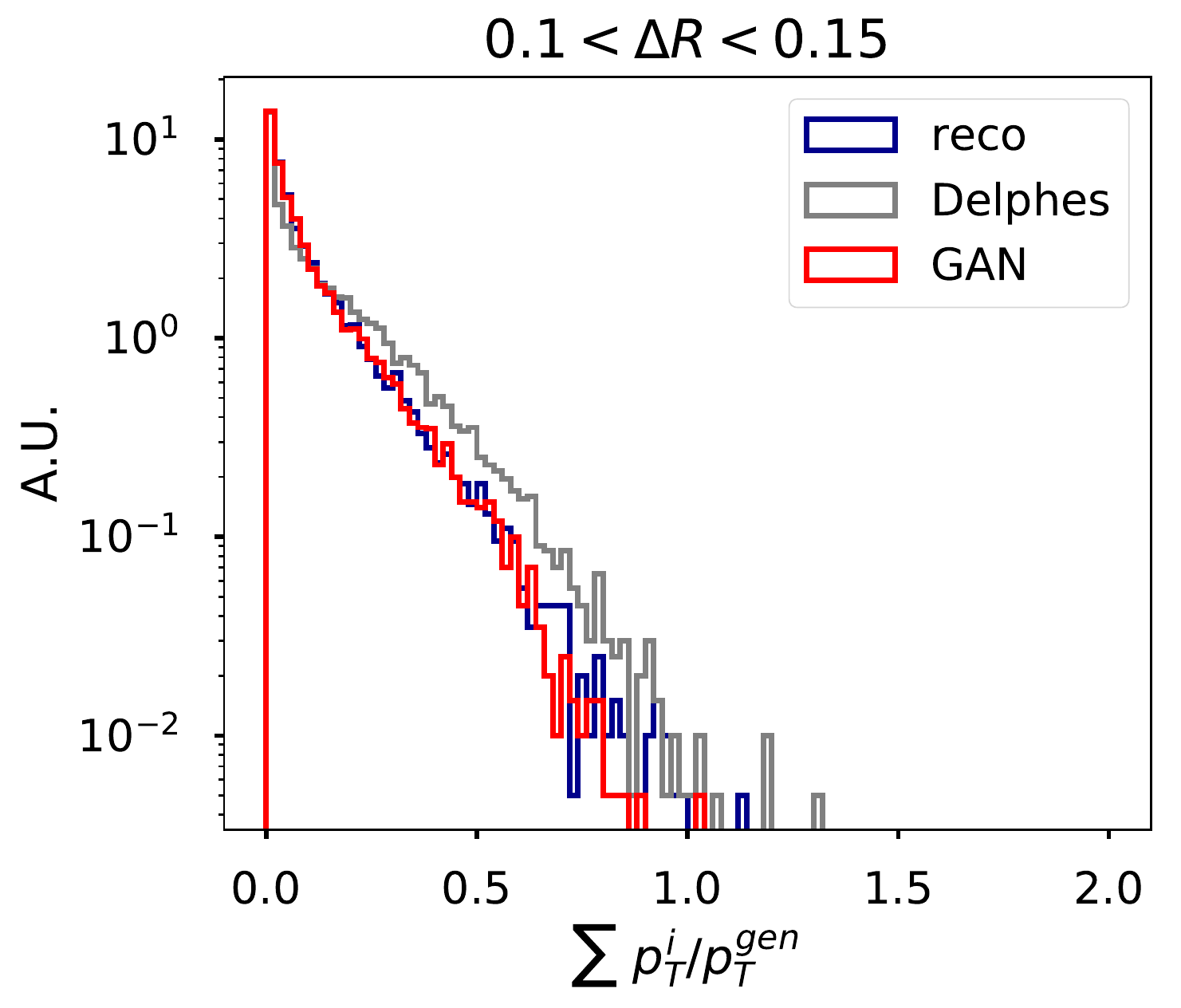} 
\includegraphics[width=0.2\textwidth]{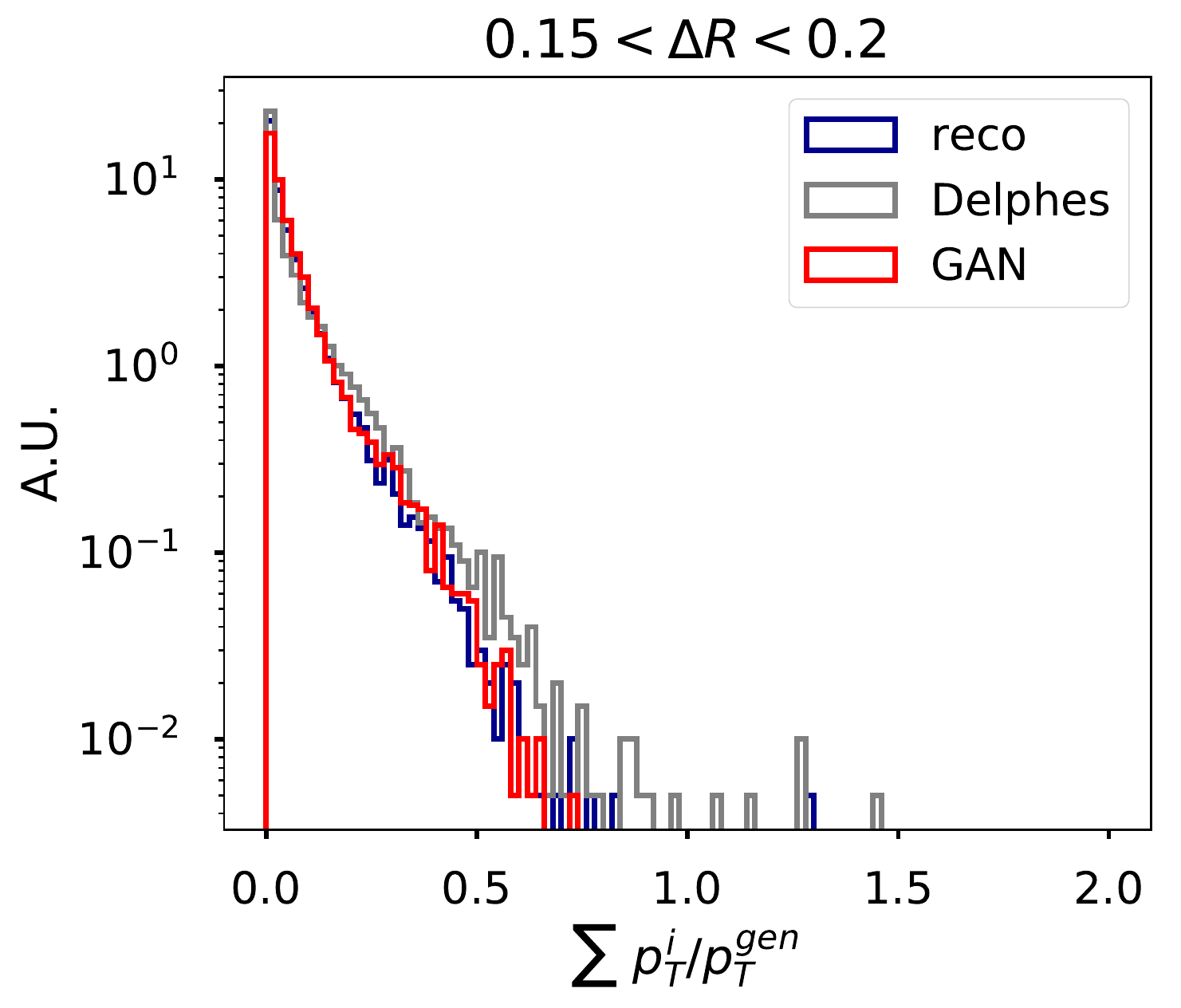} \\
\includegraphics[width=0.2\textwidth]{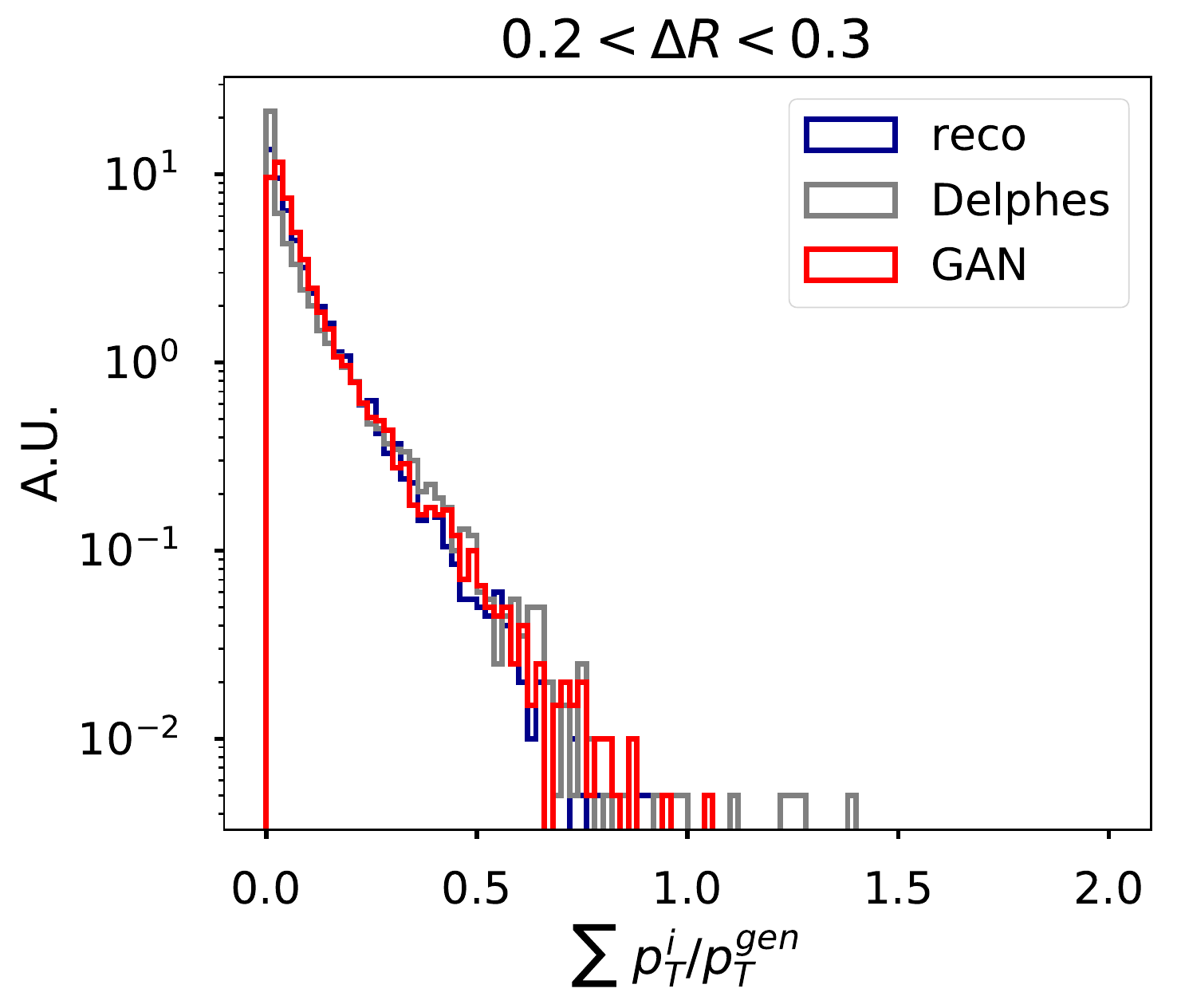} 
\includegraphics[width=0.2\textwidth]{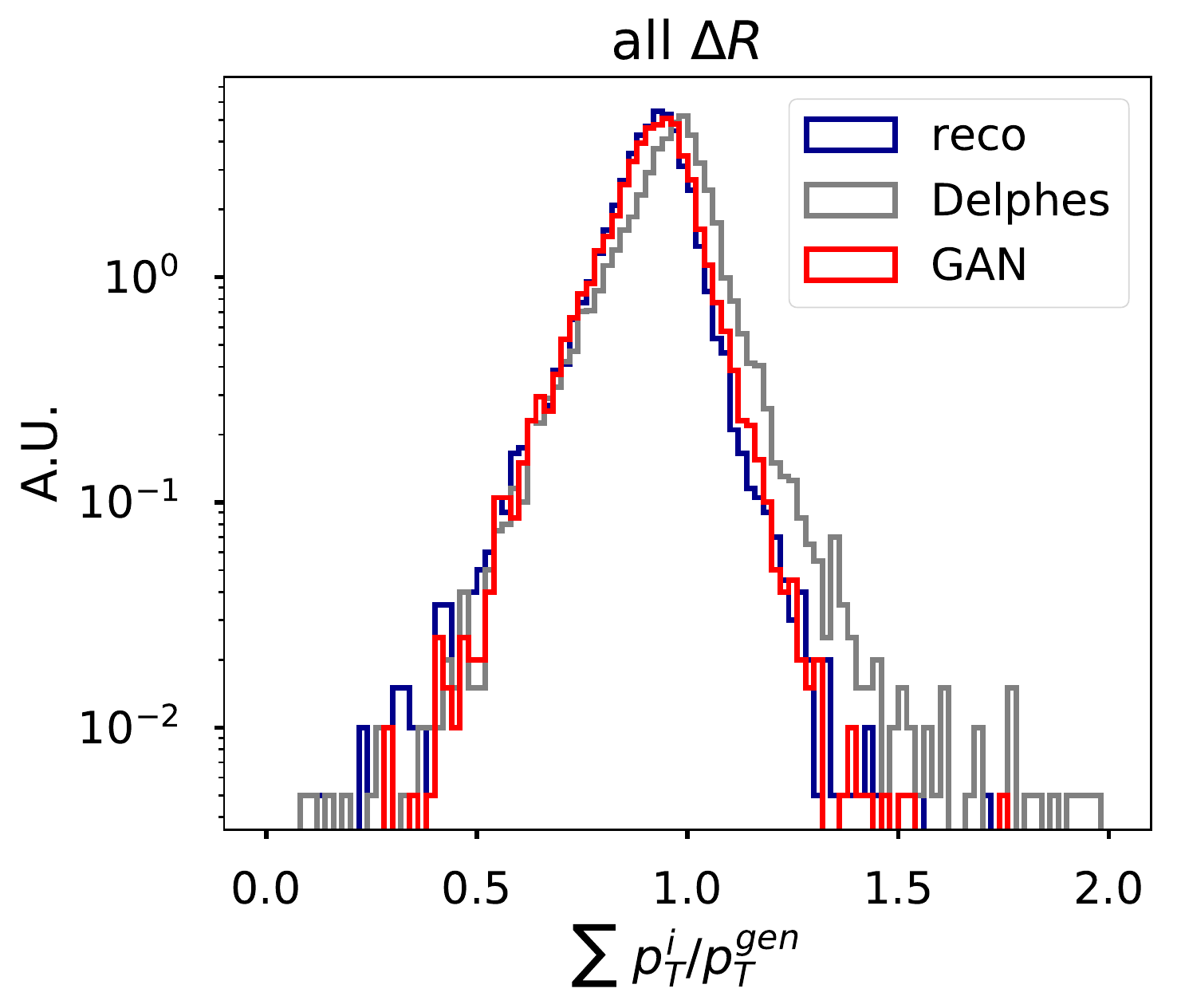}
\caption{\label{fig:sumpred_delphes}
Aggregated pixel intensities for different rings in $\Delta \eta$--$\Delta \phi$. Blue
histograms are obtained from the input data; red ones are obtained using the
generative model; gray histograms are obtained with a gaussian-smearing model.
}
\end{figure}

\begin{figure}[ht]
\centering
\includegraphics[width=0.2\textwidth]{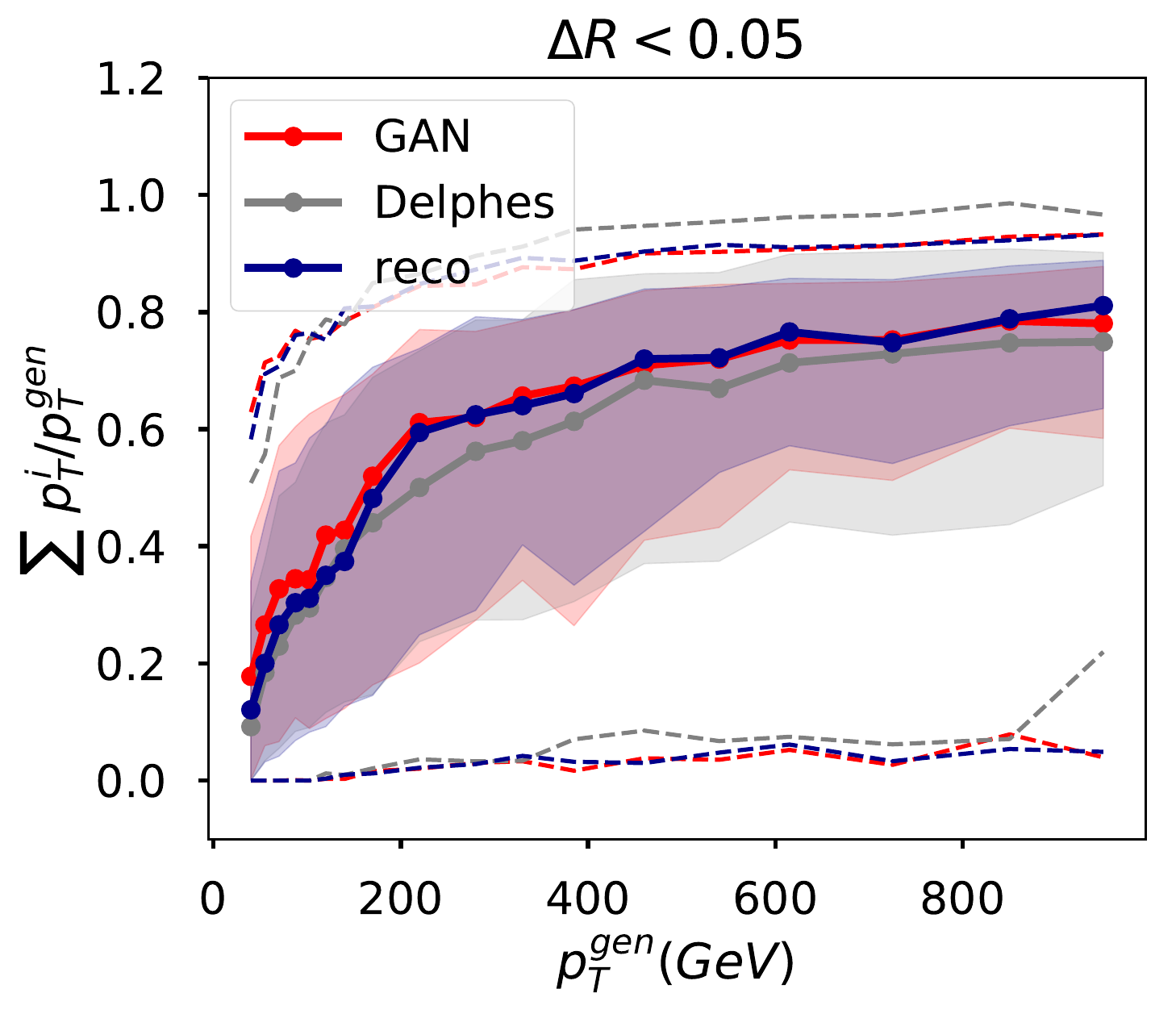} 
\includegraphics[width=0.2\textwidth]{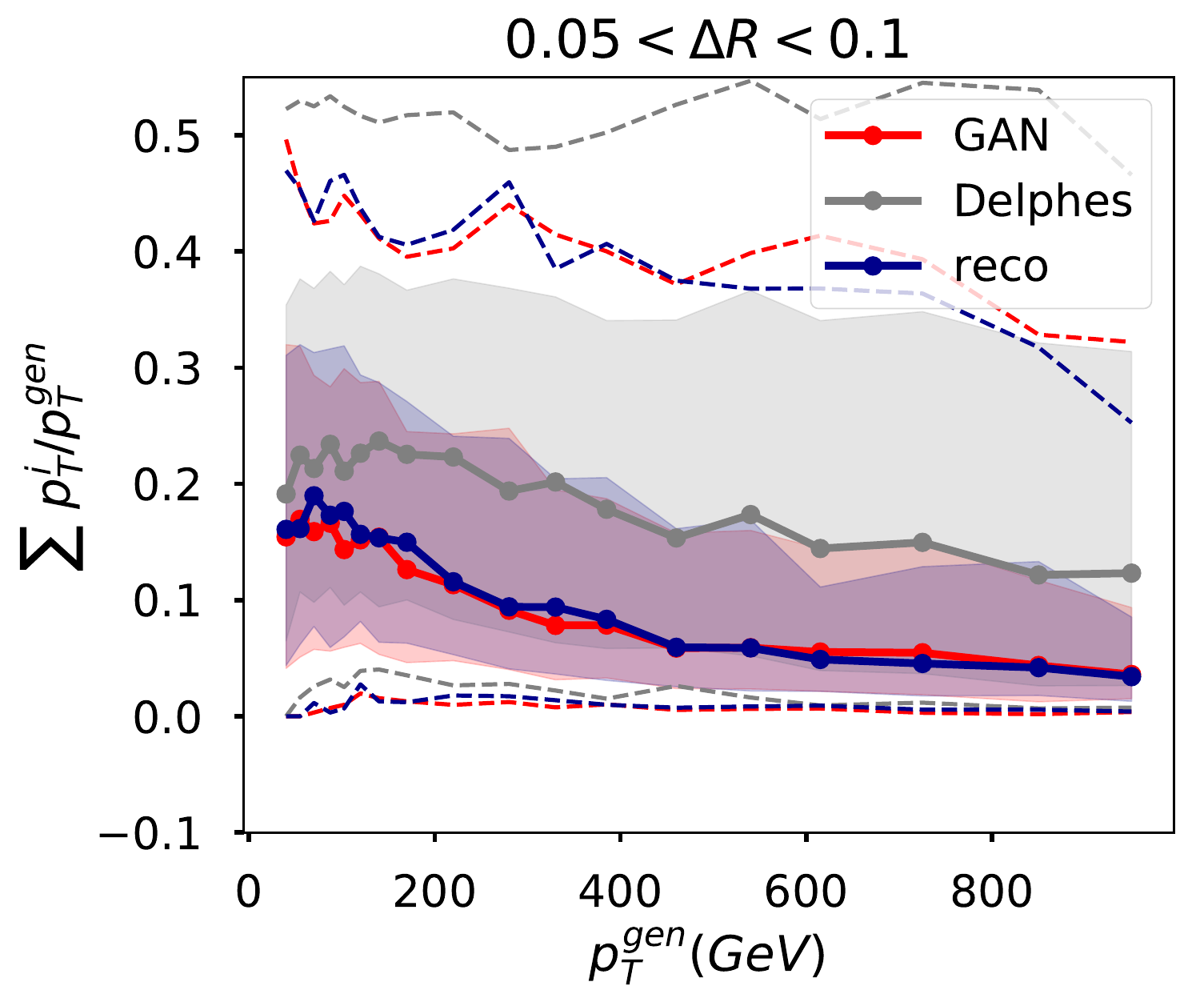} \\
\includegraphics[width=0.2\textwidth]{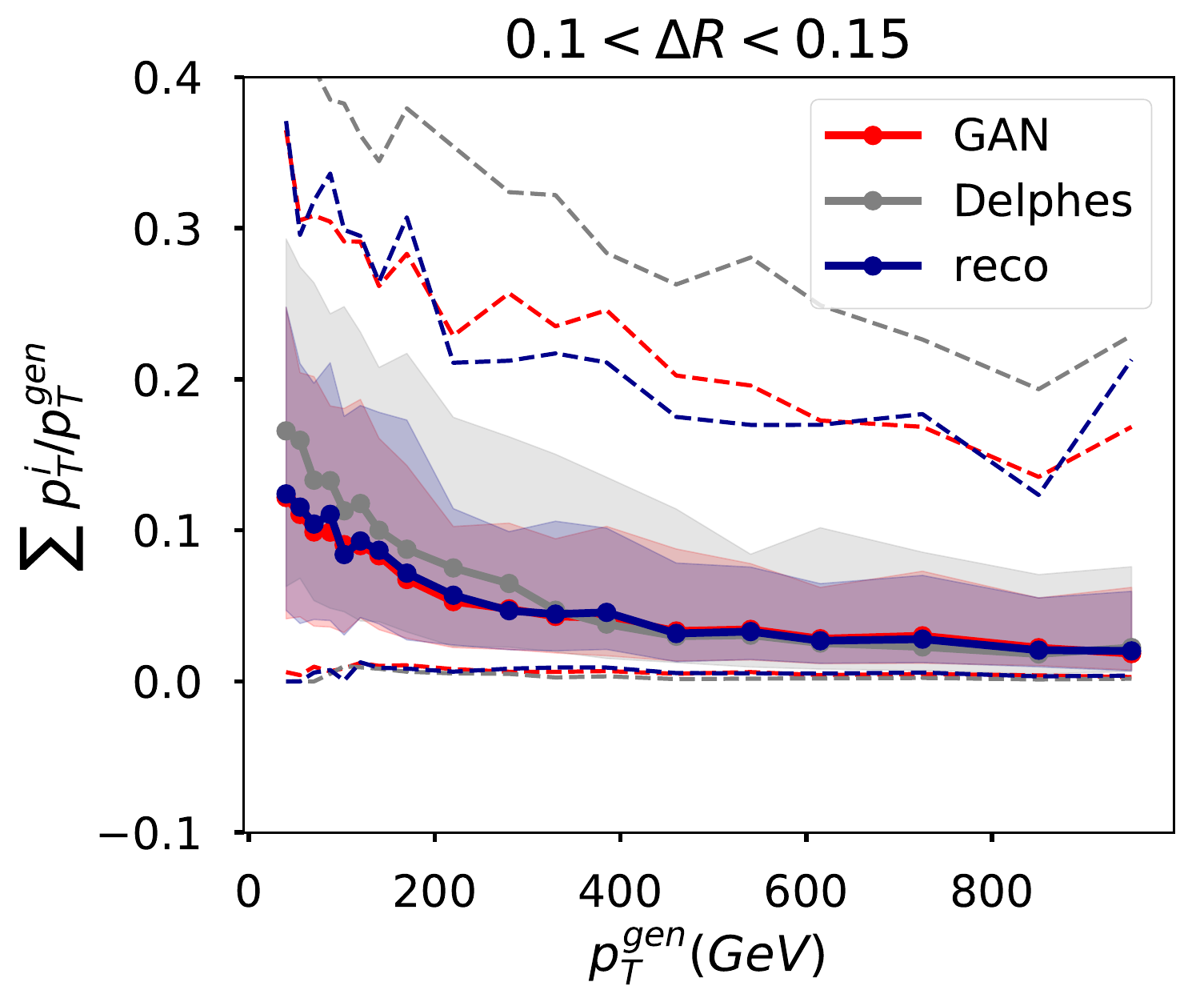} 
\includegraphics[width=0.2\textwidth]{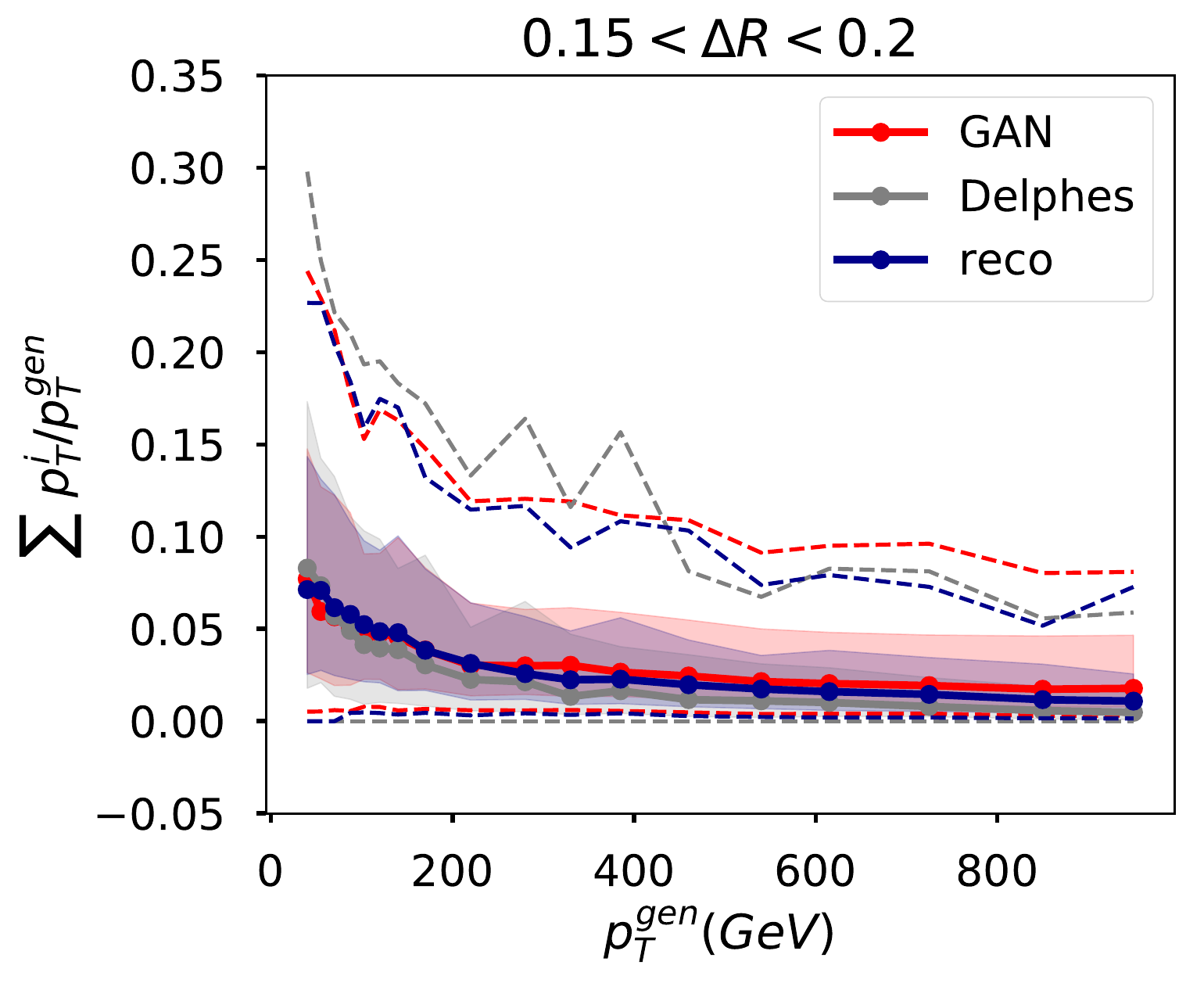} \\ 
\includegraphics[width=0.2\textwidth]{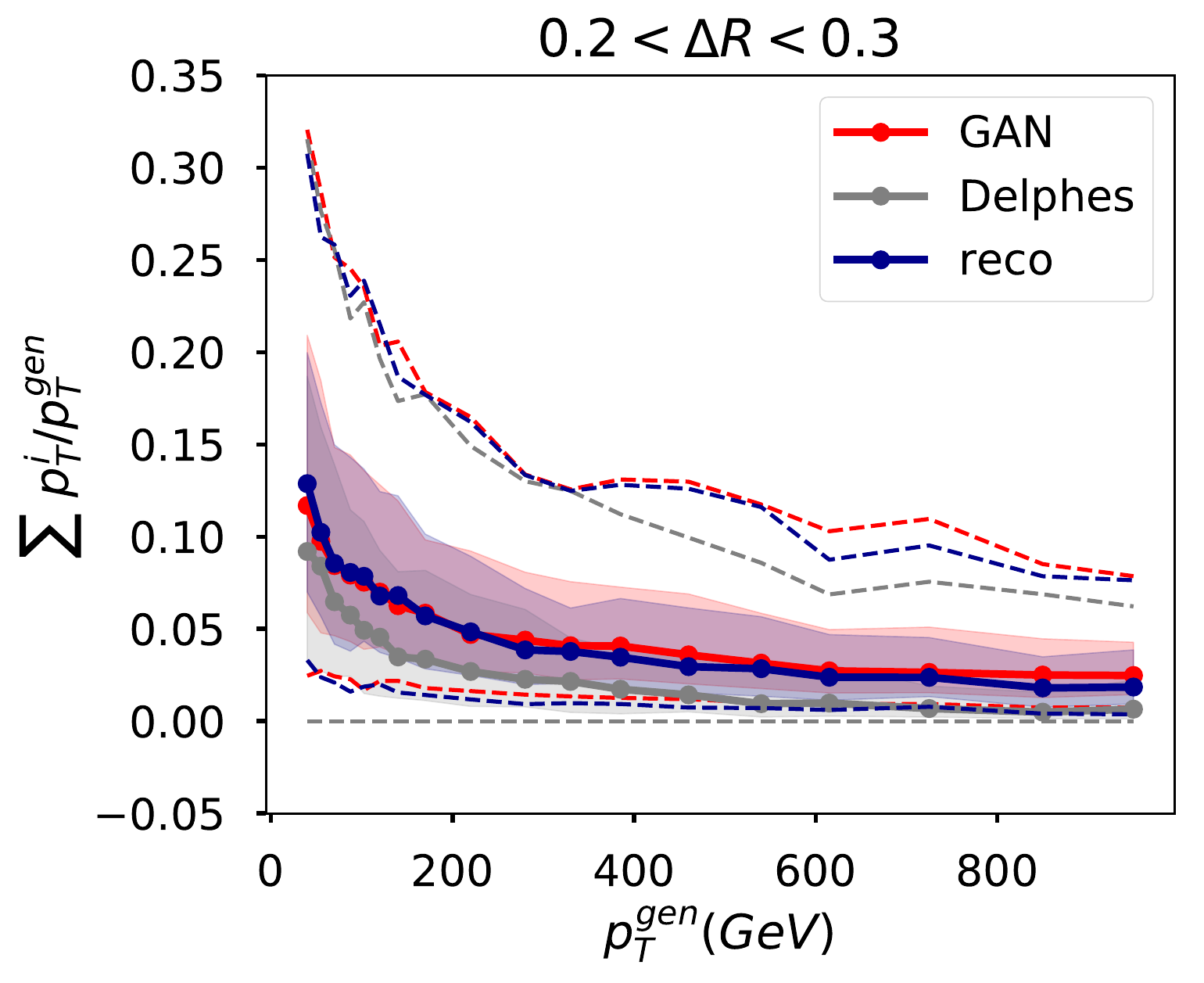} 
\includegraphics[width=0.2\textwidth]{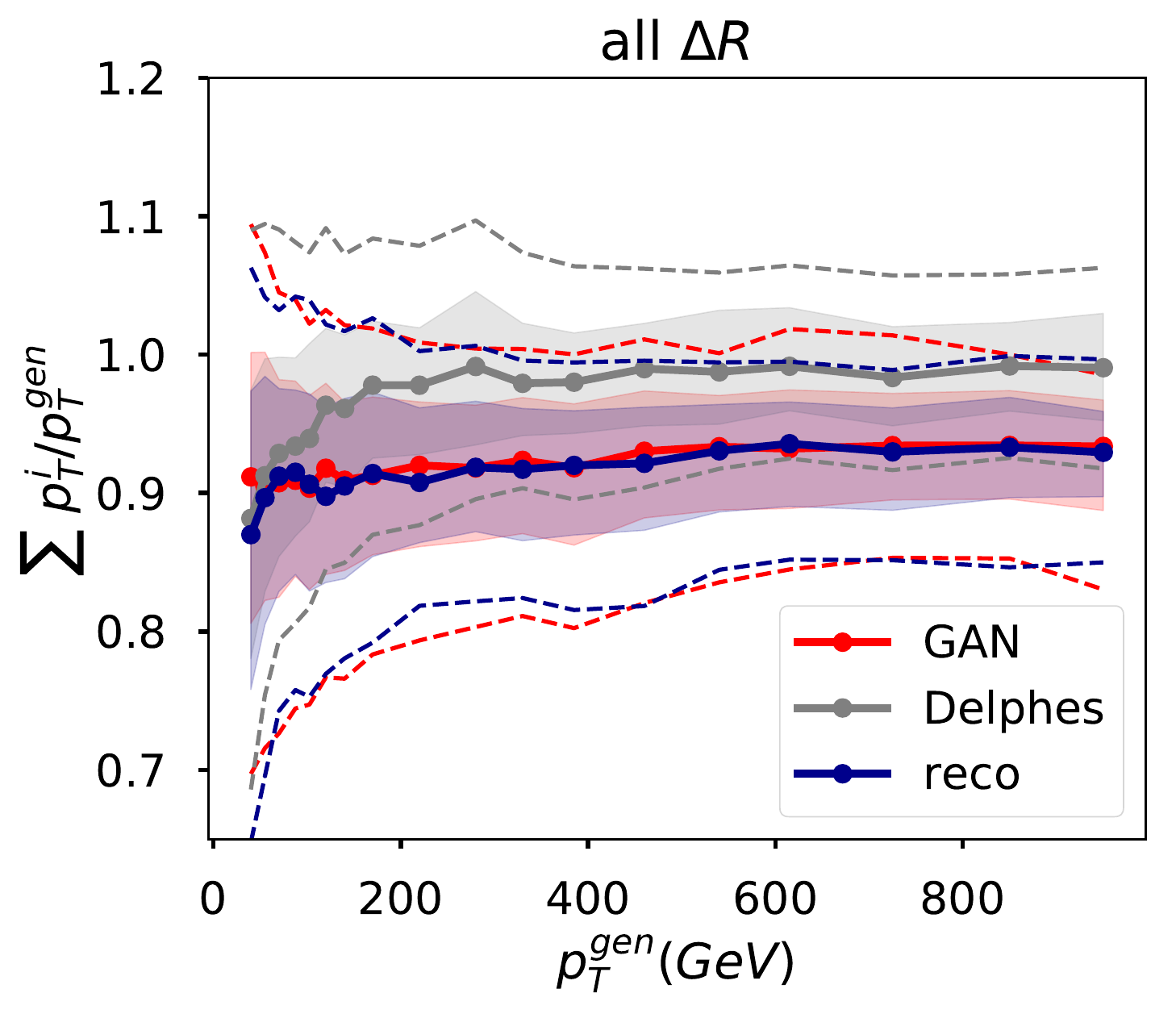}
\caption{\label{fig:ptpred_delphes}
Evolution of the aggregated pixel intensities for different rings in $\Delta
\eta$--$\Delta \phi$ as a function of the particle level jet transverse momentum. Solid
lines represent the median of the distribution, filled regions show the inter-quartile
range, while dashed lines mark the 10\% and 90\% quantiles. 
Blue lines are obtained from the input data; red ones are obtained using the
generative model; gray ones using the gaussian-smearing model.
}
\end{figure}

In figures~\ref{fig:sumpred_delphes} and \ref{fig:ptpred_delphes} we show the 
same quantities as shown in figures~\ref{fig:sumpred} and \ref{fig:ptpred},
respectively, but with the addition of the results obtained with the 
DELPHES gaussian-smearing model (shown in gray). As can be seen the simple
gaussian-smearing approach fails to describe the fine structure of particle jets,
and shows that our neural network based approach provides results of superior quality,
compared to traditional fast simulation methods.

\clearpage

\section{Differential characterisation of the image translation}

\begin{figure}[ht]
\centering
\includegraphics[width=0.2\textwidth]{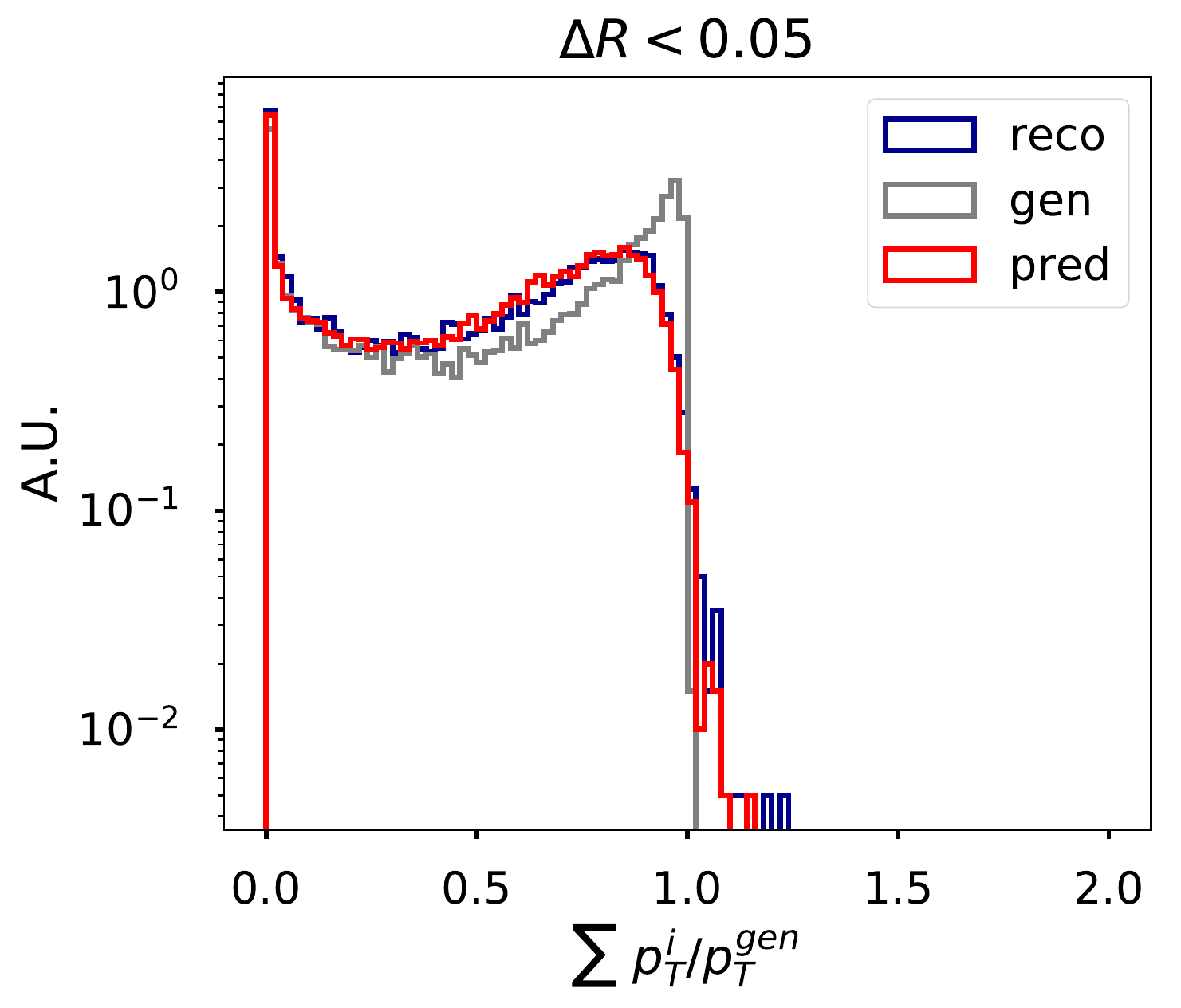} 
\includegraphics[width=0.2\textwidth]{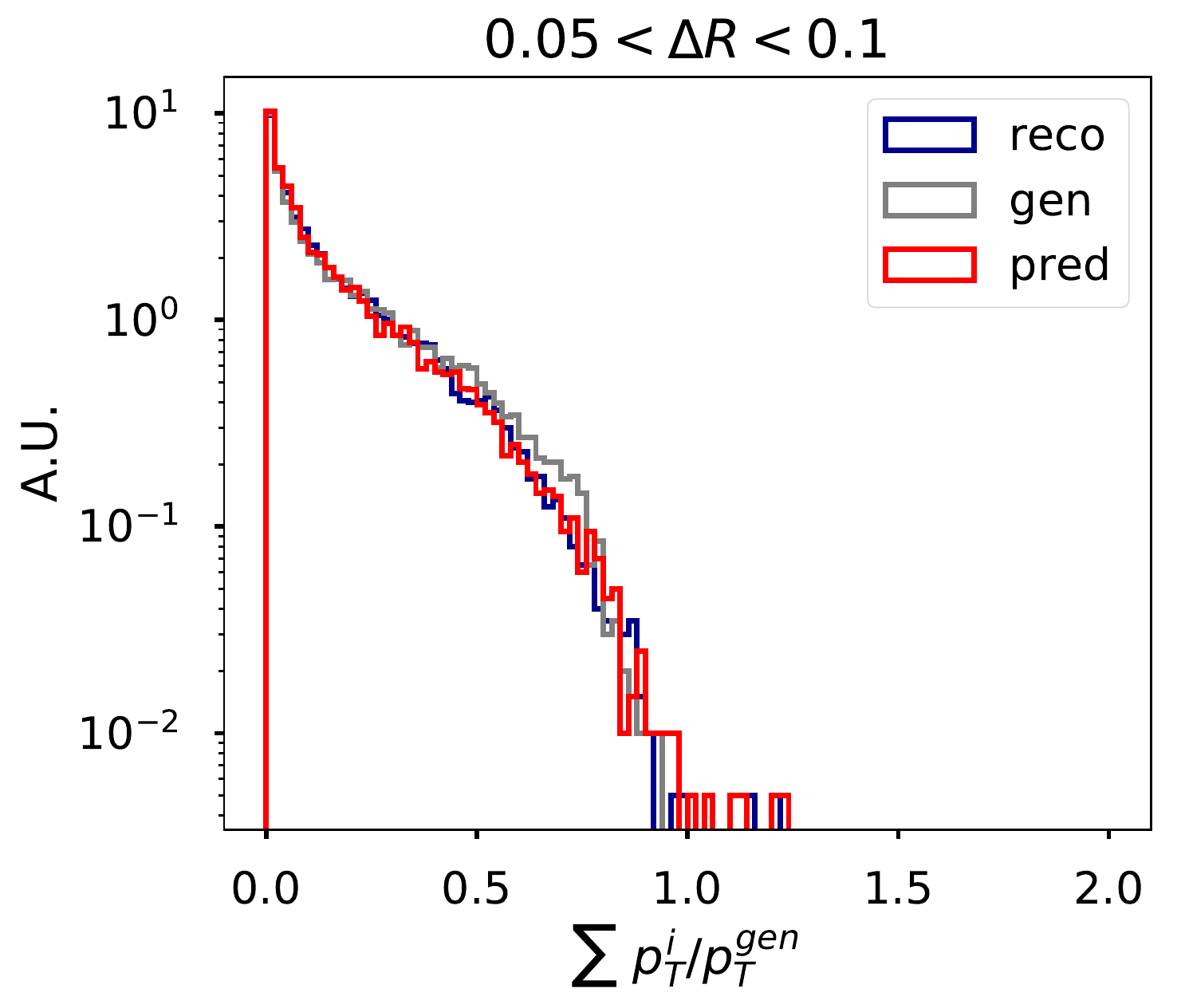} \\
\includegraphics[width=0.2\textwidth]{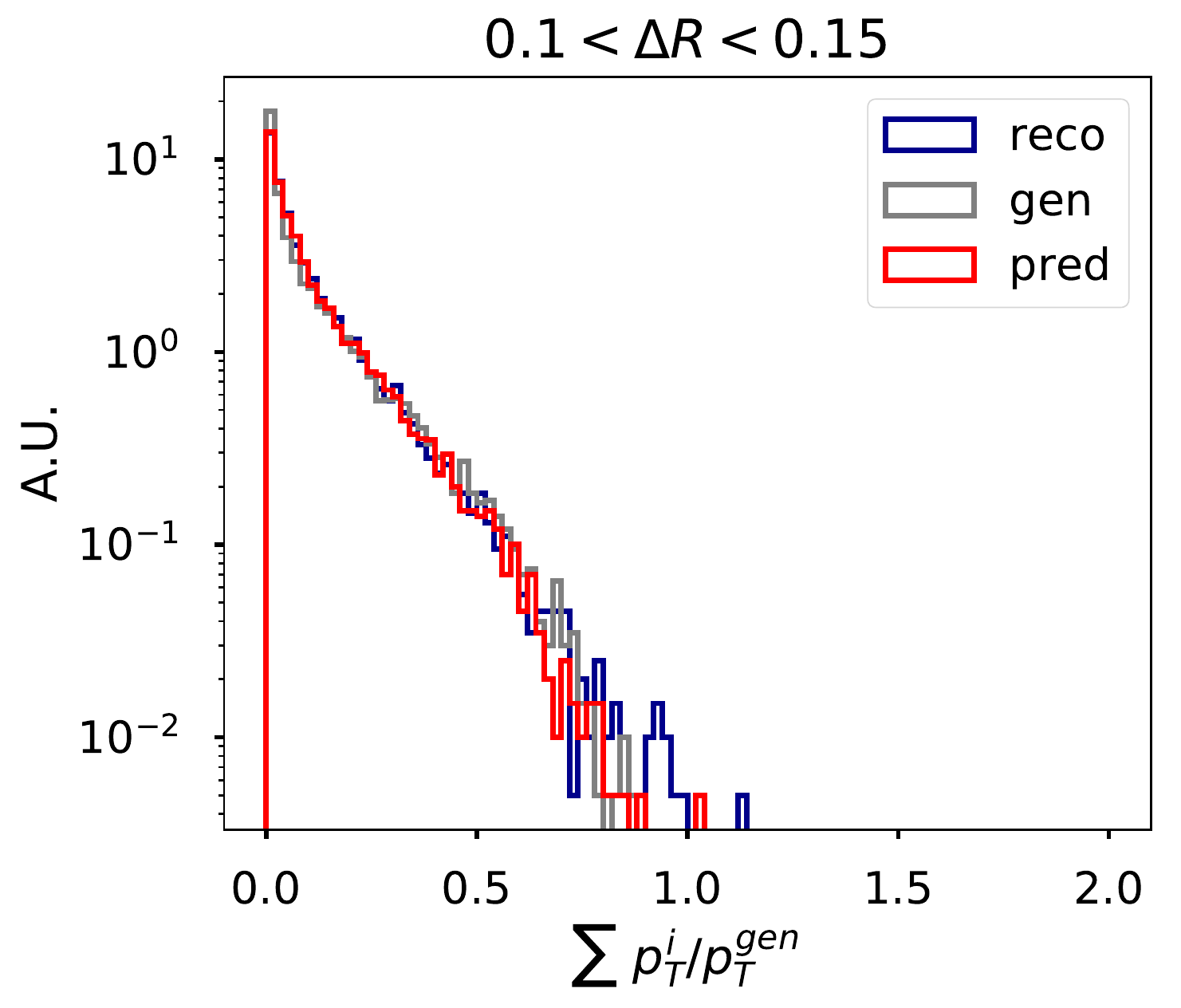} 
\includegraphics[width=0.2\textwidth]{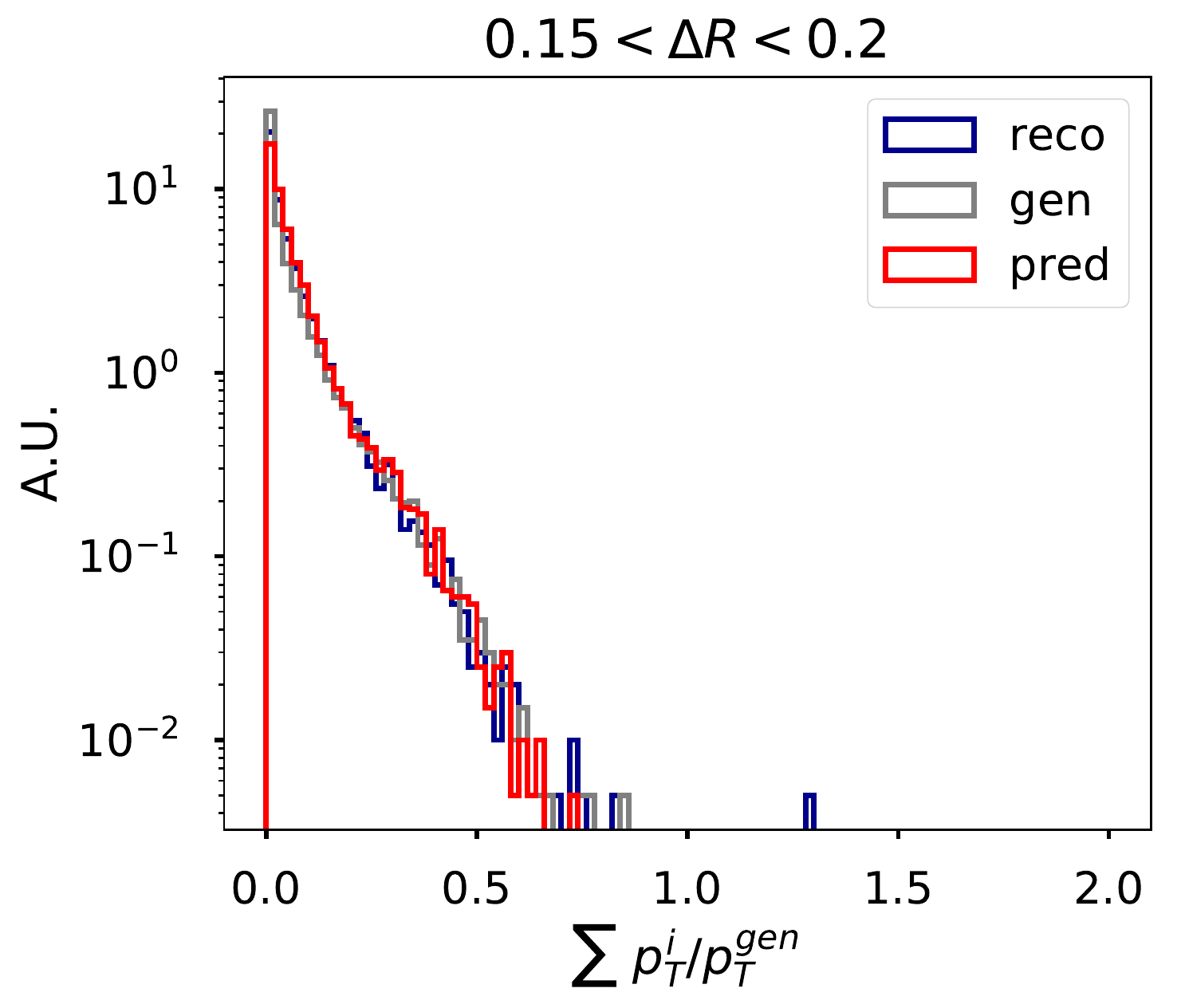} \\
\includegraphics[width=0.2\textwidth]{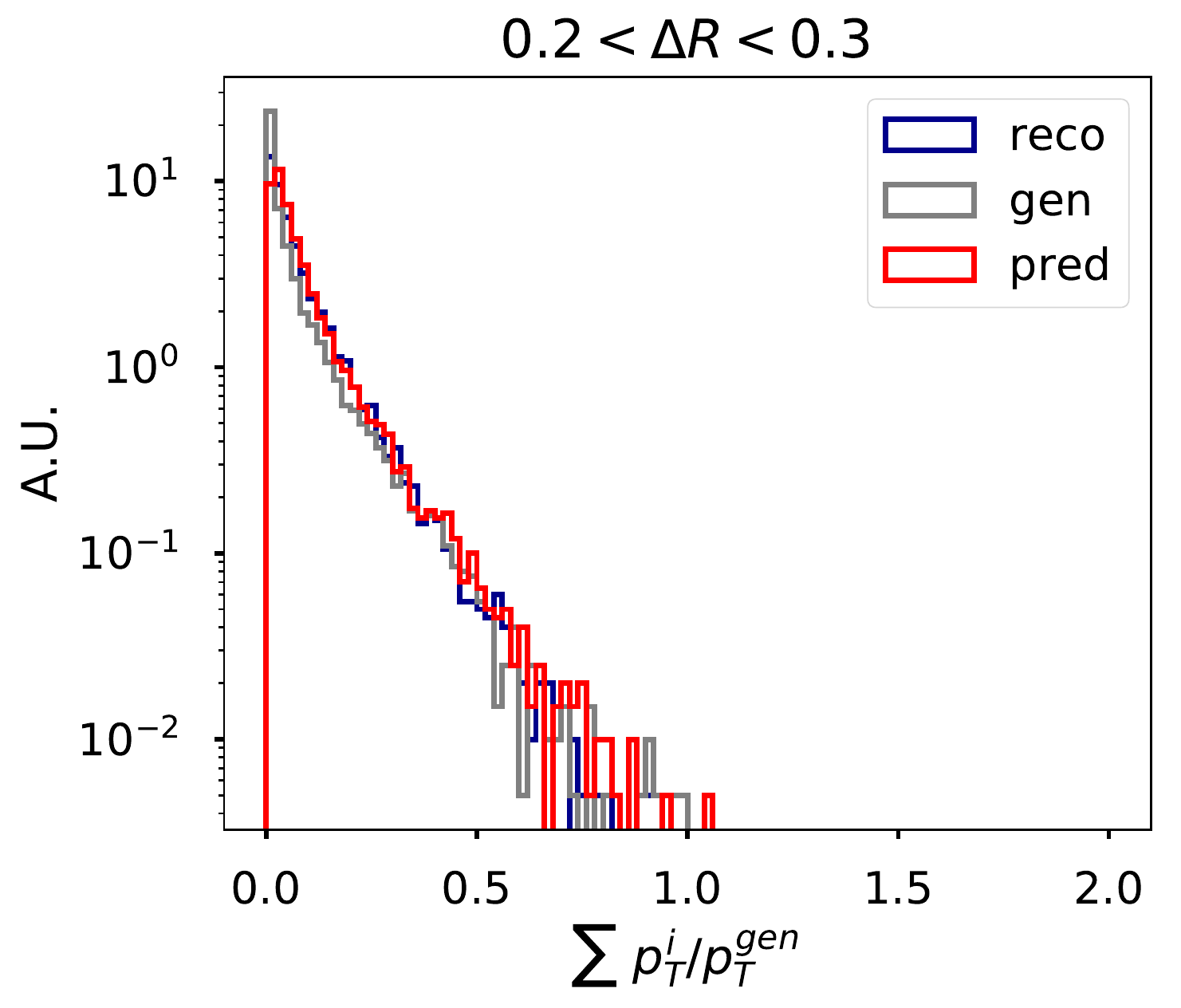} 
\includegraphics[width=0.2\textwidth]{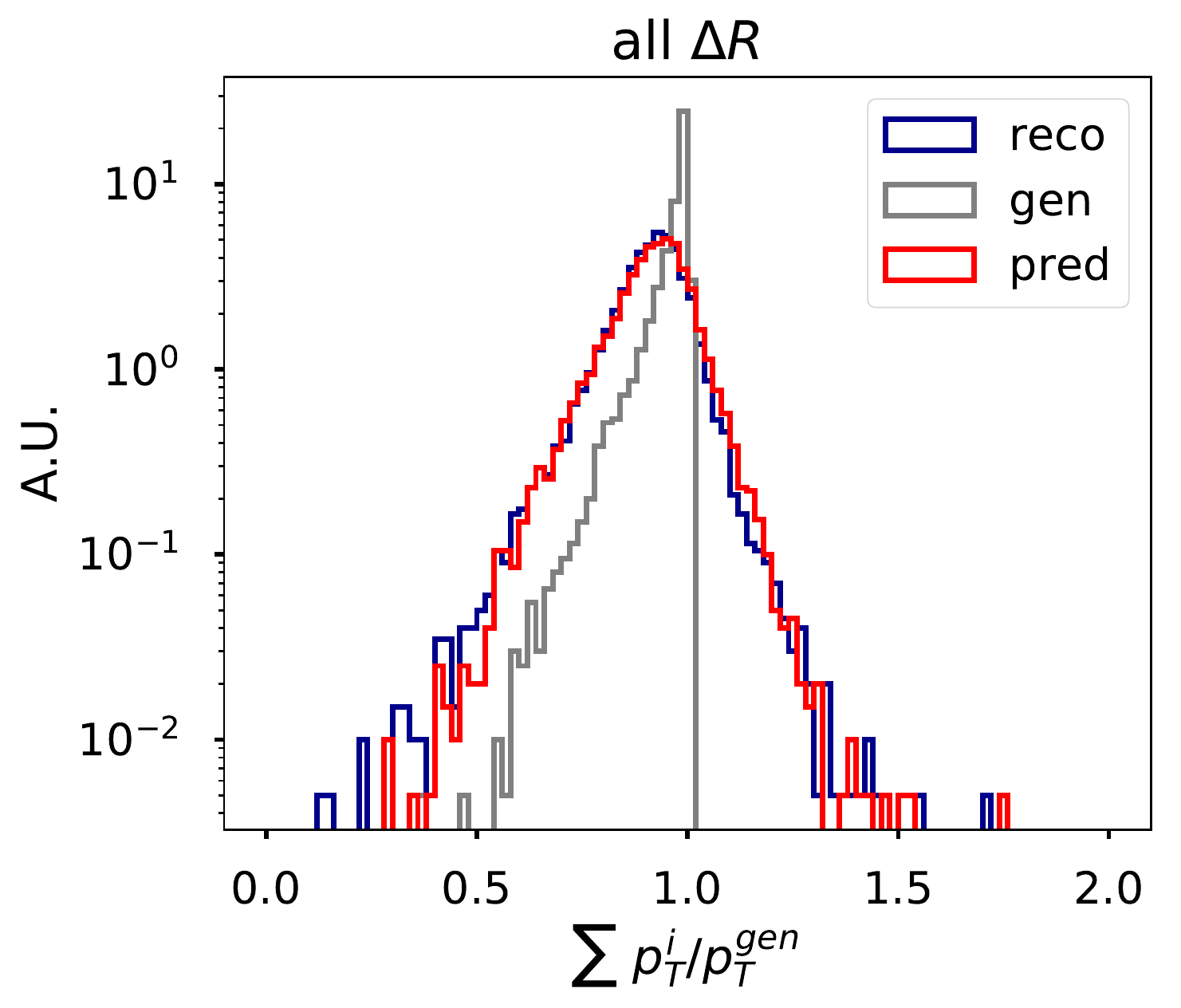}
\caption{\label{fig:sumpred_gen}
Aggregated pixel intensities for different rings in $\Delta \eta$--$\Delta \phi$. Blue
histograms are obtained from the full simulation and reconstruction chain; red ones are
obtained using the generative model; gray histograms show the quantities obtained before
detector simulation. 
}
\end{figure}

\begin{figure}[hb]
\centering
\includegraphics[width=0.2\textwidth]{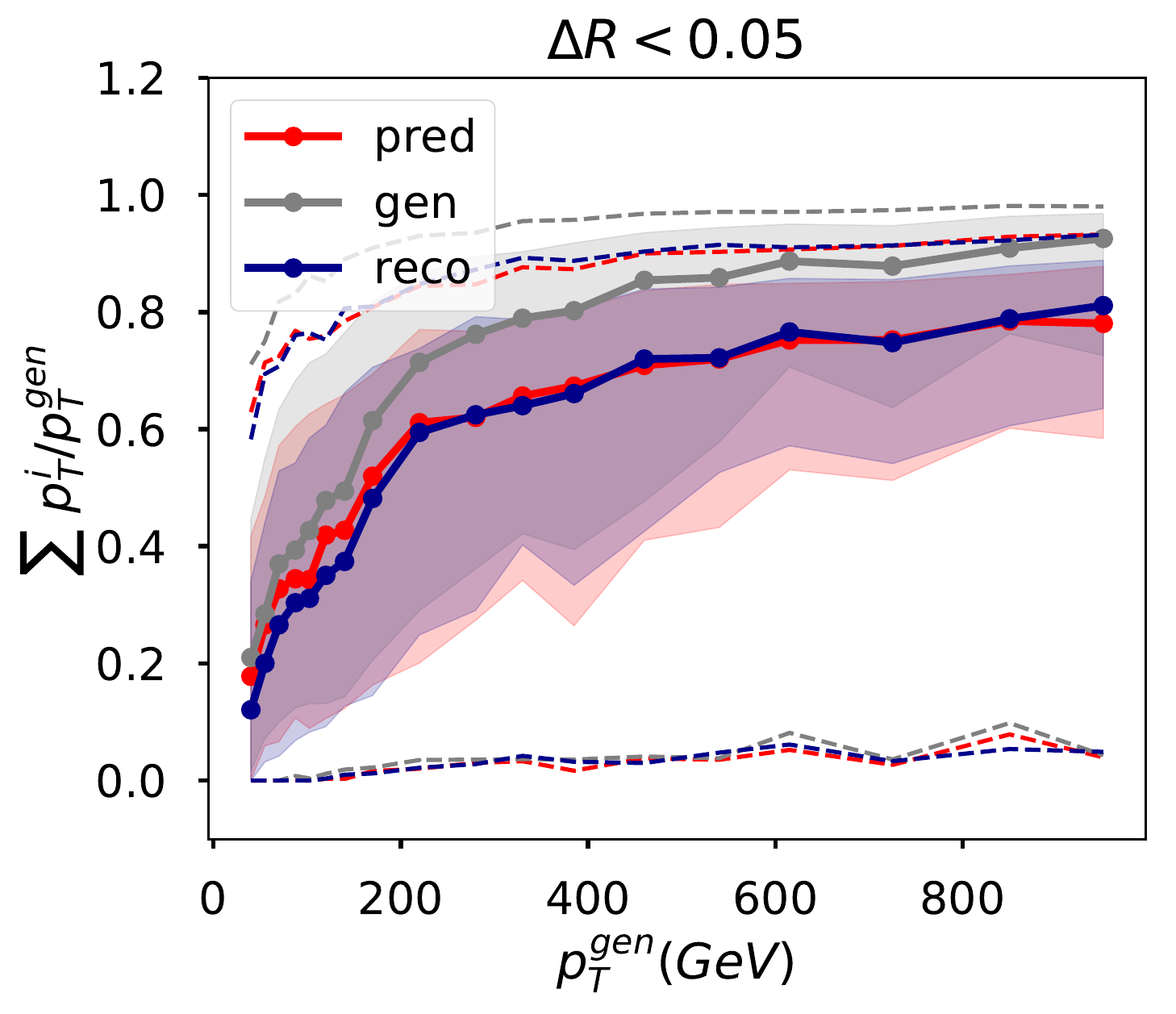} 
\includegraphics[width=0.2\textwidth]{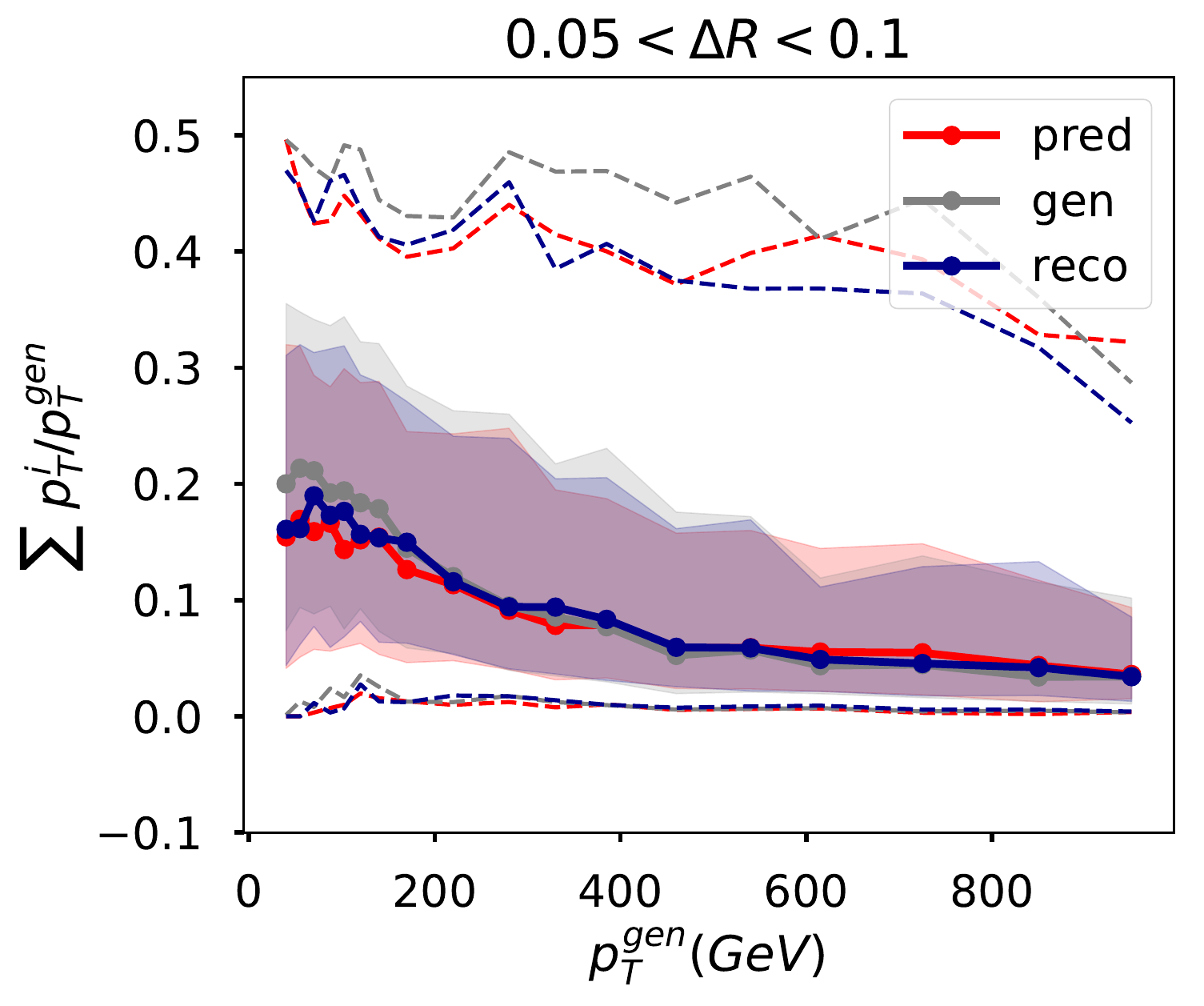} \\
\includegraphics[width=0.2\textwidth]{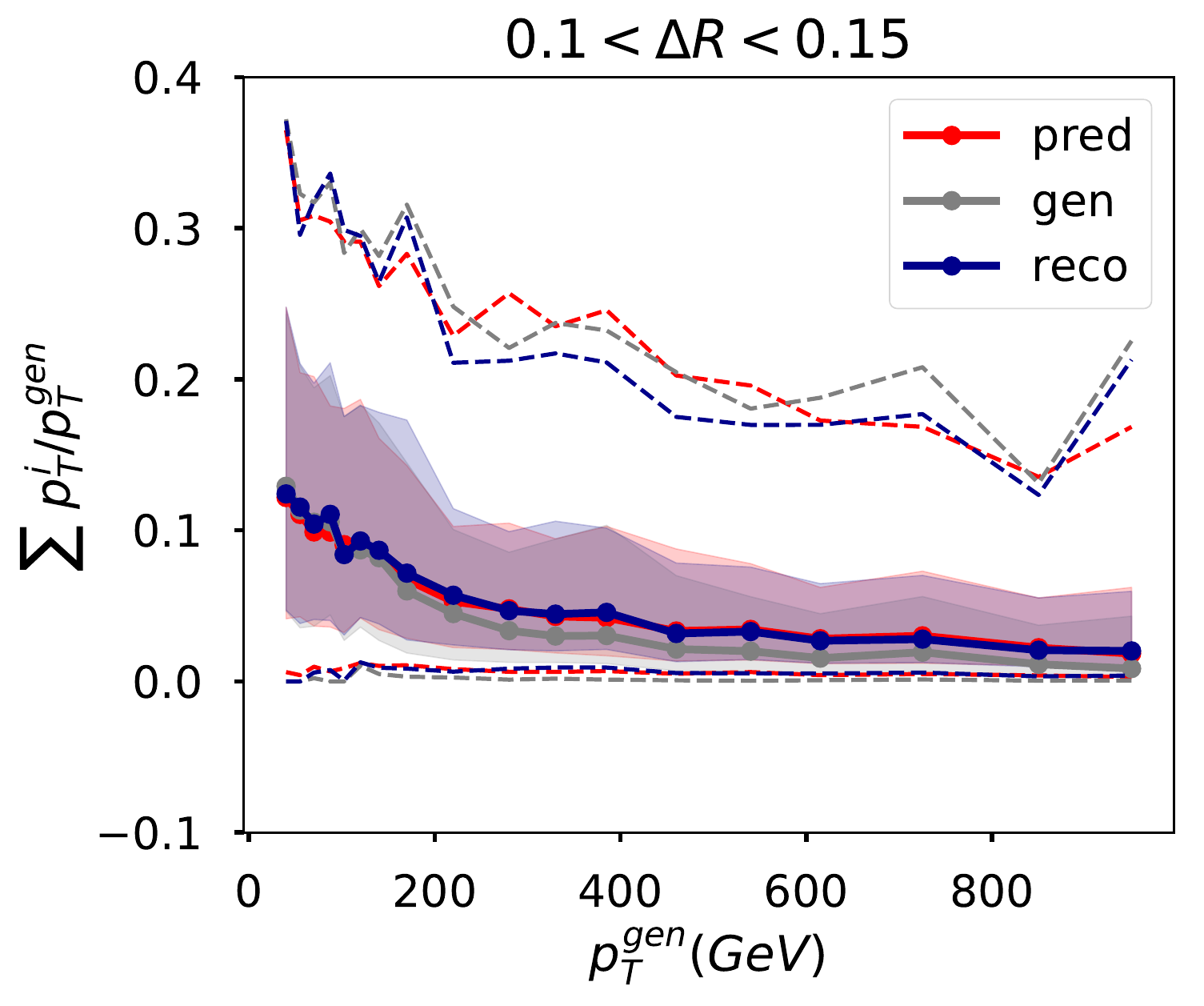} 
\includegraphics[width=0.2\textwidth]{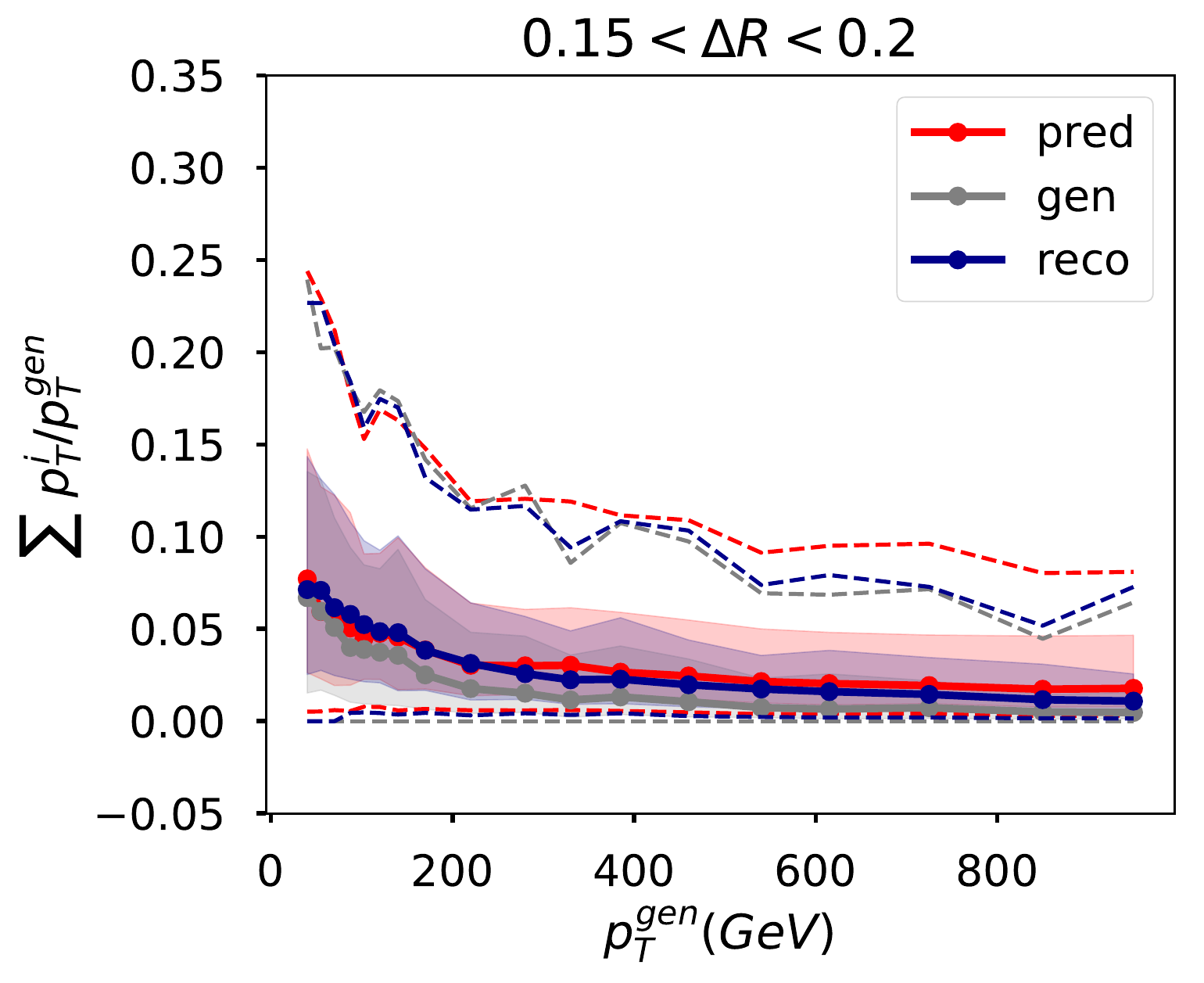} \\ 
\includegraphics[width=0.2\textwidth]{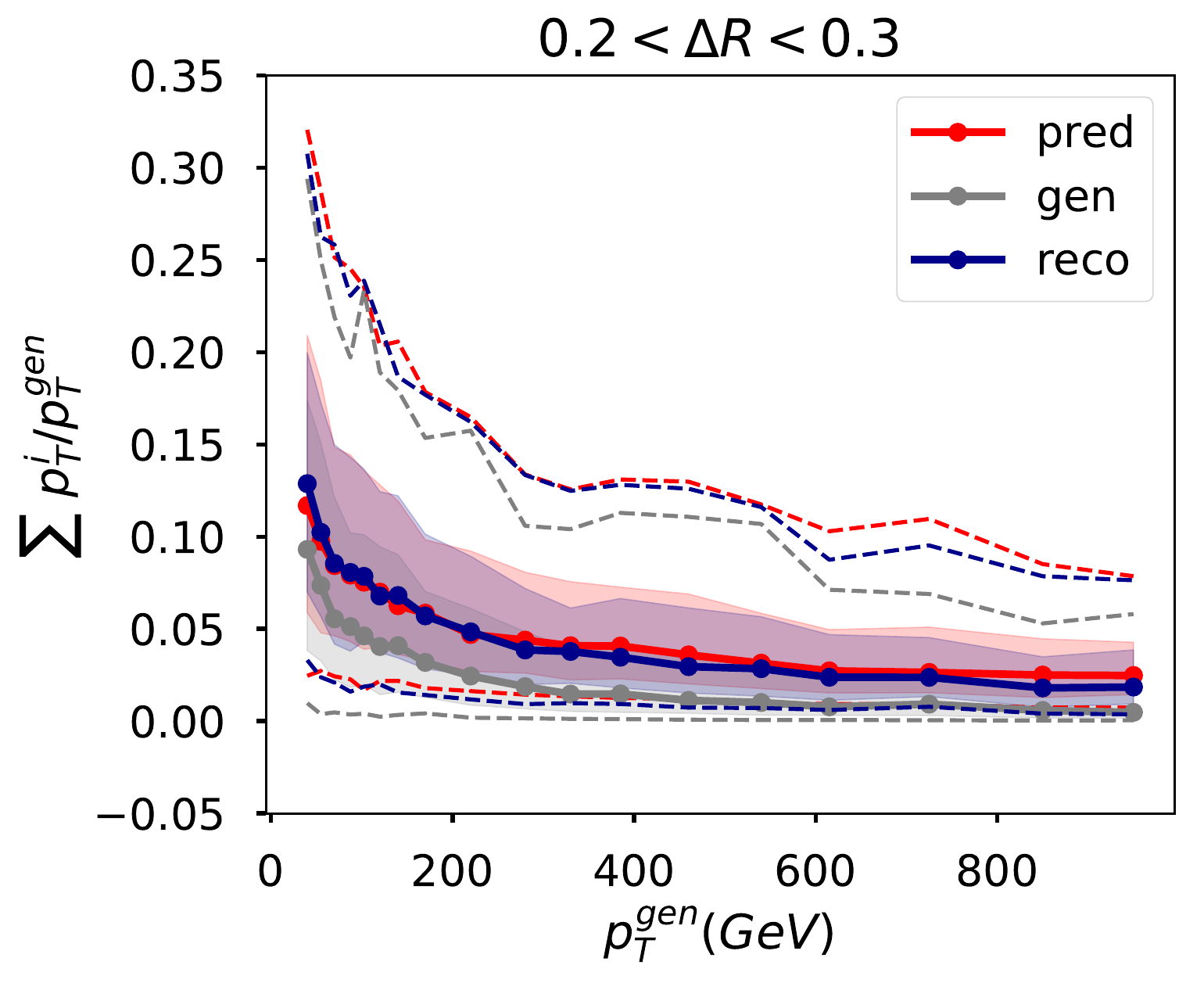} 
\includegraphics[width=0.2\textwidth]{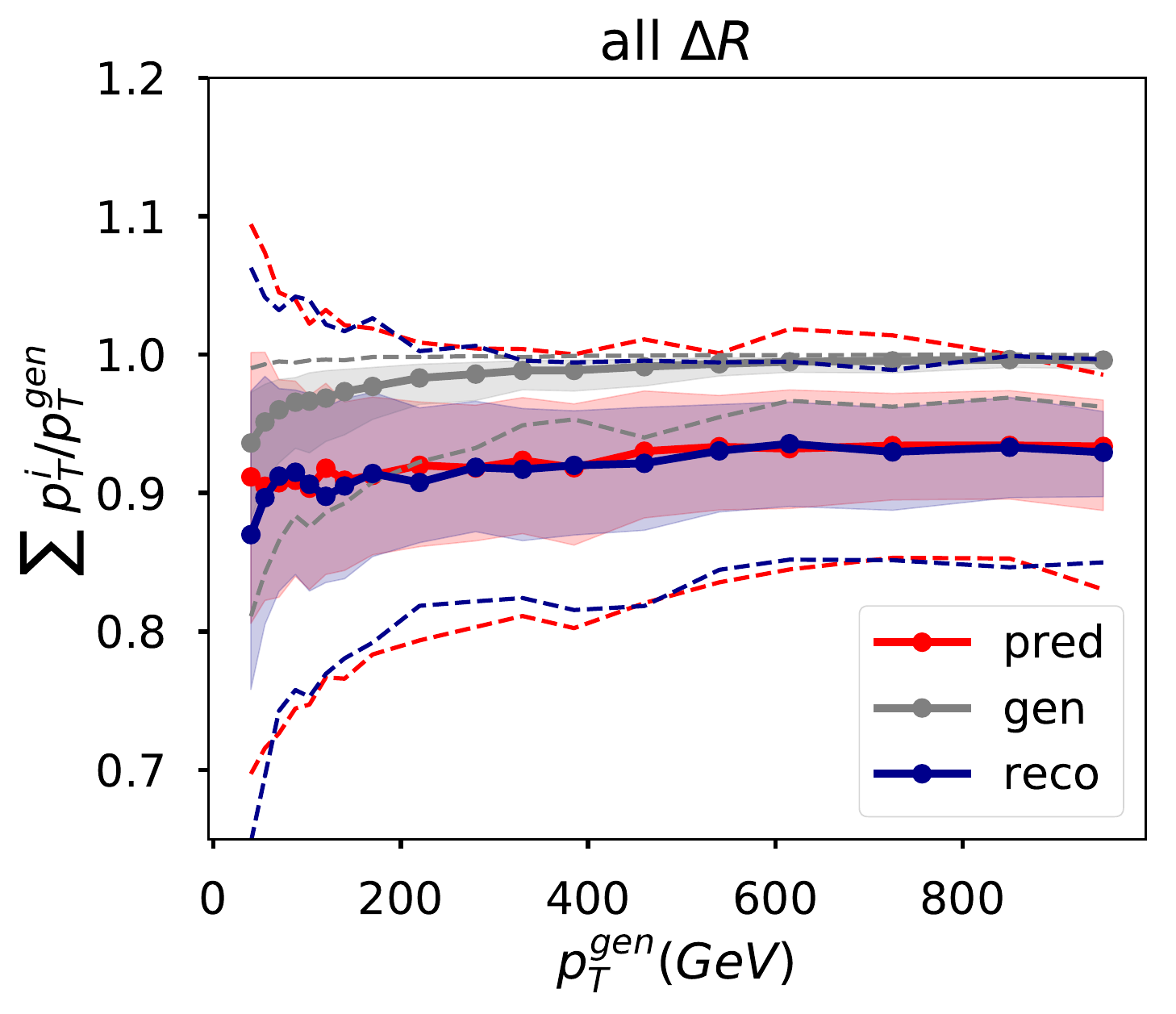}
\caption{\label{fig:ptpred_gen}
Evolution of the aggregated pixel intensities for different rings in $\Delta
\eta$--$\Delta \phi$ as a function of the particle level jet transverse momentum. Solid
lines represent the median of the distribution, filled regions show the inter-quartile
range, while dashed lines mark the 10\% and 90\% quantiles. 
Blue
histograms are obtained from the full simulation and reconstruction chain; red ones are
obtained using the generative model; gray histograms show the quantities obtained before
detector simulation.
}
\end{figure}

In this appendix we report a differential comparison of the jet images obtained at
particle level, reconstruction level, and through our generative model.
In figures~\ref{fig:sumpred_gen} and \ref{fig:ptpred_gen} we show the 
same quantities as shown in figures~\ref{fig:sumpred} and \ref{fig:ptpred},
respectively, but with the addition of results obtained using particle level jets.

%% file: main.bbl
\begin{thebibliography}{10}
\providecommand{\url}[1]{{#1}}
\providecommand{\urlprefix}{URL }
\expandafter\ifx\csname urlstyle\endcsname\relax
  \providecommand{\doi}[1]{DOI \discretionary{}{}{}#1}\else
  \providecommand{\doi}{DOI \discretionary{}{}{}\begingroup
  \urlstyle{rm}\Url}\fi

\bibitem{geant4_2003}
S.~Agostinelli, et~al., {\sc Geant4}-a simulation toolkit, Nuclear Instruments
  and Methods in Physics Research Section A \textbf{506}(3), 250  (2003).
\newblock \doi{10.1016/S0168-9002(03)01368-8}

\bibitem{ATLAS_exp}
G.~Aad, et~al., {The ATLAS Experiment at the CERN Large Hadron Collider}, JINST
  \textbf{3}, S08003 (2008).
\newblock \doi{10.1088/1748-0221/3/08/S08003}

\bibitem{CMS_exp}
S.~Chatrchyan, et~al., {The CMS Experiment at the CERN LHC}, JINST \textbf{3},
  S08004 (2008).
\newblock \doi{10.1088/1748-0221/3/08/S08004}

\bibitem{HL_LHC}
G.~Apollinari, et~al., {High-Luminosity Large Hadron Collider (HL-LHC)}
  (2017).
\newblock \doi{10.23731/CYRM-2017-004}

\bibitem{HSF_whitepaper}
A.A. Alves, Jr, et~al., {A Roadmap for HEP Software and Computing R\&D for the
  2020s}.
\newblock Tech. rep. (2017).
\newblock \urlprefix\url{https://arxiv.org/abs/1712.06982}

\bibitem{qcd30}
{\-/QCD\_Pt-30to50\_TuneZ2\_7TeV\_pythia6\-/Summer11LegDR\-PU\_S13\_START53\_LV6-v1\-/AODSIM}.
\newblock \doi{10.7483\/OPENDATA.CMS.Q3BX.69VQ}

\bibitem{qcd50}
{\-/QCD\_Pt-50to80\_TuneZ2\_7TeV\_pythia6\-/Summer11LegDR\-PU\_S13\_START53\_LV6-v1\-/AODSIM}.
\newblock \doi{10.7483\/OPENDATA.CMS.84VC.RU8W}

\bibitem{qcd80}
{\-/QCD\_Pt-80to120\_TuneZ2\_7TeV\_pythia6\-/Summer11LegDR\-PU\_S13\_START53\_LV6-v1\-/AODSIM}.
\newblock \doi{10.7483\/OPENDATA.CMS.PUTE.7H2H}

\bibitem{qcd120}
{\-/QCD\_Pt-120to170\_TuneZ2\_7TeV\_pythia6\-/Summer11LegDR\-PU\_S13\_START53\_LV6-v1\-/AODSIM}.
\newblock \doi{10.7483\/OPENDATA.CMS.QJND.HA88}

\bibitem{qcd170}
{\-/QCD\_Pt-170to300\_TuneZ2\_7TeV\_pythia6\-/Summer11LegDR\-PU\_S13\_START53\_LV6-v1\-/AODSIM}.
\newblock \doi{10.7483\/OPENDATA.CMS.WKRR.DCJP}

\bibitem{qcd300}
{\-/QCD\_Pt-300to470\_TuneZ2\_7TeV\_pythia6\-/Summer11LegDR\-PU\_S13\_START53\_LV6-v1\-/AODSIM}.
\newblock \doi{10.7483\/OPENDATA.CMS.X3XQ.USQR}

\bibitem{qcd470}
{\-/QCD\_Pt-470to600\_TuneZ2\_7TeV\_pythia6\-/Summer11LegDR\-PU\_S13\_START53\_LV6-v1\-/AODSIM}.
\newblock \doi{10.7483\/OPENDATA.CMS.BKTD.SGJX}

\bibitem{qcd600}
{\-/QCD\_Pt-600to800\_TuneZ2\_7TeV\_pythia6\-/Summer11LegDR\-PU\_S13\_START53\_LV6-v1\-/AODSIM}.
\newblock \doi{10.7483\/OPENDATA.CMS.EJT7.KSAY}

\bibitem{qcd800}
{\-/QCD\_Pt-800to1000\_TuneZ2\_7TeV\_pythia6\-/Summer11LegDR\-PU\_S13\_START53\_LV6-v1\-/AODSIM}.
\newblock \doi{10.7483\/OPENDATA.CMS.S3D5.KF2C}

\bibitem{qcd1000}
{\-/QCD\_Pt-1000to1400\_TuneZ2\_7TeV\_pythia6\-/Summer11LegDR\-PU\_S13\_START53\_LV6-v1\-/AODSIM}.
\newblock \doi{10.7483\/OPENDATA.CMS.96U2.3YAH}

\bibitem{qcd1400}
{\-/QCD\_Pt-1400to1800\_TuneZ2\_7TeV\_pythia6\-/Summer11LegDR\-PU\_S13\_START53\_LV6-v1\-/AODSIM}.
\newblock \doi{10.7483\/OPENDATA.CMS.RC9V.B5KX}

\bibitem{qcd1800}
{\-/QCD\_Pt-1800\_TuneZ2\_7TeV\_pythia6\-/Summer11LegDR\-PU\_S13\_START53\_LV6-v1\-/AODSIM}.
\newblock \doi{10.7483\/OPENDATA.CMS.CX2X.J3KW}

\bibitem{DPHEP}
R.~Mount, et~al., {Data Preservation in High Energy Physics}, Intermediate
  report of the ICFA-DPHEP Study Group  (2009).
\newblock \urlprefix\url{https://arxiv.org/abs/0912.0255}

\bibitem{goodfellow_generative_2014}
I.J. Goodfellow, J.~Pouget-Abadie, M.~Mirza, B.~Xu, D.~Warde-Farley, S.~Ozair,
  A.~Courville, Y.~Bengio, Generative {Adversarial} {Networks}  (2014).
\newblock \urlprefix\url{https://arxiv.org/abs/1406.2661}

\bibitem{gan_mix}
S.~Arora, R.~Ge, Y.~Liang, T.~Ma, Y.~Zhang, in \emph{Proceedings of the 34th
  International Conference on Machine Learning}, \emph{Proceedings of Machine
  Learning Research}, vol.~70, ed. by D.~Precup, Y.W. Teh (PMLR, International
  Convention Centre, Sydney, Australia, 2017), \emph{Proceedings of Machine
  Learning Research}, vol.~70, pp. 224--232.
\newblock \urlprefix\url{http://proceedings.mlr.press/v70/arora17a.html}

\bibitem{f_gan}
S.~{Nowozin}, B.~{Cseke}, R.~{Tomioka}, {f-GAN: Training Generative Neural
  Samplers using Variational Divergence Minimization}  (2016).
\newblock \urlprefix\url{https://arxiv.org/abs/1606.00709}

\bibitem{w_gan1}
I.~Gulrajani, F.~Ahmed, M.~Arjovsky, V.~Dumoulin, A.C. Courville, Improved
  training of wasserstein gans, CoRR \textbf{abs/1704.00028} (2017).
\newblock \urlprefix\url{http://arxiv.org/abs/1704.00028}

\bibitem{w_gan2}
M.~{Arjovsky}, S.~{Chintala}, L.~{Bottou}, {Wasserstein GAN}  (2017).
\newblock \urlprefix\url{https://arxiv.org/abs/1701.07875}

\bibitem{Louppe:2016ylz}
G.~Louppe, M.~Kagan, K.~Cranmer, Learning to pivot with adversarial networks
  (2016).
\newblock \urlprefix\url{https://arxiv.org/abs/1611.01046}

\bibitem{Shimmin:2017mfk}
C.~Shimmin, P.~Sadowski, P.~Baldi, E.~Weik, D.~Whiteson, E.~Goul, A.~Søgaard,
  {Decorrelated Jet Substructure Tagging using Adversarial Neural Networks},
  Phys. Rev. \textbf{D96}(7), 074034 (2017).
\newblock \doi{10.1103/PhysRevD.96.074034}

\bibitem{nips17_systs}
V.~Estrade, et~al., in \emph{{NIPS 2017 - workshop Deep Learning for Physical
  Sciences}} (Long Beach, United States, 2017), pp. 1--5

\bibitem{deOliveira:2017pjk}
L.~de~Oliveira, M.~Paganini, B.~Nachman, {Learning Particle Physics by Example:
  Location-Aware Generative Adversarial Networks for Physics Synthesis},
  Comput. Softw. Big Sci. \textbf{1}(1), 4 (2017).
\newblock \doi{10.1007/s41781-017-0004-6}

\bibitem{Paganini:2017hrr}
M.~Paganini, L.~de~Oliveira, B.~Nachman, {Accelerating Science with Generative
  Adversarial Networks: An Application to 3D Particle Showers in Multilayer
  Calorimeters}, Phys. Rev. Lett. \textbf{120}(4), 042003 (2018).
\newblock \doi{10.1103/PhysRevLett.120.042003}

\bibitem{nips17_calo}
F.~Carminati, et~al., in \emph{{NIPS 2017 - workshop Deep Learning for Physical
  Sciences}} (Long Beach, United States, 2017), pp. 1--5

\bibitem{Erdmann:2018kuh}
M.~Erdmann, L.~Geiger, J.~Glombitza, D.~Schmidt, {Generating and refining
  particle detector simulations using the Wasserstein distance in adversarial
  networks}  (2018).
\newblock \urlprefix\url{https://arxiv.org/abs/1802.03325}

\bibitem{pix2pix}
P.~Isola, J.~Zhu, T.~Zhou, A.A. Efros, Image-to-image translation with
  conditional adversarial networks, CoRR \textbf{abs/1611.07004} (2016).
\newblock \urlprefix\url{http://arxiv.org/abs/1611.07004}

\bibitem{improved_gans}
T.~Salimans, I.J. Goodfellow, W.~Zaremba, V.~Cheung, A.~Radford, X.~Chen,
  Improved techniques for training gans, CoRR \textbf{abs/1606.03498} (2016).
\newblock \urlprefix\url{http://arxiv.org/abs/1606.03498}

\bibitem{cgan}
M.~Mirza, S.~Osindero, Conditional generative adversarial nets, CoRR
  \textbf{abs/1411.1784} (2014).
\newblock \urlprefix\url{http://arxiv.org/abs/1411.1784}

\bibitem{deOliveira:2017rwa}
L.~de~Oliveira, M.~Paganini, B.~Nachman, in \emph{{18th International Workshop
  on Advanced Computing and Analysis Techniques in Physics Research (ACAT 2017)
  Seattle, WA, USA, August 21-25, 2017}} (2017)

\bibitem{delphes}
J.~de~Favereau, C.~Delaere, P.~Demin, A.~Giammanco, V.~Lemaître, A.~Mertens,
  M.~Selvaggi, {DELPHES 3, A modular framework for fast simulation of a generic
  collider experiment}, JHEP \textbf{02}, 057 (2014).
\newblock \doi{10.1007/JHEP02(2014)057}

\bibitem{pythia6}
T.~Sjostrand, S.~Mrenna, P.Z. Skands, {PYTHIA 6.4 Physics and Manual}, JHEP
  \textbf{05}, 026 (2006).
\newblock \doi{10.1088/1126-6708/2006/05/026}

\bibitem{antikt}
M.~Cacciari, G.P. Salam, G.~Soyez, The anti- k t jet clustering algorithm,
  Journal of High Energy Physics \textbf{2008}(04), 063 (2008).
\newblock \urlprefix\url{http://stacks.iop.org/1126-6708/2008/i=04/a=063}

\bibitem{fastjet}
M.~Cacciari, G.P. Salam, G.~Soyez, {FastJet User Manual}, Eur. Phys. J.
  \textbf{C72}, 1896 (2012).
\newblock \doi{10.1140/epjc/s10052-012-1896-2}

\bibitem{data}
The data used in this work is available through the following electronic
  reference.
\newblock \doi{10.5281/zenodo.1467678}.
\newblock \urlprefix\url{http://doi.org/10.5281/zenodo.1467678}

\bibitem{hinton_bsn}
H.~Geoffry, et~al.
\newblock Neural networks for machine learning.
\newblock \url{www.coursera.org}

\bibitem{lecun-98}
Y.~LeCun, L.~Bottou, Y.~Bengio, P.~Haffner, Gradient-based learning applied to
  document recognition, Proceedings of the IEEE \textbf{86}(11), 2278 (1998)

\bibitem{Unet}
O.~Ronneberger, P.~Fischer, T.~Brox, U-net: Convolutional networks for
  biomedical image segmentation, CoRR \textbf{abs/1505.04597} (2015).
\newblock \urlprefix\url{http://arxiv.org/abs/1505.04597}

\bibitem{BN}
S.~Ioffe, C.~Szegedy, Batch normalization: Accelerating deep network training
  by reducing internal covariate shift, CoRR \textbf{abs/1502.03167} (2015).
\newblock \urlprefix\url{http://arxiv.org/abs/1502.03167}

\bibitem{LReLU}
A.L. Maas, A.Y. Hannun, A.Y. Ng, in \emph{Proc. icml}, vol.~30 (2013), vol.~30,
  p.~3

\bibitem{code}
A software package related to this work is available through the following
  electronic reference.
\newblock \doi{10.5281/zenodo.1467665}.
\newblock \urlprefix\url{http://doi.org/10.5281/zenodo.1467665}

\bibitem{tf}
M.~Abadi, P.~Barham, J.~Chen, Z.~Chen, A.~Davis, J.~Dean, M.~Devin,
  S.~Ghemawat, G.~Irving, M.~Isard, M.~Kudlur, J.~Levenberg, R.~Monga,
  S.~Moore, D.G. Murray, B.~Steiner, P.A. Tucker, V.~Vasudevan, P.~Warden,
  M.~Wicke, Y.~Yu, X.~Zhang, Tensorflow: {A} system for large-scale machine
  learning, CoRR \textbf{abs/1605.08695} (2016).
\newblock \urlprefix\url{http://arxiv.org/abs/1605.08695}

\bibitem{keras}
F.~Chollet, et~al.
\newblock Keras.
\newblock \url{https://keras.io} (2015)

\bibitem{adam}
D.P. Kingma, J.~Ba, Adam: {A} method for stochastic optimization, CoRR
  \textbf{abs/1412.6980} (2014).
\newblock \urlprefix\url{http://arxiv.org/abs/1412.6980}

\bibitem{Pandolfi2013}
F.~Pandolfi, {Search for the Standard Model Higgs boson in the H $\to$ ZZ $\to
  \ell ^ + \ell ^ - q\bar q$ decay channel at CMS on 4.6 fb$^{-1}$ of 7 TeV
  proton-proton collision data}, The European Physical Journal Plus
  \textbf{128}(10), 117 (2013).
\newblock \doi{10.1140/epjp/i2013-13117-x}

\bibitem{subjett}
J.~Thaler, K.~Van~Tilburg, Identifying boosted objects with n-subjettiness,
  Journal of High Energy Physics \textbf{2011}(3), 15 (2011).
\newblock \doi{10.1007/JHEP03(2011)015}

\bibitem{fastjet1}
M.~Cacciari, G.P. Salam, G.~Soyez, {FastJet User Manual}, Eur. Phys. J.
  \textbf{C72}, 1896 (2012).
\newblock \doi{10.1140/epjc/s10052-012-1896-2}

\bibitem{fastjet2}
M.~Cacciari, G.P. Salam, {Dispelling the $N^{3}$ myth for the $k_t$
  jet-finder}, Phys. Lett. \textbf{B641}, 57 (2006).
\newblock \doi{10.1016/j.physletb.2006.08.037}

\bibitem{cms_fast_sim}
R.~Rahmat, R.~Kroeger, A.~Giammanco, The fast simulation of the cms experiment,
  Journal of Physics: Conference Series \textbf{396}(6), 062016.
\newblock \doi{10.1088/1742-6596/396/6/062016}

\end{thebibliography}
